\documentclass[aps,floats,pre]{revtex4-2}
\usepackage{amsmath,amssymb}
\usepackage{graphicx,epsfig}
\usepackage[greek,english]{babel}
\usepackage{bbold}

\makeatletter\AtBeginDocument{\let\@elt\relax}\makeatother

\DeclareMathOperator{\erfc}{erfc}

\begin{document}
\bibliographystyle {plain}

\pdfoutput=1
\def\oppropto{\mathop{\propto}} 
\def\opsimeq{\mathop{\simeq}}
\def\opoverderline{\mathop{\overline}}
\def\operarrow{\mathop{\longrightarrow}}
\def\opsim{\mathop{\sim}}

\def\opmin{\mathop{\min}} 
\def\opmax{\mathop{\max}} 
\def\oplim{\mathop{\lim}}

\title{ Joint distribution of two Local Times for diffusion processes \\
with the application to the construction of various conditioned processes} 

\author{Alain Mazzolo}
\affiliation{Universit\'e Paris-Saclay, CEA, Service d'\'Etudes des R\'eacteurs et de Math\'ematiques Appliqu\'ees, 91191, Gif-sur-Yvette, France}

\author{C\'ecile Monthus}
\affiliation{Universit\'e Paris-Saclay, CNRS, CEA, Institut de Physique Th\'eorique, 91191 Gif-sur-Yvette, France}


\begin{abstract}
For a diffusion process $X(t)$ of drift $\mu(x)$ and of diffusion coefficient $D=1/2$, we study the joint distribution of the two local times $A(t)= \int_{0}^{t} d\tau \delta(X(\tau))  $ and $B(t)= \int_{0}^{t} d\tau \delta(X(\tau)-L) $ at positions $x=0$ and $x=L$, as well as the simpler statistics of their sum $ \Sigma(t)=A(t)+B(t)$. Their asymptotic statistics for large time $t \to + \infty$ involves two very different cases : (i) when the diffusion process $X(t)$ is transient, the two local times $[A(t);B(t)]$ remain finite random variables $[A^*(\infty),B^*(\infty)]$ and we analyze their limiting joint distribution ; (ii) when the diffusion process $X(t)$ is recurrent, we describe the large deviations properties of the two intensive local times $a = \frac{A(t)}{t}$ and $b = \frac{B(t)}{t}$ and of their intensive sum $\sigma = \frac{\Sigma(t)}{t}=a+b$. These properties are then used to construct various conditioned processes $[X^*(t),A^*(t),B^*(t)]$ satisfying certain constraints involving the two local times, thereby generalizing our previous work [J. Stat. Mech. (2022) 103207] concerning the conditioning with respect to a single local time $A(t)$. In particular for the infinite time horizon $T \to +\infty$, we consider the conditioning towards the finite asymptotic values $[A^*(\infty),B^*(\infty)]$ or $\Sigma^*(\infty) $, as well as the conditioning towards the intensive values $[a^*,b^*] $ or $\sigma^*$, that can be compared with the appropriate 'canonical conditioning' based on the generating function of the local times in the regime of large deviations. This general construction is then applied to the simplest case where the unconditioned diffusion is the Brownian motion of uniform drift $\mu$.

\end{abstract}

\maketitle

\section{ Introduction }


\subsection{ Local Times as the basic time-additive observables for diffusion processes  }

For a one-dimensional diffusion process $X(t)$, two essential time-additive observables 
of the stochastic trajectory $X(0 \leq \tau \leq t)$ are :

(i) the occupation time $O_{[x_L,x_R]}(t) $ 
of the space interval $[x_L,x_R]$ during the time window $[0,t]$ 
\begin{eqnarray}
O_{[x_L,x_R]}(t)= \int_{0}^{t} d\tau \theta(x_L \leq X(\tau) \leq x_R )
\label{occupation}
\end{eqnarray}
that belongs to the interval $0 \leq O_{[x_L,x_R]}(t) \leq t$.

(ii) the local time $A_{x}(t) $ at the position $x$ during the time window $[0,t]$
 (see the mathematical review \cite{bjork} and references therein) 
\begin{eqnarray}
A_{x}(t)= \int_{0}^{t} d\tau \delta(X(\tau)-x) 
\label{localtimex}
\end{eqnarray}
that has for physical dimension $\frac{Time}{Length}$ (in contrast to the physical dimension Time for Eq. \ref{occupation})
and that belongs to $[0,+\infty[$ with no upper bound.
The direct links with the occupation times of Eq. \ref{occupation}
can be summarized as follows.
On one hand, the local time $A_{x}(t) $ of Eq. \ref{localtimex} can be constructed from the occupation time $O_{[x-\epsilon,x+\epsilon]}(t) $
of the space interval $ [x-\epsilon,x+\epsilon]$ of size $(2 \epsilon)>0$ centered at the position $x$ in the limit $\epsilon \to 0^+$
\begin{eqnarray}
A_{x}(t)= \int_{0}^{t} d\tau \oplim_{\epsilon \to 0^+} 
\left( \frac{\theta(x-\epsilon \leq X(\tau) \leq x+\epsilon )}{2 \epsilon } \right) 
=  \oplim_{\epsilon \to 0^+} 
\left( \frac{ O_{[x-\epsilon,x+\epsilon]}(t)}{2 \epsilon } \right) 
\label{localtimexeps}
\end{eqnarray}
On the other hand, the occupation time $O_{[x_L,x_R]}(t) $ 
can be reconstructed from the local time $A_{x}(t) $
 for all the internal positions $x \in [x_L,x_R]$
 \begin{eqnarray}
O_{[x_L,x_R]}(t)=  \int_{0}^{t} d\tau \int_{x_L}^{x_R} dx \delta(X(\tau)-x)  = \int_{x_L}^{x_R} dx A_{x}(t)
\label{occupationA}
\end{eqnarray}
As a consequence, both the occupation times and the local times have attracted a lot of interest recently
in the physics literature for many different contexts (see
\cite{occupationsinai,bressloff_occupation,grebenkov_paradigm,grebenkov_surface,grebenkov_2local,grebenkov_planar,grebenkov_encounter,bressloff_kac,bressloff_resetting,bressloff_surface,grebenkov_depletion,grebenkov_smalltarget,bressloff_spectral,bressloff_accumulation} and references therein).

More generally, the local times $A_{x}(t) $ of Eq. \ref{localtimex}
allow to reconstruct any time-additive observable involving an arbitrary function $f(.)$
 \begin{eqnarray}
  \int_{0}^{t} d\tau f(X(\tau) )  = \int_{-\infty}^{+\infty} dx f(x) A_{x}(t)
\label{arbitrary}
\end{eqnarray}
and can be thus considered as the basic time-additive observables.
Time-additive observables like Eq. \ref{arbitrary} are of course interesting for their own statistical properties, 
but they can also be used to construct conditioned processes as we now recall.


\subsection{ Reminder on the conditioning of stochastic processes with respect to time-additive observables}

Since its introduction by Doob \cite{refDoob,refbookDoob},
the conditioning of stochastic processes (see the mathematical books \cite{refbookKarlin,refbookRogers,Borodin}
and the physics recent review \cite{refMajumdarOrland}) have found many applications
in various fields like ecology \cite{refHorne}, finance \cite{refBrody} or
nuclear engineering \cite{refMulatier,refbookPazsit}. 
Over the years, many different conditioning constraints have been considered, including 
 the Brownian excursion \cite{refMajumdarExcursion,refChung}, the Brownian meander \cite{refMajumdarMeander},
the taboo processes \cite{refKnight,refPinsky,refKorzeniowski,refGarbaczewski,refAdorisio,refAlainTaboo},
or non-intersecting Brownian bridges \cite{grela}.
 Let us also mention the conditioning in the presence of killing rates \cite{refbookKarlin,Karlin1982,Karlin1983,Frydman,Steinsaltz,kolb,Evans2019,tryphon_killing,us_DoobKilling}
 or when the killing occurs only via an absorbing boundary condition
 \cite{refBaudoin,refMultiEnds,us_DoobFirstPassage,us_DoobFirstEncounter}.
Stochastic bridges have been also studied for many other Markov processes, 
including various diffusions processes \cite{henri,refSzavits,delarue},
discrete-time random walks and L\'evy flights \cite{refGarbaczewski_Levy,bruyne_discrete,Aguilar}, 
continuous-time Markov jump processes \cite{Aguilar},
run-and-tumble trajectories \cite{bruyne_run}, or
processes with resetting \cite{refdeBruyne2022}.

For the present work, the recent important generalization 
concerns the conditioning with respect to global dynamical constraints
involving time-additive observables of the stochastic trajectories. In particular, the conditioning 
on the area has been studied via various methods
for Brownian processes or bridges \cite{refMazzoloJstat} and for Ornstein-Uhlenbeck bridges \cite{Alain_OU}.
The conditioning on the area and on other time-additive observables has been then analyzed 
both for the Brownian motion and for discrete-time random walks \cite{refdeBruyne2021}.
This approach has been generalized \cite{c_microcanonical} to various types of discrete-time or continuous-time Markov processes, while the time-additive observable
can involve both the time spent in each configuration and the increments of the Markov process.
This general reformulation of the 'microcanonical conditioning', where the time-additive observable is constrained
to reach a given value after the finite time window $T$, allows to make the link \cite{c_microcanonical} 
with the 'canonical conditioning' based on generating functions of additive observables
that has been much studied recently in the field of dynamical large deviations
of Markov processes over a large time-window $T$ \cite{peliti,derrida-lecture,tailleur,sollich_review,lazarescu_companion,lazarescu_generic,jack_review,vivien_thesis,lecomte_chaotic,lecomte_thermo,lecomte_formalism,lecomte_glass,kristina1,kristina2,jack_ensemble,simon1,simon2,simon3,Gunter1,Gunter2,Gunter3,Gunter4,chetrite_canonical,chetrite_conditioned,chetrite_optimal,chetrite_HDR,touchette_circle,touchette_langevin,touchette_occ,touchette_occupation,derrida-conditioned,derrida-ring,bertin-conditioned,garrahan_lecture,Vivo,chemical,touchette-reflected,touchette-reflectedbis,c_lyapunov,previousquantum2.5doob,quantum2.5doob,quantum2.5dooblong,c_ruelle,lapolla,chabane,chabane_thesis}.
The equivalence between the 'microcanonical conditioning'
and the 'canonical conditioning' at the level of the large deviations for large time $T$
is explained in detail in the two complementary papers \cite{chetrite_conditioned,chetrite_optimal}
and in the HDR thesis \cite{chetrite_HDR}.


\subsection{ Goals and organization of the present paper }

As recalled in the previous subsection, 
the methods to construct conditioned processes with respect to time-additive observables
are now well-established, both in the 'microcanonical' and in the 'canonical' perspectives. 
However, the concrete application to specific time-additive observables like Eq. \ref{arbitrary}
for given Markov processes remains often challenging. 
It is thus important to identify the cases where conditioned processes can be
explicitly constructed and what level of technicality is required to perform the computations.

At first sight, the delta function that enters the definition of the local time in Eq. \ref{localtimex}
 might appear as very singular.
However, as in quantum mechanics where delta impurities are well-known 
to be much simpler than smoother potentials, the delta function in Eq. \ref{localtimex} is actually a huge technical simplification with respect to the arbitrary general additive observable of Eq. \ref{arbitrary}.
 Indeed, the exact Dyson equation associated to a single delta impurity allows to analyze the statistics of a single local time in terms of the properties of the propagator $G_t(x \vert x_0)$ of the diffusion process $X(t)$ alone, as recalled in detail in \cite{us_LocalTime}.
In the present paper, it will be thus interesting to use similarly the 
exact Dyson equation associated to two single delta impurities in order to characterize the
joint statistics of the two Local Times $A(t)=A_{x=0}(t)$ and $B(t)=A_{x=L}(t)$ at positions $x=0$ and $x=L$
\begin{eqnarray}
A(t) && \equiv \int_0^t d\tau \delta ( X(\tau) ) 
\nonumber \\
B(t) && \equiv \int_0^t d\tau \delta ( X(\tau)-L ) 
\label{defab}
\end{eqnarray}
for a diffusion process $X(t)$ of diffusion coefficient $D=1/2$ with an arbitrary position-dependent drift $\mu(x)$,
thereby generalizing the explicit joint statistics of the two local boundary times for the Brownian motion 
with no drift $\mu(x)=0$
on the interval $[0,L]$ with reflecting boundary conditions at $x=0$ and $x=L$ computed recently \cite{grebenkov_2local}.
The paper is organized as follows.

In section \ref{sec_xab}, we focus on the joint dynamics of the position $X(t)$ and of the two local times $A(t)$ and $B(t)$ of Eq. \ref{defab}
satisfying the Ito Stochastic Differential System involving the Wiener process $W(t)$
\begin{eqnarray}
dX(t) && =  \mu( X(t) ) dt + dW(t)
\nonumber \\
dA(t) && = \delta ( X(t) ) dt
\nonumber \\
dB(t) && = \delta ( X(t)-L ) dt
\label{itoab}
\end{eqnarray}
We use the Feynman-Kac formula and the Dyson equation in order
 to compute explicitly the time-Laplace-transform of parameter $s$
of the joint distribution $P_t(x,A,B \vert x_0) $ to see $[X(t)=x;A(t)=A,B(t)=B$ when starting at 
$[X(0)=x_0;A(0)=0;B(0)=0]$
\begin{eqnarray}
 {\hat P}_{s} (x,A,B \vert x_0) 
 \equiv \int_0^{+\infty} dt e^{-st }  P_t (x,A,B  \vert x_0)
\label{laplacesingledef}
\end{eqnarray}

In section \ref{sec_ab}, we focus on the joint distribution $\Pi_t(A,B \vert x_0)  $ 
of the two local times $[X(t)=x;A(t)=A,B(t)=B$ at time $t$ when starting at 
$[X(0)=x_0;A(0)=0;B(0)=0]$
\begin{eqnarray}
 \Pi_t(A,B \vert x_0)  \equiv \int_{-\infty}^{+\infty} dx  P_t(x,A,B \vert x_0)
\label{propagABalonedef}
\end{eqnarray}
and we compute explicitly the time-Laplace-transform of parameter $s$
\begin{eqnarray}
 {\hat \Pi}_{s} (A,B  \vert x_0) && \equiv \int_0^{+\infty} dt e^{-st }  \Pi_t(A,B  \vert x_0)
=  \int_{-\infty}^{+\infty} dx  {\hat P}_{s} (x,A,B \vert x_0) 
\label{laplacePidef}
\end{eqnarray}

In section \ref{sec_larget}, we analyse the behavior for large time $t$
 of the joint distribution $\Pi_t(A,B \vert x_0)  $.
When the diffusion process $X(t)$ is transient, the two local times $(A,B)$
remain finite random variables for $t \to +\infty$ and we compute the limit joint distribution 
$\Pi_{\infty}(A,B \vert x_0)  $. When the diffusion process $X(t)$ is recurrent, then 
it is more appropriate to introduce the two intensive local times
\begin{eqnarray}
a && \equiv \frac{A}{t}
 \nonumber \\
b && \equiv \frac{B}{t}
\label{additiveIntensive}
\end{eqnarray}
and to analyze their joint large deviations properties
\begin{eqnarray}
 \Pi_t(A=ta,B=tb \vert x_0)  && \opsimeq_{t \to +\infty} \aleph_t(a,b,x_0) e^{- t I (a, b )} 
\label{rateab}
\end{eqnarray}
where the positive rate function $ I (a, b ) \geq 0 $ governs the leading exponential decay in time,
while the prefactor $\aleph_t(a,b,x_0) $ that contains the dependence with
respect to the initial position $x_0$ will be also computed explicitly.
In section \ref{sec_sum}, we describe the corresponding 
simpler statistical properties of the sum $\Sigma(t)=A(t)+B(t)$ of the two local times.

In section \ref{sec_doob}, we construct various conditioned joint processes $[X^*(t),A^*(t),B^*(t)] $
satisfying certain constraints involving the two local times, thereby generalizing our previous work 
\cite{us_LocalTime} concerning the conditioning with respect to the single local time $A(t)$.
 The conditioned Bridge towards the given local times $[A^*_T,B^*_T] $ 
at the finite time horizon $T$ can be constructed as follows : the conditioned process $[X^*(t),A^*(t),B^*(t)]$ 
satisfies the following Ito SDE system analog to Eq. \ref{itoab}
\begin{eqnarray}
dX^*(t) && =  \mu^{[A_T^*,B_T^*]}_T( X^*(t),A^*(t),B^*(t),t ) dt + dW(t)
\nonumber \\
dA^*(t) && = \delta ( X^*(t) ) dt
\nonumber \\
dB^*(t) && = \delta ( X^*(t)-L ) dt
\label{itoabstar}
\end{eqnarray}
where the conditioned drift $ \mu^{[A_T^*,B_T^*]}_T( x,A,B,t ) $ involves the unconditioned drift $\mu(x)$
and the logarithmic derivative of the unconditioned distribution $\Pi_{T-t}( A_T^*-A,B_T^*-B \vert  x) $  with respect to its initial position $x$
\begin{eqnarray}
\mu^{[A_T^*,B_T^*]}_T( x,A,B,t ) = \mu(x) +  \partial_x    \ln \Pi_{T-t}( A_T^*-A,B_T^*-B \vert  x) 
\label{mustarbridgepi}
\end{eqnarray}
In the limit of the infinite time horizon $T \to +\infty$, we will analyze the conditioning towards the finite asymptotic values $[A^*(\infty),B^*(\infty)]$, as well as the conditioning towards the intensive values $[a^*,b^*] $.

In section \ref{sec_brown}, the general framework of the previous sections written 
for the diffusion process with an arbitrary unconditioned drift $\mu(x)$
is applied to the simplest example of the Brownian motion with the uniform drift $\mu(x)=\mu$ on the whole line $]-\infty,+\infty[$ in order to construct various conditioned processes involving its two local times.
Our conclusions are summarized in section \ref{sec_conclusion}.
The two appendices \ref{sec_canonical} and \ref{sec_canonicalbrown}
are devoted to the canonical conditioned processes $X^*_{p,q}(t)$ of parameters $(p,q)$
conjugated to the two local times $A$ and $B$,
in order to compare with the microcanonical conditioning described in the main text. 
Two other Appendices \ref{app_laplace} \ref{app_taboo} contain more technical computations.


\section{ Propagator $P_t(x,A,B \vert x_0) $ for the position $x$ and the two local times $A$ and $B$}

\label{sec_xab}

In this section, we analyze the joint propagator $P_t(x,A,B \vert x_0) $ 
associated to the Ito system of Eq. \ref{itoab} that satisfies the Fokker-Planck dynamics
\begin{eqnarray}
\partial_t P_t(x,A,B \vert x_0)
 = - \delta(x) \partial_A P_t(x,A,B \vert x_0)
- \delta(x-L) \partial_B P_t(x,A,B \vert x_0)
 -    \partial_x \left[ \mu(x) P_t(x,A,B \vert x_0) \right] +  \frac{1}{2} \partial_x^2  P_t(x,A,B \vert x_0) 
 \ \ \ \ \ \ 
\label{forwardjoint}
\end{eqnarray}
For clarity, it will be convenient to denote by another letter
the propagator $G_t(x \vert x_0)$ of the position $x$ alone 
 \begin{eqnarray}
 G_t(x \vert x_0)
 =  \int_0^{+\infty} dA  \int_0^{+\infty} dB P_t(x,A,B \vert x_0)
\label{GfromPjoint}
\end{eqnarray}
that satisfies
 \begin{eqnarray}
\partial_t G_t(x \vert x_0)
 =   -    \partial_x \left[ \mu(x) G_t(x \vert x_0) \right] 
 +  \frac{1}{2} \partial_x^2  G_t(x \vert x_0) 
\label{forwardx}
\end{eqnarray}


\subsection{ Decomposition of the joint propagator $P_t(x,A,B \vert x_0) $ into four contributions}

The joint propagator $P_t(x,A,B \vert x_0) $ can be decomposed into the four contributions
based on the two delta functions $\delta(A)$ and $\delta(B)$
and on the two Heaviside functions $\theta(A>0)$ and $\theta(B>0)$
\begin{eqnarray}
 P_t(x,A,B \vert x_0)
&& =  \delta(A) \delta(B) G^{abs(0,L)}_t (x \vert x_0)
+ \theta(A>0) \delta(B) {\cal A}^{abs(L)}_t (x,A \vert x_0)
 + \delta(A)\theta(B>0)  {\cal B}^{abs(0)}_t (x,B \vert x_0)
 \nonumber \\ &&
+ \theta(A>0) \theta(B>0)  {\cal C}_t (x,A,B \vert x_0)
\label{four}
\end{eqnarray}
with the following meaning.

(1) The first contribution containing the two delta functions $\delta(A)\delta(B)$
means that the two local times have kept their initial values $A=0$ and $B=0$,
i.e. the diffusion process has not been able to visit the positions $x=0$ and $x=L$.
As a consequence, the amplitude is given
by the propagator $G^{abs(0,L)}_t(x \vert x_0) $ 
in the presence of two absorbing boundary conditions at position $0$ and at position $L$,
so that it is non-vanishing only if the the final position $x $ and the initial position $x_0$ 
belong both to the left region $]-\infty,0[$, or belong both to the middle region $]0,L[$ 
or belong both to the right region $]L,+\infty[$, 
and in the two external regions, only one absorbing boundary condition
is sufficient
  \begin{eqnarray}
 G^{abs(0,L)}_t(x \vert x_0)  && = \theta(x<0) \theta(x_0 <0)G^{abs(0)}_t(x \vert x_0) 
\nonumber \\
&& +  \theta(0<x<L) \theta(0<x_0 <L) G^{abs(0,L)}_t(x \vert x_0)
\nonumber \\
&& +  \theta(x>L) \theta(x_0 >L) G^{abs(L)}_t(x \vert x_0)
\label{Gabsab}
\end{eqnarray}

(2) The second contribution containing $\theta(A>0) \delta(B) $ means that the diffusion process has not been able to visit $x=L$ but has visited $x=0$, so that it is non-vanishing only for $x < L$ and $x_0 <L$
 \begin{eqnarray}
{\cal A}^{abs(L)}_t (x,A \vert x_0) =\theta(x<L) \theta(x_0 <L) {\cal A}^{abs(L)}_t (x,A \vert x_0)
\label{calAabsb}
\end{eqnarray}

(3) The third contribution containing $\delta(A)\theta(B>0) $ means that the diffusion process has not been able to visit $x=0$ but has visited $x=L$, so that it is non-vanishing only for $x >0$ and $x_0 >0$
 \begin{eqnarray}
{\cal B}^{abs(0)}_t (x,B \vert x_0) =\theta(x>0) \theta(x_0 >0) {\cal B}^{abs(0)}_t (x,B \vert x_0) 
\label{calBabsa}
\end{eqnarray}

(4) The fourth contribution containing $\theta(A>0)\theta(B>0) $ means that the diffusion process has visited
both positions $x=0$ and $x=L$. 

\begin{figure}[h]
\centering
\includegraphics[width=5.5in,height=4in]{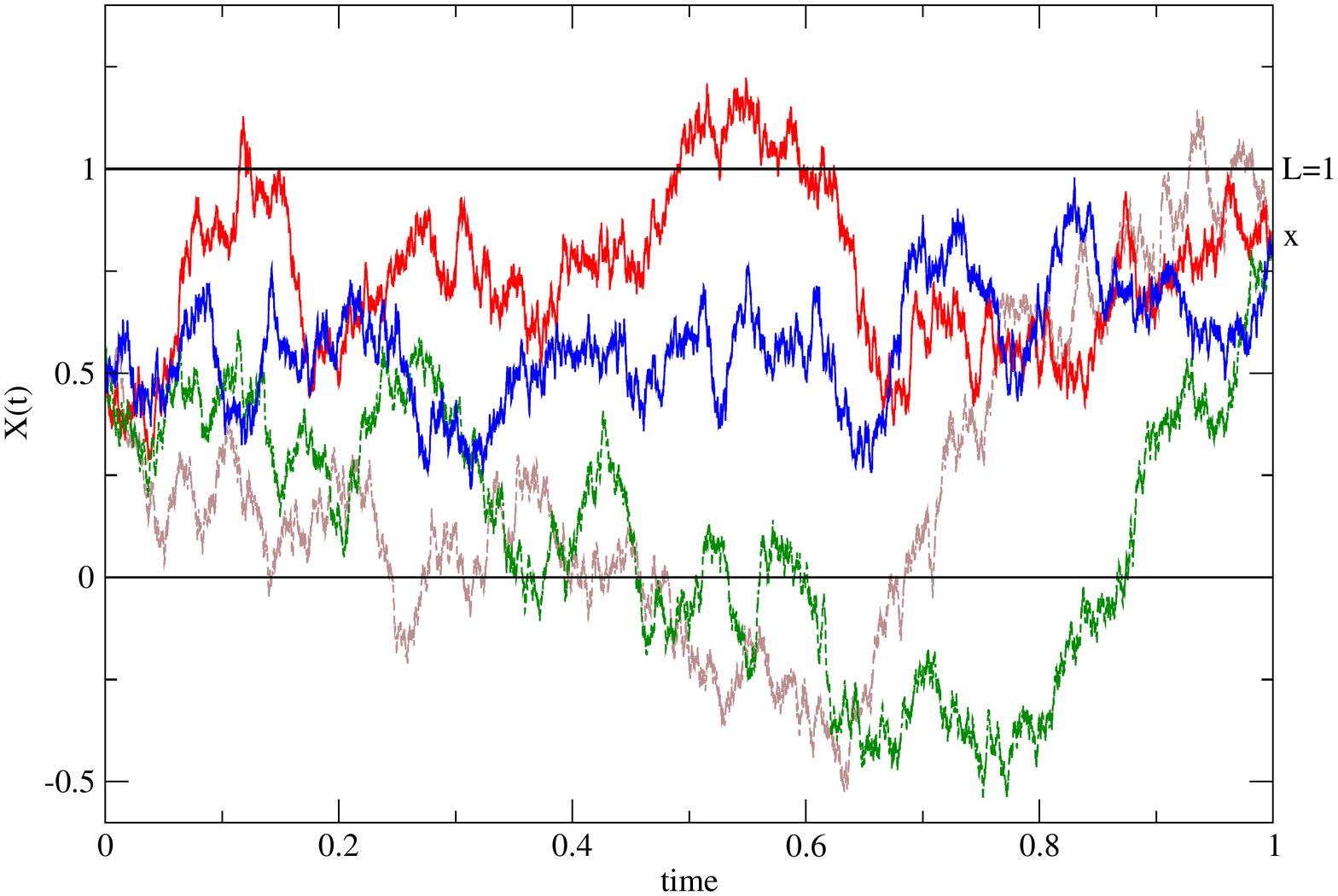}
\setlength{\abovecaptionskip}{15pt}  
\caption{Examples of simulated trajectories $x(t)$ of the Brownian bridge on the interval $t \in [0,1]$ that illustrate the four contributions of the joint propagator $P_t(x,A,B \vert x_0)$ in Eq. \ref{four}. 
The four trajectories $x(t)$ start from $x_0=0.5$ at time $t=0$ and end at $x=0.8$ at time $t=1$, but have different properties with respect to the two local times $A$ and $B$: 
the blue trajectory that has not been able to visit the position $x=0$ and $x=L=1$ contributes to the first 
term of Eq. \ref{four} involving $\delta(A) \delta(B)$; 
the green trajectory that has not been able to visit $x=L=1$ but has visited $x=0$ contributes to the second term of Eq. \ref{four} involving $\theta(A>0) \delta(B)$; 
the red trajectory that has not been able to visit $x=0$ but has visited $x=L=1$ contributes to the third 
term of Eq. \ref{four} involving $\delta(A)\theta(B>0) $ ; 
the brown trajectory that has visited both positions $x=0$ and $x=L=1$ contributes to the fourth 
term of Eq. \ref{four} involving $\theta(A>0)\theta(B>0) $. The time step used in the discretization is $dt = 10^{-4}$}
\label{fig1}
\end{figure}

Figure \ref{fig1} shows examples of trajectories of the Brownian bridge that illustrate these four contributions of the propagator $P_t(x,A,B \vert x_0) $ in Eq. \ref{four}.

\subsection{ Laplace transform  ${\tilde P}_{t,p,q} (x  \vert x_0)$ with respect to the two local times $A$ and $B$ : Feynman-Kac formula }

For the Laplace transform $ {\tilde P}_{t,p,q} (x  \vert x_0) $ of 
the propagator $P_t(x,A,B \vert x_0) $ of Eq. \ref{four}
with respect to the two local times $A$ and $B$
\begin{eqnarray}
 {\tilde P}_{t,p,q} (x  \vert x_0) && \equiv 
\int_0^{+\infty} dA e^{-p A } \int_0^{+\infty} dB e^{-q B } P_t(x,A,B \vert x_0)
\nonumber \\
&& =  G^{abs(0,L)}_t (x \vert x_0)
+  {\tilde {\cal A}^{abs(L)}}_{t,p} (x \vert x_0)
 +  {\tilde {\cal B}}^{abs(0)}_{t,q} (x \vert x_0)
+ {\tilde {\cal C}}_{t,p,q} (x \vert x_0)
\label{laplacefour}
\end{eqnarray}
Eq. \ref{forwardjoint}
translates into 
\begin{eqnarray}
\partial_t  {\tilde P}_{t,p,q} (x  \vert x_0)
&& = -  p \delta(x)  {\tilde P}_{t,p,q} (x  \vert x_0)
-  q \delta(x-L)  {\tilde P}_{t,p,q} (x  \vert x_0)
 -    \partial_x \left[ \mu(x)  {\tilde P}_{t,p,q} (x  \vert x_0) \right] +  \frac{1}{2} \partial_x^2   {\tilde P}_{t,p,q} (x  \vert x_0) 
\label{feynmankac}
\end{eqnarray}
For $p=0=q$, one recovers the propagator $G_t(x \vert x_0)$ of the position alone 
 \begin{eqnarray}
{\tilde P}_{t,p=0,q=0} (x  \vert x_0) = G_t(x \vert x_0)
\label{pqzeroG}
\end{eqnarray}
Eq. \ref{feynmankac} is an example of the Feynman-Kac formula, where the initial Fokker-Planck dynamics 
of Eq. \ref{forwardx} is now supplemented by the two additional terms involving $p \delta(x)$ and $q \delta(x-L) $.


\subsection{ Explicit solution for the further time-Laplace-transform ${\hat {\tilde P}}_{s,p,q} (x \vert x_0) $ via the Dyson Equation }

For the further Laplace transform ${\hat {\tilde P}}_{s,p,q} (x \vert x_0) $ of Eq. \ref{laplacefour}
with respect to the time $t$
\begin{eqnarray}
{\hat {\tilde P}}_{s,p,q} (x \vert x_0) && \equiv \int_0^{+\infty} dt e^{-st }  {\tilde P}_{t,p,q} (x  \vert x_0)
\nonumber \\
&& = {\hat G}^{abs(0,L)}_s (x \vert x_0)
+  {\hat {\tilde {\cal A}}}^{abs(L)}_{s,p} (x \vert x_0)
 +  {\hat {\tilde {\cal B}}}^{abs(0)}_{s,q} (x \vert x_0)
+ {\hat {\tilde {\cal C}}}_{s,p,q} (x \vert x_0)
\label{laplacetriple}
\end{eqnarray}
Eq. \ref{feynmankac}
translates into
\begin{eqnarray}
- \delta(x-x_0) + s {\hat {\tilde P}}_{s,p,q}  (x \vert x_0)
= - p \delta(x)  {\hat {\tilde P}}_{s,p,q}  (0 \vert x_0)
- q \delta(x-L)  {\hat {\tilde P}}_{s,p,q}  (L \vert x_0)
 -       \partial_x \left[ \mu(x) {\hat {\tilde P}}_{s,p,q}  \right] +  \frac{1}{2} \partial_x^2  {\hat {\tilde P}}_{s,p,q} 
\label{forward1dlaplacep}
\end{eqnarray}
The knowledge of the solution for $p=q=0$ of Eq. \ref{pqzeroG}
 \begin{eqnarray}
{\hat {\tilde P}}_{s,p=0,q=0}  (x \vert x_0) = {\hat G}_s (x \vert x_0)
\label{pzeroG}
\end{eqnarray}
allows to rewrite the solution for arbitrary Laplace parameters $(p,q)$ as 
\begin{eqnarray}
{\hat {\tilde P}}_{s,p,q} (x \vert x_0) = {\hat G}_s (x \vert x_0)
- p {\hat G}_s (x \vert 0){\hat {\tilde P}}_{s,p,q} (0 \vert x_0)
- q {\hat G}_s (x \vert L){\hat {\tilde P}}_{s,p,q} (L \vert x_0)
\label{dyson}
\end{eqnarray}
where ${\hat {\tilde P}}_{s,p,q} (0 \vert x_0) $ and ${\hat {\tilde P}}_{s,p,q} (L \vert x_0) $ 
should satisfy the following system, where Eq. \ref{dyson} is written for $x=0$ and for $x=L$
respectively
\begin{eqnarray}
\left[ 1+ p {\hat G}_s (0 \vert 0) \right] {\hat {\tilde P}}_{s,p,q} (0 \vert x_0) 
+ q {\hat G}_s (0 \vert L){\hat {\tilde P}}_{s,p,q} (L \vert x_0)
&&  = {\hat G}_s (0 \vert x_0)
\nonumber \\
 p {\hat G}_s (L \vert 0){\hat {\tilde P}}_{s,p,q} (0 \vert x_0)
+\left[ 1+ q {\hat G}_s (L \vert L) \right] {\hat {\tilde P}}_{s,p,q} (L \vert x_0) && = {\hat G}_s (L \vert x_0)
\label{dysonsystem}
\end{eqnarray}
The solution of this system
\begin{eqnarray}
 {\hat {\tilde P}}_{s,p,q} (0 \vert x_0) 
&&  = \frac{ \left[ 1+ q {\hat G}_s (L \vert L) \right]{\hat G}_s (0 \vert x_0)- q {\hat G}_s (0 \vert L){\hat G}_s (L \vert x_0) } 
{\left[ 1+ p {\hat G}_s (0 \vert 0) \right]\left[ 1+ q {\hat G}_s (L \vert L) \right] - p q{\hat G}_s (L \vert 0) {\hat G}_s (0 \vert L) } 
\nonumber \\
 {\hat {\tilde P}}_{s,p,q} (L \vert x_0) 
&& = \frac{\left[ 1+ p {\hat G}_s (0 \vert 0) \right] {\hat G}_s (L \vert x_0) -  p {\hat G}_s (L \vert 0) {\hat G}_s (0 \vert x_0)}
{\left[ 1+ p {\hat G}_s (0 \vert 0) \right]\left[ 1+ q {\hat G}_s (L \vert L) \right] - p q{\hat G}_s (L \vert 0) {\hat G}_s (0 \vert L) } 
\label{dysonsystemsol}
\end{eqnarray}
 can be plugged into Eq. \ref{dyson} to obtain the general solution 
 \begin{eqnarray}
{\hat {\tilde P}}_{s,p,q} (x \vert x_0) && 
= {\hat G}_s (x \vert x_0)
 -   \frac{pq   \Omega_s (x \vert x_0)   +p  \frac{ {\hat G}_s (x \vert 0) {\hat G}_s (0 \vert x_0)}{\Delta_s } +q  \frac{{\hat G}_s (x \vert L) {\hat G}_s (L \vert x_0)}{\Delta_s } }
{ p q + p \frac{ {\hat G}_s (0 \vert 0) }{\Delta_s }  + q \frac{{\hat G}_s (L \vert L)}{\Delta_s }  + \frac{1}{\Delta_s } } 
\label{dysonpoly}
\end{eqnarray}
  where we have introduced the two following notations 
    \begin{eqnarray}
   \Delta_s && \equiv {\hat G}_s (0 \vert 0)  {\hat G}_s (L \vert L)  -  {\hat G}_s (L \vert 0) {\hat G}_s (0 \vert L) 
\label{nota}  \\
 \Omega_s (x \vert x_0) && \equiv 
 \frac{  {\hat G}_s (L \vert L) {\hat G}_s (x \vert 0) {\hat G}_s (0 \vert x_0)
    -  {\hat G}_s (x \vert 0)  {\hat G}_s (0 \vert L) {\hat G}_s (L \vert x_0)
 +   {\hat G}_s (0 \vert 0)  {\hat G}_s (x \vert L) {\hat G}_s (L \vert x_0)  
-    {\hat G}_s (x \vert L){\hat G}_s (L \vert 0) {\hat G}_s (0 \vert x_0) }{ \Delta_s}
  \nonumber 
\end{eqnarray}
in order to see more clearly the rational fraction structure of Eq. \ref{dysonpoly}
with respect to the two Laplace parameters $(p,q)$.


\subsection{ Explicit Laplace inversion of ${\hat {\tilde P}}_{s,p,q} (x \vert x_0) $ with respect to $p$ and $q$ to obtain ${\hat P}_{s} (x,A,B \vert x_0)  $  }

The goal is now to obtain 
the time-Laplace-transform of parameter $s$ of the propagator of Eq. \ref{four}
with its four contributions
\begin{eqnarray}
 {\hat P}_{s} (x,A,B \vert x_0) 
&& \equiv \int_0^{+\infty} dt e^{-st }  P_t (x,A,B  \vert x_0)
\nonumber \\
&& =    \delta(A) \delta(B) {\hat G}^{abs(0,L)}_s (x\vert x_0)
+ \theta(A>0) \delta(B) {\hat {\cal A}}^{abs(L)}_s (x,A \vert x_0)
 + \delta(A)\theta(B>0)  {\hat {\cal B}}^{abs(0)}_s (x,B \vert x_0)
 \nonumber \\ &&
+ \theta(A>0) \theta(B>0) {\hat  {\cal C}}_s (x,A,B \vert x_0)
\label{laplacesingle}
\end{eqnarray}
from the Laplace inversion with respect to the two Laplace parameters $(p,q)$ of
 \begin{eqnarray}
{\hat {\tilde P}}_{s,p,q} (x \vert x_0) && 
= \int_0^{+\infty} dA e^{-p A } \int_0^{+\infty} dB e^{-q B } {\hat P}_{s} (x,A,B \vert x_0) 
\nonumber \\
&& =  {\hat G}^{abs(0,L)}_s (x\vert x_0)
+  \int_0^{+\infty} dA e^{-p A } {\hat {\cal A}}^{abs(L)}_s (x,A \vert x_0)
 +  \int_0^{+\infty} dB e^{-q B } {\hat {\cal B}}^{abs(0)}_s (x,B \vert x_0)
 \nonumber \\ &&
+ \int_0^{+\infty} dA e^{-p A } \int_0^{+\infty} dB e^{-q B } {\hat  {\cal C}}_s (x,A,B \vert x_0)
\nonumber \\
&& =  {\hat G}^{abs(0,L)}_s (x\vert x_0)
+  {\hat {\tilde {\cal A}}}^{abs(L)}_{s,p} (x \vert x_0)
 +  {\hat {\tilde {\cal B}}}^{abs(0)}_{s,q} (x \vert x_0)
+ {\hat {\tilde {\cal C}}}_{s,p,q} (x \vert x_0)
\label{dysonpolypq}
\end{eqnarray}
whose explicit expression was computed in Eq. \ref{dysonpoly}.


\subsubsection{ First contribution of amplitude ${\hat G}^{abs(0,L)}_s (x\vert x_0) $  }
  
Equation \ref{dysonpolypq} yields that the first contribution ${\hat G}^{abs(0,L)}_s (x\vert x_0) $ 
can be found by considering the double limit $(p \to +\infty,q \to +\infty)$ of 
${\hat {\tilde P}}_{s,p,q} (x \vert x_0) $ given by Eq. \ref{dysonpoly}
\begin{eqnarray}
  {\hat G}^{abs(0,L)}_s (x\vert x_0)  = {\hat {\tilde P}}_{s,p=+\infty,q=+\infty} (x \vert x_0)
  = {\hat G}_s (x \vert x_0)
 -  \Omega_s (x \vert x_0)
 \label{pqinfty}
\end{eqnarray}


\subsubsection{ Second contribution of amplitude ${\hat {\cal A}}^{abs(L)}_s (x,A \vert x_0) $  }

Equation \ref{dysonpolypq} yields that the second contribution ${\hat {\tilde {\cal A}}}^{abs(L)}_{s,p} (x \vert x_0)$ 
can be found by considering the limit $q \to + \infty$ of 
${\hat {\tilde P}}_{s,p,q} (x \vert x_0) $ given by Eq. \ref{dysonpoly}
  \begin{eqnarray}
{\hat G}^{abs(0,L)}_s (x\vert x_0)
+   {\hat {\tilde {\cal A}}}^{abs(L)}_{s,p} (x \vert x_0) && ={\hat {\tilde P}}_{s,p,q=+\infty} (x \vert x_0)
 =  {\hat G}_s (x \vert x_0)
  -   \frac{p   \Omega_s (x \vert x_0)    +  \frac{{\hat G}_s (x \vert L) {\hat G}_s (L \vert x_0)}{\Delta_s } }
{ p   +  \frac{{\hat G}_s (L \vert L)}{\Delta_s }   } 
\label{qinfty}
\end{eqnarray}  
The difference with the first contribution of Eq. \ref{pqinfty}
 yields
 \begin{eqnarray}
 {\hat {\tilde {\cal A}}}^{abs(L)}_{s,p} (x \vert x_0)
 =  \Omega_s (x \vert x_0)
  -   \frac{p   \Omega_s (x \vert x_0)    +  \frac{{\hat G}_s (x \vert L) {\hat G}_s (L \vert x_0)}{\Delta_s } }
{ p   +  \frac{{\hat G}_s (L \vert L)}{\Delta_s }   } 
= \frac{     \alpha_s^{[00]}(x \vert x_0) }
{ p   +  \frac{1}{ {\hat G}^{abs(L)}_s (0 \vert 0) }   } 
 \label{calA}
\end{eqnarray}
where we have introduced the Laplace transform ${\hat G}_s^{abs(L)} (x \vert x_0) $ of
the propagator $G^{abs(L)}_t(x \vert x_0) $ in the presence of a single absorbing boundary at $x=L$
 \begin{eqnarray}
  {\hat G}_s^{abs(L)} (x \vert x_0) 
  \equiv   \int_0^{+\infty} dt e^{-st }G^{abs(L)}_t(x \vert x_0)
  =  {\hat G}_s (x \vert x_0) 
  - \frac{{\hat G}_s (x \vert L) {\hat G}_s (L \vert x_0)}{{\hat G}_s (L \vert L)}
\label{Gabsbonly}
\end{eqnarray}
in order to simplify the denominator of Eq. \ref{calA} in terms of
 \begin{eqnarray}
  {\hat G}_s^{abs(L)} (0 \vert 0) 
  =  {\hat G}_s (0 \vert 0) 
  - \frac{{\hat G}_s (0 \vert L) {\hat G}_s (L \vert 0)}{{\hat G}_s (L \vert L)} = \frac{\Delta_s}{{\hat G}_s (L \vert L)}
\label{Gabsbonlyaa}
\end{eqnarray}
and in order to simplify the numerator of Eq. \ref{calA} 
  \begin{eqnarray}
  \alpha_s^{[00]}(x \vert x_0) && \equiv 
   \frac{\Omega_s (x \vert x_0) {\hat G}_s (L \vert L) - {\hat G}_s (x \vert L) {\hat G}_s (L \vert x_0)}{\Delta_s } 
  = \left(\frac{    {\hat G}^{abs(L)}_s (x \vert 0)   }{{\hat G}^{abs(L)}_s (0 \vert 0)}\right)
 \left( \frac{   {\hat G}^{abs(L)}_s (0 \vert x_0)  }{ {\hat G}^{abs(L)}_s (0 \vert 0)} \right)
\label{alpha}
\end{eqnarray}
and to make obvious that it is non-vanishing only for $x < L$ and $x_0 <L$ as Eq. \ref{calAabsb}
 \begin{eqnarray}
 \alpha_s^{[00]}(x \vert x_0) =\theta(x<L) \theta(x_0 <L)  \alpha_s^{[00]}(x \vert x_0)
\label{alphavanish}
\end{eqnarray}

The Laplace inversion of $ {\hat {\tilde {\cal A}}}^{abs(L)}_{s,p} (x \vert x_0) $ in Eq. \ref{calA}
with respect to $p$ gives the following exponential form with respect to the local time $A$
\begin{eqnarray}
 {\hat {\cal A}}^{abs(L)}_s (x,A \vert x_0)  = \alpha_s^{[00]}(x \vert x_0)
 e^{ - \frac{A}{ {\hat G}^{abs(L)}_s (0 \vert 0)} }
\label{calApinv}
\end{eqnarray}


\subsubsection{ Third contribution of amplitude $  {\hat {\cal B}}^{abs(0)}_s (x,B \vert x_0)$ }

  Similarly, Eq. \ref{dysonpolypq} yields that the third contribution $ {\hat {\tilde {\cal B}}}^{abs(0)}_{s,q} (x \vert x_0) $  
can be found by considering the limit $p \to +\infty$ 
${\hat {\tilde P}}_{s,p,q} (x \vert x_0) $ given by Eq. \ref{dysonpoly}
  \begin{eqnarray}
{\hat G}^{abs(0,L)}_s (x\vert x_0)
 +   {\cal B}^{abs(0)}_{s,q} (x \vert x_0) && ={\hat {\tilde P}}_{s,p=+\infty,q} (x \vert x_0) 
 = {\hat G}_s (x \vert x_0)
  -   \frac{q   \Omega_s (x \vert x_0)   +  \frac{ {\hat G}_s (x \vert 0) {\hat G}_s (0 \vert x_0)}{\Delta_s }  }
{  q +  \frac{ {\hat G}_s (0 \vert 0) }{\Delta_s }   } 
\label{pinfty}
\end{eqnarray}  
 The difference with the first contribution of Eq. \ref{pqinfty}
 yields
 \begin{eqnarray}
  {\cal B}^{abs(0)}_{s,q} (x \vert x_0) 
  = \Omega_s (x \vert x_0)
  -   \frac{q   \Omega_s (x \vert x_0)   +  \frac{ {\hat G}_s (x \vert 0) {\hat G}_s (0 \vert x_0)}{\Delta_s }  }
{  q +  \frac{ {\hat G}_s (0 \vert 0) }{\Delta_s }   } 
 =
     \frac{      \beta_s^{[LL]}(x \vert x_0)      }
{  q +   \frac{1}{ {\hat G}^{abs(0)}_s (L \vert L) }   } 
 \label{calB}
\end{eqnarray}
where we have introduced the Laplace transform of
the propagator in the presence of a single absorbing boundary at $x=0$ 
 \begin{eqnarray}
  {\hat G}_s^{abs(0)} (y \vert y_0) 
   \equiv   \int_0^{+\infty} dt e^{-st }G^{abs(L)}_t(x \vert x_0)
  = {\hat G}_s (y \vert y_0) 
  - \frac{{\hat G}_s (y \vert 0) {\hat G}_s (0 \vert y_0)}{{\hat G}_s (0 \vert 0)}
\label{Gabsaonly}
\end{eqnarray}
in order to simplify the denominator of Eq. \ref{calB} in terms of
 \begin{eqnarray}
  {\hat G}_s^{abs(0)} (L \vert L) 
  = {\hat G}_s (L \vert L) 
  - \frac{{\hat G}_s (L \vert 0) {\hat G}_s (L \vert y_0)}{{\hat G}_s (0 \vert 0)} 
  = \frac{\Delta_s}{{\hat G}_s (0 \vert 0)}
\label{Gabsaonlybb}
\end{eqnarray}
and in order to simplify the numerator of Eq. \ref{calB} 
  \begin{eqnarray}
  \beta_s^{[LL]}(x \vert x_0)  \equiv 
   \frac{\Omega_s (x \vert x_0) {\hat G}_s (0 \vert 0) - {\hat G}_s (x \vert 0) {\hat G}_s (0 \vert x_0)}{\Delta_s }  
     =\left( \frac{    {\hat G}^{abs(0)}_s (x \vert L)
  }{ {\hat G}^{abs(0)}_s (L \vert L)} \right)
  \left( \frac{    {\hat G}^{abs(0)}_s (L \vert x_0)
  }{ {\hat G}^{abs(0)}_s (L \vert L)} \right)
\label{beta}
\end{eqnarray}
and to make obvious that it is non-vanishing only for $x >0$ and $x_0 >0$ as in Eq. \ref{calBabsa}
 \begin{eqnarray}
 \beta_s^{[LL]}(x  \vert x_0) =\theta(x>0) \theta(x_0 >0)  \beta_s^{[LL]}(x  \vert x_0)
\label{betavanish}
\end{eqnarray}
The Laplace inversion of $  {\cal B}^{abs(0)}_{s,q} (x \vert x_0)  $ in Eq. \ref{calB}
with respect to $q$ gives the following exponential form with respect to the local time $B$
  \begin{eqnarray}
{\hat {\cal B}}^{abs(0)}_s (x,B \vert x_0)  =
 \beta_s^{[LL]}(x \vert x_0) e^{ - \frac{B}{ {\hat G}^{abs(0)}_s (L \vert L)} }
 \label{calBqinv}
\end{eqnarray}


\subsubsection{ Fourth contribution of amplitude $ {\hat  {\cal C}}_s (x,A,B \vert x_0)   $  }

The fourth contribution ${\hat {\tilde {\cal C}}}_{s,p,q} (x \vert x_0)$ of Eq. \ref{dysonpolypq} 
can be obtained from the difference between 
the full solution of Eq. \ref{dysonpoly}
and the three previous contributions of Eqs \ref{pqinfty}, \ref{calA} and \ref{calB}
 \begin{eqnarray}
&& {\hat {\tilde {\cal C}}}_{s,p,q} (x \vert x_0)  = {\hat {\tilde P}}_{s,p,q} (x \vert x_0)  
-  {\hat G}^{abs(0,L)}_s (x\vert x_0)  
- {\hat {\tilde {\cal A}}}^{abs(L)}_{s,p} (x \vert x_0)
 -  {\hat {\tilde {\cal B}}}^{abs(0)}_{s,q} (x \vert x_0)\nonumber \\
&& =  \Omega_s (x \vert x_0)
 -   \frac{pq   \Omega_s (x \vert x_0)   +p  \frac{ {\hat G}_s (x \vert 0) {\hat G}_s (0 \vert x_0)}{\Delta_s } +q  \frac{{\hat G}_s (x \vert L) {\hat G}_s (L \vert x_0)}{\Delta_s } }
{\left(p  +   \frac{1}{ {\hat G}^{abs(L)}_s (0 \vert 0) } \right) 
\left( q + \frac{1}{ {\hat G}^{abs(0)}_s (L \vert L) }  \right)  
- \frac{{\hat G}_s (0 \vert L){\hat G}_s (L \vert 0) }{\Delta_s^2 } } 
-  \frac{     \alpha_s^{[00]}(x \vert x_0) }
{ p   +  \frac{1}{ {\hat G}^{abs(L)}_s (0 \vert 0) }   } 
-   \frac{      \beta_s^{[LL]}(x \vert x_0)      }
{  q +   \frac{1}{ {\hat G}^{abs(0)}_s (L \vert L) }   } 
\label{calCdef}
\end{eqnarray}

The Laplace inversion with respect to $p$ and $q$
is described in Appendix \ref{app_laplace}.
The final result
\begin{eqnarray}
 {\hat  {\cal C}}_{s} (x,A,B \vert x_0)
 && =  {\hat  {\cal C}}^{[\alpha]}_{s} (x,A,B \vert x_0)  +  {\hat  {\cal C}}^{[\beta]}_{s} (x,A,B \vert x_0)  
  +  {\hat  {\cal C}}^{[\gamma]}_{s} (x,A,B \vert x_0)
\label{laplaceCabcinvthree}
\end{eqnarray}
involves the three terms
\begin{eqnarray}
  {\hat  {\cal C}}^{[\alpha]}_{s} (x,A,B \vert x_0)    && 
 = \alpha_s^{[00]}(x \vert x_0)  c_s
  e^{- \frac{A}{{\hat G}^{abs(L)}_s (0 \vert 0)}}  
       e^{- \frac{B}{{\hat G}^{abs(0)}_s (L \vert L)}}       
        \frac{  \sqrt{A}}{ \sqrt{B} }  
         I_1 \left( 2 c_s  \sqrt{AB} \right) 
  \nonumber \\
 {\hat  {\cal C}}^{[\beta]}_{s} (x,A,B \vert x_0)
 && 
=  \beta_s^{[LL]}(x \vert x_0)  c_s e^{- \frac{A}{{\hat G}^{abs(L)}_s (0 \vert 0)}}  
       e^{- \frac{B}{{\hat G}^{abs(0)}_s (L \vert L)}} 
             \frac{  \sqrt{B}}{ \sqrt{A} }  
              I_1 \left( 2 c_s  \sqrt{AB} \right) 
\nonumber \\
 {\hat  {\cal C}}^{[\gamma]}_{s} (x,A,B \vert x_0)  && 
      = \gamma_s(x \vert x_0)
        e^{- \frac{A}{{\hat G}^{abs(L)}_s (0 \vert 0)}}  
       e^{- \frac{B}{{\hat G}^{abs(0)}_s (L \vert L)}}
         I_0 \left( 2 c_s  \sqrt{AB} \right) 
\label{laplaceCabcinv}
\end{eqnarray}
where $I_0(z) $ is the modified Bessel function of order zero
 \begin{eqnarray}
  I_0 (z) =  \sum_{k=0}^{+\infty} \frac{\left( \frac{z}{2} \right)^{2k}}{( k!)^2}
\label{I0serie}
\end{eqnarray}
with its derivative
 \begin{eqnarray}
I_0'(z) =  I_1 (z) =  \sum_{k=1}^{+\infty} \frac{\left( \frac{z}{2} \right)^{2k-1}}{ k! (k-1)!}
\label{I1serie}
\end{eqnarray}
that both display the same asymptotic leading behavior for large $z$
 \begin{eqnarray}
  I_0 (z) && \opsimeq_{z \to +\infty} \frac{e^z}{\sqrt{2 \pi z} }
  \nonumber \\
  I_1 (z) && \opsimeq_{z \to +\infty} \frac{e^z}{\sqrt{2 \pi z} }  
\label{I0I1zlarge}
\end{eqnarray}
while the amplitude $\gamma_s(x \vert x_0) $ of the third term in Eq. \ref{laplaceCabcinv}
 \begin{eqnarray}
        \gamma_s(x \vert x_0) 
       =    \gamma_s^{[L0]}(x \vert x_0) +     \gamma_s^{[0L]}(x \vert x_0) 
\label{gamma2}
\end{eqnarray}
can be decomposed into the two terms
 \begin{eqnarray}
 \gamma_s^{[L0]}(x \vert x_0) &&  \equiv \left( \frac{    {\hat G}^{abs(0)}_s (x \vert L)
  }{ {\hat G}^{abs(0)}_s (L \vert L)} \right)
      \frac{    {\hat G}_s (L \vert 0)   
  }{\Delta_s  }  
  \left( \frac{   {\hat G}^{abs(L)}_s (0 \vert x_0)  }{ {\hat G}^{abs(L)}_s (0 \vert 0)} \right)
    \nonumber \\
 \gamma_s^{[0L]}(x \vert x_0) &&  \equiv   
  \left(\frac{    {\hat G}^{abs(L)}_s (x \vert 0)   }{{\hat G}^{abs(L)}_s (0 \vert 0)}\right) 
   \frac{    {\hat G}_s (0 \vert L) 
  }{\Delta_s  }  
   \left( \frac{    {\hat G}^{abs(0)}_s (L \vert x_0)
  }{ {\hat G}^{abs(0)}_s (L \vert L)} \right)
\label{gamma}
\end{eqnarray}
in order to make obvious that the first term $\gamma_s^{[L0]}(x \vert x_0) $
is non-vanishing only for $x >0$ and $x_0 <L$
and that the second term $\gamma_s^{[0L]}(x \vert x_0) $
is non-vanishing only for $x <L$ and $x_0 >0$
  \begin{eqnarray}
  \gamma_s^{[L0]}(x \vert x_0) && = \theta(x>0) \theta(x_0 <L) \gamma_s^{[L0]}(x \vert x_0)     
        \nonumber \\
   \gamma_s^{[0L]}(x \vert x_0) && = \theta(x<L) \theta(x_0 >0)     \gamma_s^{[0L]}(x \vert x_0) 
\label{gammavanish}
\end{eqnarray}
In order to understand the physical meaning of the various terms,
it is now useful to consider the following four special cases when $x$ and $x_0$ 
are either at $0$ or at $L$.


\paragraph{ Special cases when the final position $x$ and the initial position $x_0$ are either at $0$ or at $L$ }\

[00] For $x=0=x_0$, Eqs \ref{betavanish} and \ref{gammavanish} yield that
only the contribution $[\alpha]$ survives in Eq. \ref{laplaceCabcinv}
while the amplitude of Eq. \ref{alpha} reduces to $ \alpha_s^{[00]}(0 \vert 0) =1 $
\begin{eqnarray}
 {\cal C}_{s} (x=0,A,B \vert x_0=0)  =  {\cal C}^{[\alpha]}_{s} (x=0,A,B \vert x_0=0)   
 =   c_s
  e^{- \frac{A}{{\hat G}^{abs(L)}_s (0 \vert 0)}}  
       e^{- \frac{B}{{\hat G}^{abs(0)}_s (L \vert L)}}       
        \frac{  \sqrt{A}}{ \sqrt{B} }  
         I_1 \left( 2 c_s  \sqrt{AB} \right) 
\label{laplaceCabcaa}
\end{eqnarray}

[LL] For $x=L=x_0$, Eqs \ref{alphavanish} and \ref{gammavanish} yield that
only the contribution $[\beta]$ survives in Eq. \ref{laplaceCabcinv}
while the amplitude of Eq. \ref{beta} reduces to $\beta_s^{[LL]}(L \vert L)=1 $
\begin{eqnarray}
 {\cal C}_{s} (x=L,A,B \vert x_0=L)  =  {\cal C}^{[\beta]}_{s} (x=L,A,B \vert x_0=L)   
=  \beta_s^{[LL]}(L \vert L)  c_s e^{- \frac{A}{{\hat G}^{abs(L)}_s (0 \vert 0)}}  
       e^{- \frac{B}{{\hat G}^{abs(0)}_s (L \vert L)}} 
             \frac{  \sqrt{B}}{ \sqrt{A} }  
              I_1 \left( 2 c_s  \sqrt{AB} \right) 
\label{laplaceCabcbb}
\end{eqnarray}

[0L] For $x=0$ and $x_0=L$, Eqs \ref{alphavanish} and \ref{betavanish} yield that
only the contribution $[\gamma]$ survives in Eq. \ref{laplaceCabcinv}
while the amplitude of Eq. \ref{gamma} reduces to
 \begin{eqnarray}
        \gamma_s(0 \vert L) 
        =    \gamma_s^{[0L]}(0 \vert L)
  =        \frac{     {\hat G}_s (0 \vert L)     }{\Delta_s  }
\label{gammaab}
\end{eqnarray}
so that one obtains
\begin{eqnarray}
 {\cal C}_{s} (x=0,A,B \vert x_0=L) =  {\cal C}^{[\gamma]}_{s} (x=0,A,B \vert x_0=L)   
 =   \frac{     {\hat G}_s (0 \vert L)     }{\Delta_s  }
        e^{- \frac{A}{{\hat G}^{abs(L)}_s (0 \vert 0)}}  
       e^{- \frac{B}{{\hat G}^{abs(0)}_s (L \vert L)}}
         I_0 \left( 2 c_s  \sqrt{AB} \right) 
\label{laplaceCabcab}
\end{eqnarray}

[L0] For $x=L$ and $x_0=0$, Eqs \ref{alphavanish} and \ref{betavanish} yield that
only the contribution $[\gamma]$ survives in Eq. \ref{laplaceCabcinv}
while the amplitude of Eq. \ref{gamma} reduces to
\begin{eqnarray}
        \gamma_s(L \vert 0) 
        =     \gamma_s^{[L0]}(L \vert 0)           =              \frac{   {\hat G}_s (L \vert 0)      }{\Delta_s  }
\label{gammaba}
\end{eqnarray}
so that one obtains
\begin{eqnarray}
 {\cal C}_{s} (x=L,A,B \vert x_0=0) =  {\cal C}^{[\gamma]}_{s} (x=L,A,B \vert x_0=0)   
 = \frac{   {\hat G}_s (L \vert 0)      }{\Delta_s  }
        e^{- \frac{A}{{\hat G}^{abs(L)}_s (0 \vert 0)}}  
       e^{- \frac{B}{{\hat G}^{abs(0)}_s (L \vert L)}}
         I_0 \left( 2 c_s  \sqrt{AB} \right) 
\label{laplaceCabcba}
\end{eqnarray}


\paragraph{ Decomposition with respect to the first-passage at $0$ or $L$ 
and with respect to the last-passage at $0$ or  $L$ }\

The contribution ${\hat  {\cal C}}_{s} (x,A,B \vert x_0) $ of Eq. \ref{laplaceCabcinv} 
involve the stochastic trajectories
that visit the two positions $0$ and $L$, so 

(i) either $0$ or $L$ is visited first when starting at $x_0$

(ii) either $0$ or $L$ is visited last when ending at $x$

It is thus interesting to decompose Eq. \ref{laplaceCabcinv} according to the
these last-passage and first-passage properties 
\begin{eqnarray}
 {\hat  {\cal C}}_{s} (x,A,B \vert x_0)
 && =  {\hat  {\cal C}}^{[00]}_{s} (x,A,B \vert x_0)  +  {\hat  {\cal C}}^{[LL]}_{s} (x,A,B \vert x_0)  
  +  {\hat  {\cal C}}^{[0L]}_{s} (x,A,B \vert x_0)  +  {\hat  {\cal C}}^{[L0]}_{s} (x,A,B \vert x_0)
\label{laplaceCabcinvfirstlast}
\end{eqnarray}
where the four possibilities are directly related to the four special cases described above :

[00] If the last-passage is at $x_{last}=0$ and the first-passage is at $x_{first}=0$,
 the corresponding contribution $ {\cal C}^{[00]}_{s} (x,A,B \vert x_0) $ 
 can be rewritten as a product of three terms using the special case of Eq. \ref{laplaceCabcaa}
\begin{eqnarray}
 {\cal C}^{[00]}_{s} (x,A,B \vert x_0)  
 = \left( \frac{    {\hat G}^{abs(L)}_s (x \vert 0)   }{ {\hat G}^{abs(L)}_s (0 \vert 0)} \right)
  {\cal C}_{s} (x=0,A,B \vert x_0=0)
 \left( \frac{    {\hat G}^{abs(L)}_s (0 \vert x_0)  }{ {\hat G}^{abs(L)}_s (0 \vert 0)} \right)
 \label{laplaceC00}
\end{eqnarray}

[LL] If the last-passage is at $x_{last}=L$ and the first-passage is at $x_{first}=L$, 
the corresponding contribution $ {\cal C}^{[LL]}_{s} (x,A,B \vert x_0) $ can be rewritten 
as a product of three terms using the special case of Eq. \ref{laplaceCabcbb}
\begin{eqnarray}
 {\cal C}^{[LL]}_{s} (x,A,B \vert x_0) 
  = \left(  \frac{    {\hat G}^{abs(0)}_s (x \vert L)   }{ {\hat G}^{abs(0)}_s (L \vert L)}\right)
   {\cal C}_{s} (x=L,A,B \vert x_0=L)
\left(  \frac{    {\hat G}^{abs(0)}_s (L \vert x_0)  }{ {\hat G}^{abs(0)}_s (L \vert L)}\right)
\label{laplaceCLL}
\end{eqnarray}

[0L] If the last-passage is at $x_{last}=0$ and the first-passage is at $x_{first}=L$, 
the corresponding contribution ${\cal C}^{[0L]}_{s} (x,A,B \vert x_0) $  can be rewritten 
as a product of three terms using the special case of Eq. \ref{laplaceCabcab} 
\begin{eqnarray}
{\cal C}^{[0L]}_{s} (x,A,B \vert x_0)
= \left(\frac{    {\hat G}^{abs(L)}_s (x \vert 0)   }{ {\hat G}^{abs(L)}_s (0 \vert 0)}\right)
{\cal C}_{s} (x=0,A,B \vert x_0=L) 
\left( \frac{    {\hat G}^{abs(0)}_s (L \vert x_0)  }{ {\hat G}^{abs(0)}_s (L \vert L)} \right)
\label{laplaceC0L}
\end{eqnarray}

[L0] If the last-passage is at $x_{last}=L$ and the first-passage is at $x_{first}=0$, 
the corresponding contribution ${\cal C}^{[L0]}_{s} (x,A,B \vert x_0) $  can be rewritten 
as a product of three terms using the special case of Eq. \ref{laplaceCabcba} 
\begin{eqnarray}
{\cal C}^{[L0]}_{s} (x,A,B \vert x_0)
= \left( \frac{    {\hat G}^{abs(0)}_s (x \vert L)   }{ {\hat G}^{abs(0)}_s (L \vert L)}\right)
 {\cal C}_{s} (x=L,A,B \vert x_0=0) 
 \left(  \frac{    {\hat G}^{abs(L)}_s (0 \vert x_0)  }{ {\hat G}^{abs(L)}_s (0 \vert 0)}\right)
\label{laplaceCL0}
\end{eqnarray}

\begin{figure}[h]
\centering
\includegraphics[width=5.5in,height=4in]{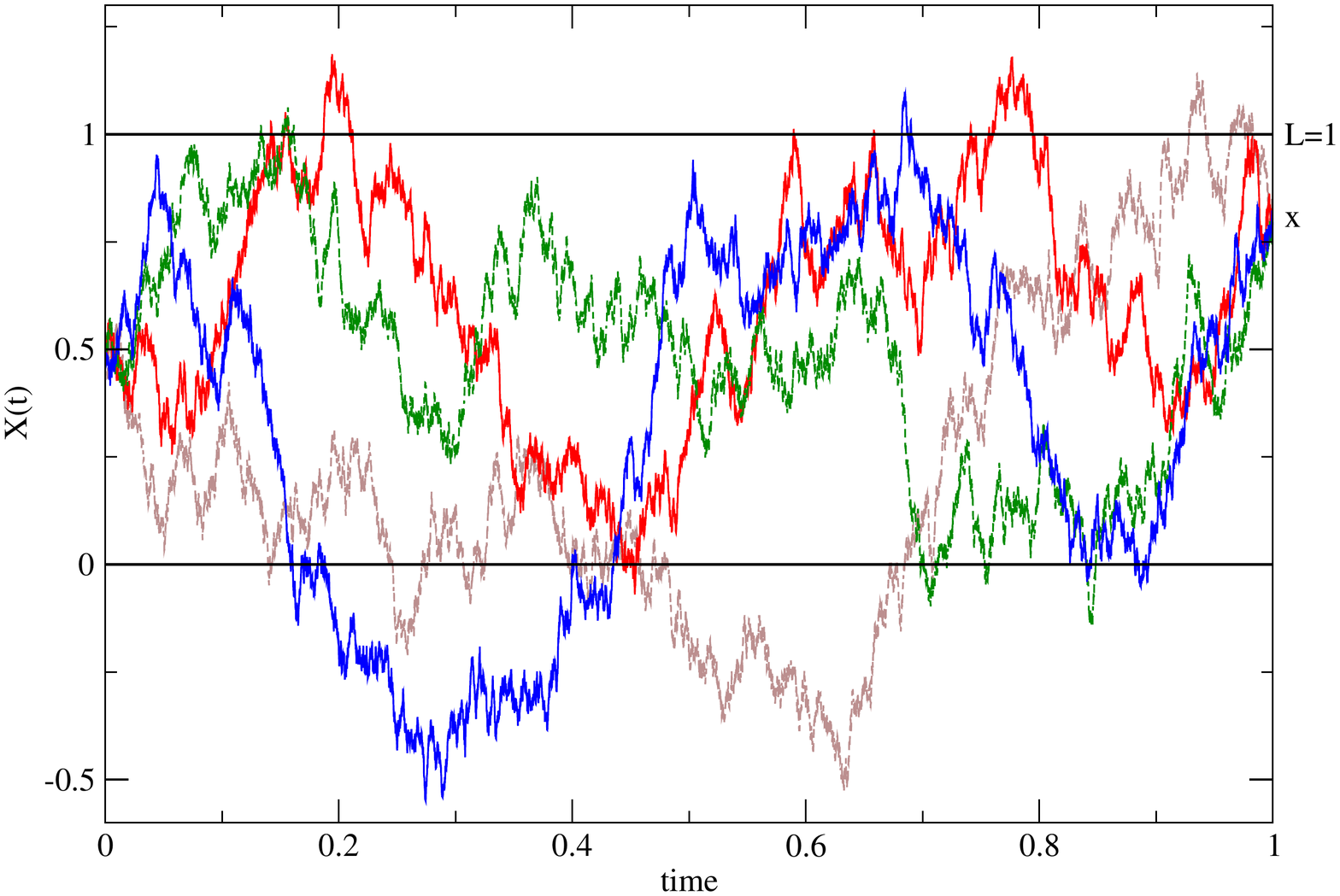}
\setlength{\abovecaptionskip}{15pt}  
\caption{Examples of simulated trajectories $x(t)$ of the Brownian bridge on the interval $t \in [0,1]$ that illustrate the four contributions of ${\hat  {\cal C}}_{s} (x,A,B \vert x_0)$ in Eq \ref{laplaceCabcinvfirstlast}. 
The four trajectories $x(t)$ start from $x_0=0.5$ at time $t=0$ and end at $x=0.8$ at time $t=1$ with two non-vanishing local times $A>0$ and $B>0$, but have different last-passage and first-passage properties:
 the blue trajectory corresponding to $x_{last}=0=x_{first}$ contributes to the first term [00] of Eq \ref{laplaceCabcinvfirstlast} given by Eq. \ref{laplaceC00}; 
  the red trajectory corresponding to $x_{last}=L=x_{first}$ contributes to the second term [LL] of Eq \ref{laplaceCabcinvfirstlast} given by Eq. \ref{laplaceCLL}; 
  the green trajectory corresponding to $x_{last}=0$ and $x_{first}=L$ contributes to the third term [0L] of Eq \ref{laplaceCabcinvfirstlast} given by Eq. \ref{laplaceC0L};
  the brown trajectory corresponding to $x_{last}=L$ and $x_{first}=0$ contributes to the fourth term [L0] of Eq \ref{laplaceCabcinvfirstlast} given by Eq. \ref{laplaceCL0}.
The time step used in the discretization is $dt = 10^{-4}$}
\label{fig2}
\end{figure}

Figure \ref{fig2} shows examples of trajectories of the Brownian bridge that illustrate these four contributions of ${\hat  {\cal C}}_{s} (x,A,B \vert x_0)$ in Eq. \ref{laplaceCabcinvfirstlast}.

\section{ Distribution $\Pi_t(A,B \vert x_0)= \int dx P_t(x,A,B \vert x_0)$ of the two local times $A$ and $B$ at time $t$}

\label{sec_ab}

In this section, we focus on 
the joint distribution $\Pi_t(A,B \vert x_0)  $ of the local times $A$ and $B$ at time $t$ if one starts at position $x_0$.

\subsection{ Decomposition of the distribution $\Pi_t(A,B \vert x_0)= \int dx P_t(x,A,B \vert x_0)$ into four contributions}

The integration over the final position $x$
of the joint propagator $P_t(x,A,B \vert x_0) $ of Eq. \ref{four}
\begin{eqnarray}
 \Pi_t(A,B \vert x_0) && \equiv \int_{-\infty}^{+\infty} dx  P_t(x,A,B \vert x_0)
 \nonumber \\
 && =   \delta(A) \delta(B) S^{abs(0,L)}_t (  x_0)
+ \theta(A>0) \delta(B) {\cal A}^{]-\infty,L[}_t (A \vert x_0)
 + \delta(A)\theta(B>0)  {\cal B}^{]0,+\infty[}_t (B\vert x_0)
\nonumber \\
&&+ \theta(A>0) \theta(B>0)  {\cal C}_t (A,B \vert x_0)
\label{propagABalone}
\end{eqnarray}
involves the four contributions  :

(1) the first contribution corresponding to the integration over $x$ of $G^{abs(0,L)}_t (x \vert x_0) $ of Eq. \ref{Gabsab}
 represents the survival probability at time $t$ when starting at $x_0$
in the presence of absorbing boundaries at positions $x=0$ and $x=L$
\begin{eqnarray}
S^{abs(0,L)}_t (  x_0) && = \int_{-\infty}^{+\infty} dx  G^{abs(0,L)}_t (x \vert x_0)
\nonumber \\
&& =  \theta(x_0 <0)S^{]-\infty,0[}_t( x_0) 
 +  \theta(0<x_0 <L) S^{]0,L[}_t( x_0)
 +  \theta(x_0 >L) S^{]L,+\infty[}_t( x_0)
\label{survivalab}
\end{eqnarray}
so that it can be decomposed into the survival probabilities in the three regions $]-\infty,0[ $, $]0,L[ $ and $]L,+\infty[ $.

(2) the second contribution corresponding to the integration over $x$ of ${\cal A}^{abs(L)}_t (x,A \vert x_0) $
of Eq. \ref{calAabsb} is non-vanishing only for $x_0 <L$
 \begin{eqnarray}
{\cal A}^{]-\infty,L[}_t (A \vert x_0)  \equiv \int_{-\infty}^{+\infty} dx{\cal A}^{abs(L)}_t (x,A \vert x_0) 
= \theta(x_0 <L) {\cal A}^{]-\infty,L[}_t (A \vert x_0)
\label{calAx0}
\end{eqnarray}

(3) the third contribution corresponding to the integration over $x$ of $ {\cal B}^{abs(0)}_t (x,B \vert x_0)$
of Eq. \ref{calBabsa}
 is non-vanishing only for $x_0 >0$
 \begin{eqnarray}
 {\cal B}^{]0,+\infty[}_t (B\vert x_0)  \equiv \int_{-\infty}^{+\infty} dx
 {\cal B}^{abs(0)}_t (x,B \vert x_0) = \theta(x_0 >0) {\cal B}^{]0,+\infty[}_t (B \vert x_0) 
\label{calBx0}
\end{eqnarray}

(4) the fourth contribution reads
\begin{eqnarray}
{\cal C}_t (A,B \vert x_0)  \equiv \int_{-\infty}^{+\infty} dx  {\cal C}_t (x,A,B \vert x_0)
\label{calCx0}
\end{eqnarray}


\subsection{ Time-Laplace transform ${\hat \Pi}_{s} (A,B \vert x_0)  $ with its four contributions }

The time-Laplace transform  of Eq. \ref{propagABalone}
\begin{eqnarray}
 {\hat \Pi}_{s} (A,B  \vert x_0) && \equiv \int_0^{+\infty} dt e^{-st }  \Pi_t(A,B  \vert x_0)
\nonumber \\
&& =    \delta(A) \delta(B) {\hat S}^{abs(0,L)}_s ( x_0)
+ \theta(A>0) \delta(B) {\hat {\cal A}}^{]-\infty,L[}_s (A \vert x_0)
 + \delta(A)\theta(B>0)  {\hat {\cal B}}^{]0,+\infty[}_s (B \vert x_0)
\nonumber \\
&&+ \theta(A>0) \theta(B>0)  {\hat {\cal C}}_s (A,B \vert x_0)  
\label{laplacePi}
\end{eqnarray}
can be computed via the integration over $x$ of ${\hat P}_s(x,A,B \vert x_0)  $ of Eq. \ref{laplacesingle}
\begin{eqnarray}
{\hat \Pi}_{s} (A,B  \vert x_0)  = \int_{-\infty}^{+\infty} dx  {\hat P}_s(x,A,B \vert x_0) 
\label{laplacesinglePi}
\end{eqnarray}
so that one needs to integrate over $x$ the various contributions computed in the previous section.
The normalization of the propagator $G_t(x \vert x_0) $
 \begin{eqnarray}
\int_{-\infty}^{+\infty} dx G_t(x \vert x_0) = 1
\label{normaGpropagator}
\end{eqnarray}
translates for its Laplace transform into
\begin{eqnarray}
\int_{-\infty}^{+\infty} dx  {\hat G}_s (x \vert x_0)
=  \int_0^{+\infty} dt e^{-st }
\int_{-\infty}^{+\infty} dx G_t(x \vert x_0) = \int_0^{+\infty} dt e^{-st } = \frac{1}{s}
\label{hatsnormaintegx}
\end{eqnarray}


\subsection{ First contribution $ {\hat S}^{abs(0,L)}_s (x_0) $ }

Using Eq. \ref{hatsnormaintegx}, the integration over $x$ of Eq. \ref{pqinfty} 
corresponding to the Laplace transform ${\hat S}^{abs(0,L)}_s (x_0) $ of the survival probability 
$S^{abs(0,L)}_t (x_0) $ of Eq. \ref{survivalab}
reads
\begin{eqnarray}
 {\hat S}^{abs(0,L)}_s (x_0) && \equiv   
   \int_{-\infty}^{+\infty} dx {\hat G}^{abs(0,L)}_s (x\vert x_0)
 \nonumber \\
 && =  \frac{1}{s} \left[ 1 
 -  \frac{  {\hat G}_s (L \vert L)  {\hat G}_s (0 \vert x_0)
    -    {\hat G}_s (0 \vert L) {\hat G}_s (L \vert x_0)
 +   {\hat G}_s (0 \vert 0)  {\hat G}_s (L \vert x_0)  
-   {\hat G}_s (L \vert 0)  {\hat G}_s (0 \vert x_0) }{ \Delta_s} \right]
 \nonumber \\
 && =  \frac{1}{s} \left[ 1
  - \frac{ {\hat G}^{abs(L)}_s (0 \vert x_0)}{{\hat G}^{abs(L)}_s (0 \vert 0)} 
  - \frac{{\hat G}^{abs(0)}_s (L \vert x_0)}{{\hat G}^{abs(0)}_s (L \vert L)}
  \right]
\label{laplacesurvival}
\end{eqnarray}
with its simplifications of Eq. \ref{survivalab}
in the two extreme regions $]-\infty,0[ $ and $]L,+\infty[ $
\begin{eqnarray}
 {\hat S}^{]-\infty,0[}_s( x_0) && =  \frac{1}{s} \left[ 1
  - \frac{ {\hat G}^{abs(L)}_s (0 \vert x_0)}{{\hat G}^{abs(L)}_s (0 \vert 0)} 
  \right]
\nonumber \\
 {\hat S}^{]L,+\infty[}_s( x_0) && = \frac{1}{s} \left[ 1
  - \frac{{\hat G}^{abs(0)}_s (L \vert x_0)}{{\hat G}^{abs(0)}_s (L \vert L)}
  \right]
\label{laplacesurvivaltwoextremes}
\end{eqnarray}


\subsection{ Second contribution $  {\hat {\cal A}}^{]-\infty,L[}_{s} (A \vert x_0) $ }

Using Eq. \ref{hatsnormaintegx}, the integration over $x$ of $\alpha_s^{[00]}(x \vert x_0)   $ of Eq. \ref{alpha}
 \begin{eqnarray}
\alpha_s^{]-\infty,L[}(x_0) && \equiv  \int_{-\infty}^{+\infty} dx  \alpha_s^{[00]}(x \vert x_0)  
  = \int_{-\infty}^{+\infty} dx \left[ {\hat G}_s (x \vert 0) - \frac{{\hat G}_s (x \vert L){\hat G}_s (L \vert 0)}{{\hat G}_s (L \vert L)} \right] \frac{    {\hat G}^{abs(L)}_s (0 \vert x_0)
  }{\left[ {\hat G}^{abs(L)}_s (0 \vert 0)\right]^2}
  \nonumber \\
&& =  \frac{1}{s} \left[1 - \frac{{\hat G}_s (L \vert 0)}{{\hat G}_s (L \vert L)} \right] \frac{    {\hat G}^{abs(L)}_s (0 \vert x_0)
  }{\left[ {\hat G}^{abs(L)}_s (0 \vert 0)\right]^2}
\label{alphainteg}
\end{eqnarray}
that is non-vanishing only for $x_0 \in  ]-\infty,0[$
allows to obtain the integration over $x$ of the contribution ${\cal A}^{abs(L)}_{s} (x,A \vert x_0) $  of Eq. \ref{calApinv}
\begin{eqnarray}
 {\hat {\cal A}}^{]-\infty,L[}_{s} (A \vert x_0)  \equiv   \int_{-\infty}^{+\infty} dx {\cal A}^{abs(L)}_{s} (x,A \vert x_0) 
  = \alpha_s^{]-\infty,L[}( x_0)
 e^{ - \frac{A}{ {\hat G}^{abs(L)}_s (0 \vert 0)} }
\label{calApinvinteg}
\end{eqnarray}


\subsection{ Third contribution ${\hat {\cal B}}^{]0,+\infty[}_{s} (B \vert x_0)  $ }

Using Eq. \ref{hatsnormaintegx}, the integration over $x$ of $ \beta_s^{[LL]}(x \vert x_0)  $ of Eq. \ref{beta}
 \begin{eqnarray}
  \beta_s^{]0,+\infty[}( x_0)  && \equiv \int_{-\infty}^{+\infty} dx  \beta_s^{[LL]}(x \vert x_0) 
     =\int_{-\infty}^{+\infty} dx \left[ {\hat G}_s (x \vert L) - \frac{{\hat G}_s (x \vert 0){\hat G}_s (0 \vert L)}{{\hat G}_s (0 \vert 0)} \right] 
      \frac{    {\hat G}^{abs(0)}_s (L \vert x_0)
  }{\left[ {\hat G}^{abs(0)}_s (L \vert L)\right]^2}
    \nonumber \\
&& =  \frac{1}{s} \left[1- \frac{{\hat G}_s (0 \vert L)}{{\hat G}_s (0 \vert 0)} \right] 
      \frac{    {\hat G}^{abs(0)}_s (L \vert x_0)
  }{\left[ {\hat G}^{abs(0)}_s (L \vert L)\right]^2}
\label{betainteg}
\end{eqnarray}
that is non-vanishing only for $x_0 \in  ]0,+\infty[$
allows to obtain the integration over $x$ of the contribution $ {\cal B}^{abs(0)}_{s} (x,B \vert x_0)$ of Eq. \ref{calBqinv}
  \begin{eqnarray}
{\hat {\cal B}}^{]0,+\infty[}_{s} (B \vert x_0)  \equiv   \int_{-\infty}^{+\infty} dx {\cal B}^{abs(0)}_{s} (x,B \vert x_0)  =
\beta_s^{]0,+\infty[}( x_0) e^{ - \frac{B}{ {\hat G}^{abs(0)}_s (L \vert L)} }
 \label{calBqinvinteg}
\end{eqnarray}


\subsection{ Fourth contribution ${\hat {\cal C}}_{s} (A,B \vert x_0)   $ }

The integration over $x$ of $\gamma_s(x \vert x_0)   $ of Eq. \ref{gamma}
  \begin{eqnarray}
   \gamma_s(x_0) &&  \equiv \int_{-\infty}^{+\infty} dx  \gamma_s(x \vert x_0)  
   = \gamma_s^{]-\infty,L[}(x_0) + \gamma_s^{]0,+\infty[}(x_0)
\label{gammainteg}
\end{eqnarray}
involves the two following terms using Eq. \ref{hatsnormaintegx}
  \begin{eqnarray}
 \gamma_s^{]-\infty,L[}(x_0) && \equiv  \int_{-\infty}^{+\infty} dx   \gamma_s^{[L0]}(x \vert x_0)   
  = \int_{-\infty}^{+\infty} dx 
  \left( \frac{   \left[ {\hat G}_s (x \vert L) - \frac{{\hat G}_s (x \vert 0){\hat G}_s (0 \vert L)}{{\hat G}_s (0 \vert 0)} \right]
  }{ {\hat G}^{abs(0)}_s (L \vert L)} \right)
      \frac{    {\hat G}_s (L \vert 0)     }{\Delta_s  }  
  \left( \frac{   {\hat G}^{abs(L)}_s (0 \vert x_0)  }{ {\hat G}^{abs(L)}_s (0 \vert 0)} \right)
  \nonumber \\
  && =  \left( \frac{   \left[ 1 - \frac{{\hat G}_s (0 \vert L)}{{\hat G}_s (0 \vert 0)} \right]
  }{ {\hat G}^{abs(0)}_s (L \vert L)} \right)
      \frac{    {\hat G}_s (L \vert 0)     }{ s \Delta_s  }  
  \left( \frac{   {\hat G}^{abs(L)}_s (0 \vert x_0)  }{ {\hat G}^{abs(L)}_s (0 \vert 0)} \right)  
    \nonumber \\
\gamma_s^{]0,+\infty[}(x_0) &&  \equiv  \int_{-\infty}^{+\infty} dx
 \gamma_s^{[0L]}(x \vert x_0)   =  \int_{-\infty}^{+\infty} dx 
  \left(\frac{    \left[ {\hat G}_s (x \vert 0) - \frac{{\hat G}_s (x \vert L){\hat G}_s (L \vert 0)}{{\hat G}_s (L \vert L)} \right]   }{{\hat G}^{abs(L)}_s (0 \vert 0)}\right) 
   \frac{    {\hat G}_s (0 \vert L) 
  }{\Delta_s  }  
   \left( \frac{    {\hat G}^{abs(0)}_s (L \vert x_0)
  }{ {\hat G}^{abs(0)}_s (L \vert L)} \right)
  \nonumber \\
  &&  = \left(\frac{    \left[ 1 - \frac{
  {\hat G}_s (L \vert 0)}{{\hat G}_s (L \vert L)} \right]   }{{\hat G}^{abs(L)}_s (0 \vert 0)}\right) 
   \frac{    {\hat G}_s (0 \vert L) 
  }{s\Delta_s  }  
   \left( \frac{    {\hat G}^{abs(0)}_s (L \vert x_0)
  }{ {\hat G}^{abs(0)}_s (L \vert L)} \right)  
\label{gammaintegtwo}
\end{eqnarray}

The integration over $x$ of the three terms of Eq. \ref{laplaceCabcinv}
\begin{eqnarray}
&&  {\hat {\cal C}}^{[\alpha]}_{s} (A,B \vert x_0)     \equiv \int_{-\infty}^{+\infty} dx        {\cal C}^{[\alpha]}_{s} (x,A,B \vert x_0)     
 = \alpha_s^{]-\infty,L[}( x_0)  c_s
  e^{- \frac{A}{{\hat G}^{abs(L)}_s (0 \vert 0)}}  
       e^{- \frac{B}{{\hat G}^{abs(0)}_s (L \vert L)}}       
        \frac{  \sqrt{A}}{ \sqrt{B} }  
         I_1 \left( 2 c_s  \sqrt{AB} \right) 
  \nonumber \\
&&  {\hat {\cal C}}^{[\beta]}_{s} (A,B \vert x_0)     \equiv \int_{-\infty}^{+\infty} dx  {\cal C}^{[\beta]}_{s} (x,A,B \vert x_0) 
=  \beta_s^{]0,+\infty[}(x_0 )  c_s e^{- \frac{A}{{\hat G}^{abs(L)}_s (0 \vert 0)}}  
       e^{- \frac{B}{{\hat G}^{abs(0)}_s (L \vert L)}} 
             \frac{  \sqrt{B}}{ \sqrt{A} }  
              I_1 \left( 2 c_s  \sqrt{AB} \right) 
\nonumber \\
&& {\hat {\cal C}}^{[\gamma]}_{s} (A,B \vert x_0)   \equiv \int_{-\infty}^{+\infty} dx {\cal C}^{[\gamma]}_{s} (x,A,B \vert x_0)   
     = \gamma_s( x_0)
        e^{- \frac{A}{{\hat G}^{abs(L)}_s (0 \vert 0)}}  
       e^{- \frac{B}{{\hat G}^{abs(0)}_s (L \vert L)}}
         I_0 \left( 2 c_s  \sqrt{AB} \right) 
\label{laplaceCabcinvintegthree}
\end{eqnarray}
yields that the total contribution ${\hat {\cal C}}_{s} (A,B \vert x_0) $ of Eq \ref{laplaceCabcinvthree} reads
\begin{eqnarray}
&& {\hat {\cal C}}_{s} (A,B \vert x_0)  =  {\hat {\cal C}}^{[\alpha]}_{s} (A,B \vert x_0) 
  +   {\hat {\cal C}}^{[\beta]}_{s} (A,B \vert x_0)  
 + {\hat {\cal C}}^{[\gamma]}_{s} (A,B \vert x_0) 
\nonumber \\
&& =    
  e^{- \frac{A}{{\hat G}^{abs(L)}_s (0 \vert 0)}}  
       e^{- \frac{B}{{\hat G}^{abs(0)}_s (L \vert L)}}       
     \left[  \left( \alpha_s^{]-\infty,L[}( x_0)   \frac{  \sqrt{A}}{ \sqrt{B} }   
         +      \beta_s^{]0,+\infty[}(x_0 )    \frac{  \sqrt{B}}{ \sqrt{A} }  \right)  c_s I_1 \left( 2 c_s  \sqrt{AB} \right) 
              +   \gamma_s( x_0)
         I_0 \left( 2 c_s  \sqrt{AB} \right) \right]
\label{laplaceCabcinvinteg}
\end{eqnarray}

In particular, the asymptotic behaviors of Eq. \ref{I0I1zlarge} for the modified Bessel functions $I_0(z)$ and $I_1(z)$
for large argument $z \to + \infty$ yield the following asymptotic behaviors for large $(AB)$ 
of the three terms
\begin{eqnarray}
  {\hat {\cal C}}^{[\alpha]}_{s} (A,B \vert x_0)   
&&  \opsimeq_{(AB) \to +\infty}    
          \frac{ \alpha_s^{]-\infty,L[}( x_0)     \sqrt{c_s} A^{\frac{1}{4}} }{2\sqrt{ \pi }    B^{\frac{3}{4}}}     
           e^{- \frac{A}{{\hat G}^{abs(L)}_s (0 \vert 0)}- \frac{B}{{\hat G}^{abs(0)}_s (L \vert L)} + 2 c_s  \sqrt{AB}}  
  \nonumber \\
  {\hat {\cal C}}^{[\beta]}_{s} (A,B \vert x_0)    &&  
 \opsimeq_{(AB) \to +\infty}  
             \frac{ \beta_s^{]0,+\infty[}(x_0 )   \sqrt{c_s }B^{\frac{1}{4}} }{2\sqrt{ \pi }   A^{\frac{3}{4}}}                
              e^{- \frac{A}{{\hat G}^{abs(L)}_s (0 \vert 0)}- \frac{B}{{\hat G}^{abs(0)}_s (L \vert L)} + 2 c_s  \sqrt{AB}}            
            \nonumber \\
 {\hat {\cal C}}^{[\gamma]}_{s} (A,B \vert x_0)    && 
  \opsimeq_{(AB) \to +\infty}            
         \frac{  \gamma_s( x_0)}{2\sqrt{ \pi  c_s }  A^{\frac{1}{4}}B^{\frac{1}{4}}}         
          e^{- \frac{A}{{\hat G}^{abs(L)}_s (0 \vert 0)}- \frac{B}{{\hat G}^{abs(0)}_s (L \vert L)} + 2 c_s  \sqrt{AB}}  
 \label{laplaceCabcinvinteglarge}
\end{eqnarray}
so that the global asymptotic behavior for large $(AB)$ is given by
\begin{eqnarray}
&& {\hat {\cal C}}_{s} (A,B \vert x_0)  =  {\hat {\cal C}}^{[\alpha]}_{s} (A,B \vert x_0) 
  +   {\hat {\cal C}}^{[\beta]}_{s} (A,B \vert x_0)  
 + {\hat {\cal C}}^{[\gamma]}_{s} (A,B \vert x_0) 
 \nonumber \\
 &&  \opsimeq_{(AB) \to +\infty}    \frac{1}{2 \sqrt{\pi} }
      \left[     \sqrt{c_s}  \left( \frac{ \alpha_s^{]-\infty,L[}( x_0)   A^{\frac{1}{4}} }{    B^{\frac{3}{4}}}   
      +  \frac{ \beta_s^{]0,+\infty[}(x_0 ) B^{\frac{1}{4}} }{   A^{\frac{3}{4}}}  \right)
      +   \frac{  \gamma_s( x_0)}{\sqrt{ c_s }  A^{\frac{1}{4}}B^{\frac{1}{4}}}   
      \right]  
           e^{- \frac{A}{{\hat G}^{abs(L)}_s (0 \vert 0)}- \frac{B}{{\hat G}^{abs(0)}_s (L \vert L)} + 2 c_s  \sqrt{AB}}  
\label{laplaceCabcinvinteglargetot}
\end{eqnarray}


\section{Joint distribution $\Pi_t( A,B \vert  x_0) $ of the local times $A(t)$ and $B(t)$ for large time $t$}

\label{sec_larget}

In this section, we analyze the joint distribution $\Pi_t( A,B \vert  x_0) $ of the two local times 
$A(t)$ and $B(t)$ in the limit of large time $t \to + \infty$.

\subsection{ When $X(t)$ is transient  : the two local times $(A,B)$
remain finite random variables for $t \to +\infty$  }

\label{subsec_trans}

When the diffusion process $X(t)$ is transient, then the Laplace transform ${\hat G}_{s} (x \vert x_0) $
of the propagator $G_t (x \vert x_0) $  remains finite for $s=0$
\begin{eqnarray}
{\hat G}_{s=0} (x \vert x_0) <+\infty
\label{Gszerofinite}
\end{eqnarray}

Then the two local times $A$ and $B$ will remain 
 finite random variables for $t \to +\infty$
with the following notation for their limit distribution obtained from Eq. \ref{propagABalone}
\begin{eqnarray}
\Pi_{\infty}( A, B \vert  x_0) && = \lim_{t \to +\infty} \Pi_t( A,B \vert  x_0)
\nonumber \\
&&   =   \delta(A) \delta(B) S^{abs(0,L)}_{\infty} (  x_0)
+ \theta(A>0) \delta(B) {\cal A}^{]-\infty,L[}_{\infty} (A \vert x_0)
 + \delta(A)\theta(B>0)  {\cal B}^{]0,+\infty[}_{\infty} (B \vert x_0)
\nonumber \\
&&+ \theta(A>0) \theta(B>0)  {\cal C}_{\infty} (A,B \vert x_0)
\label{piAlimit}
\end{eqnarray}
These contributions involving the infinite-time limit $t \to +\infty$ can be obtained 
from their Laplace transforms by considering the limit $s \to 0$ as follows :
 the first contribution can be computed from Eq. \ref{laplacesurvival}
\begin{eqnarray}
S^{abs(0,L)}_{\infty} (  x_0)  = \lim_{s \to 0} \left[ s  {\hat S}^{abs(0,L)}_s ( x_0)\right]
 =  1  - \frac{ {\hat G}^{abs(L)}_0 (0 \vert x_0)}{{\hat G}^{abs(L)}_0 (0 \vert 0)} 
  - \frac{{\hat G}^{abs(0)}_0 (L \vert x_0)}{{\hat G}^{abs(0)}_0 (L \vert L)}
 \label{survivallimitfromlaplace}
\end{eqnarray}
while the second contribution can be computed from Eq. \ref{calApinvinteg}
\begin{eqnarray}
{\cal A}^{]-\infty,L[}_{\infty} (A \vert x_0)  = \lim_{s \to 0} \left[ s {\hat {\cal A}}^{]-\infty,L[}_s (A \vert x_0)\right]
= \left[1 - \frac{{\hat G}_0 (L \vert 0)}{{\hat G}_0 (L \vert L)} \right] \frac{    {\hat G}^{abs(L)}_0 (0 \vert x_0)
  }{\left[ {\hat G}^{abs(L)}_0 (0 \vert 0)\right]^2}
 e^{ - \frac{A}{ {\hat G}^{abs(L)}_0 (0 \vert 0)} }
\label{calAlimitfromlaplace}
\end{eqnarray}
and the third contribution from Eq. \ref{calBqinvinteg}
\begin{eqnarray}
{\cal B}^{]0,+\infty[}_{\infty} (B \vert x_0)  = \lim_{s \to 0} \left[ s  {\hat {\cal B}}^{]0,+\infty[}_s (B \vert x_0)\right]
= 
\left[1- \frac{{\hat G}_0 (0 \vert L)}{{\hat G}_0 (0 \vert 0)} \right] 
      \frac{    {\hat G}^{abs(0)}_0 (L \vert x_0)
  }{\left[ {\hat G}^{abs(0)}_0 (L \vert L)\right]^2}
 e^{ - \frac{B}{ {\hat G}^{abs(0)}_0 (L \vert L)} }
\label{calBlimitfromlaplace}
\end{eqnarray}
Finally the three terms of Eq. \ref{laplaceCabcinvintegthree} allow to compute
\begin{eqnarray}
&& {\cal C}^{[\alpha]}_{\infty} (A,B \vert x_0) 
  = \lim_{s \to 0} \left[ s  {\hat {\cal C}}^{[\alpha]}_{s} (A,B \vert x_0)  \right]
 = 
\left[1 - \frac{{\hat G}_0 (L \vert 0)}{{\hat G}_0 (L \vert L)} \right] \frac{    {\hat G}^{abs(L)}_0 (0 \vert x_0)
  }{\left[ {\hat G}^{abs(L)}_0 (0 \vert 0)\right]^2}
  c_0
  e^{- \frac{A}{G^{abs(L)}_0 (0 \vert 0)}}  
       e^{- \frac{B}{G^{abs(0)}_0 (L \vert L)}}       
        \frac{  \sqrt{A}}{ \sqrt{B} }  
         I_1 \left( 2 c_0  \sqrt{AB} \right) 
  \nonumber \\
&&  {\cal C}^{[\beta]}_{\infty} (A,B \vert x_0)
  = \lim_{s \to 0} \left[ s  {\hat {\cal C}}^{[\beta]}_{s} (A,B \vert x_0)  \right]
= 
\left[1- \frac{{\hat G}_0 (0 \vert L)}{{\hat G}_0 (0 \vert 0)} \right] 
      \frac{    {\hat G}^{abs(0)}_0 (L \vert x_0)
  }{\left[ {\hat G}^{abs(0)}_0 (L \vert L)\right]^2}  c_0 e^{- \frac{A}{G^{abs(L)}_0 (0 \vert 0)}}  
       e^{- \frac{B}{G^{abs(0)}_0 (L \vert L)}} 
             \frac{  \sqrt{B}}{ \sqrt{A} }  
              I_1 \left( 2 c_0  \sqrt{AB} \right) 
\nonumber \\
&& {\cal C}^{[\gamma]}_{\infty} (A,B \vert x_0) 
  = \lim_{s \to 0} \left[ s  {\hat {\cal C}}^{[\gamma]}_{s} (A,B \vert x_0)  \right]
= e^{- \frac{A}{G^{abs(L)}_0 (0 \vert 0)}}  
       e^{- \frac{B}{G^{abs(0)}_0 (L \vert L)}}
         I_0 \left( 2 c_0  \sqrt{AB} \right) 
 \nonumber \\ &&  
\times \left[ 
\left( \frac{   \left[ 1 - \frac{{\hat G}_0 (0 \vert L)}{{\hat G}_0 (0 \vert 0)} \right]
  }{ {\hat G}^{abs(0)}_0 (L \vert L)} \right)
      \frac{    {\hat G}_0 (L \vert 0)     }{  \Delta_0  }  
  \left( \frac{   {\hat G}^{abs(L)}_0 (0 \vert x_0)  }{ {\hat G}^{abs(L)}_0 (0 \vert 0)} \right)  
  +
  \left(\frac{    \left[ 1 - \frac{
  {\hat G}_0 (L \vert 0)}{{\hat G}_0 (L \vert L)} \right]   }{{\hat G}^{abs(L)}_0 (0 \vert 0)}\right) 
   \frac{    {\hat G}_0 (0 \vert L) 
  }{\Delta_0  }  
   \left( \frac{    {\hat G}^{abs(0)}_0 (L \vert x_0)
  }{ {\hat G}^{abs(0)}_0 (L \vert L)} \right) 
\right]      
\label{calClimitfromlaplacethree}
\end{eqnarray}
so that the fourth contribution of Eq. \ref{piAlimit}
reads 
\begin{eqnarray}
 {\cal C}_{\infty} (A,B \vert x_0) 
&&  =  {\cal C}^{[\alpha]}_{\infty} (A,B \vert x_0)
  +   {\cal C}^{[\beta]}_{\infty} (A,B \vert x_0)  
 + {\cal C}^{[\gamma]}_{\infty} (A,B \vert x_0)
 \nonumber \\ 
 && = e^{- \frac{A}{G^{abs(L)}_0 (0 \vert 0)}}  
       e^{- \frac{B}{G^{abs(0)}_0 (L \vert L)}}  
       \left( \frac{   {\hat G}^{abs(L)}_0 (0 \vert x_0)  }{ {\hat G}^{abs(L)}_0 (0 \vert 0)} \right)      
 \nonumber \\
\times && 
\left[ 
\frac{  c_0  \left(1 - \frac{{\hat G}_0 (L \vert 0)}{{\hat G}_0 (L \vert L)} \right)  }{ {\hat G}^{abs(L)}_0 (0 \vert 0)}        \frac{  \sqrt{A}}{ \sqrt{B} }           I_1 \left( 2 c_0  \sqrt{AB} \right) 
+  \frac{    {\hat G}_0 (L \vert 0)  \left(1 - \frac{{\hat G}_0 (0 \vert L)}{{\hat G}_0 (0 \vert 0)} \right)}{\Delta_0 {\hat G}^{abs(0)}_0 (L \vert L)} 
I_0 \left( 2 c_0  \sqrt{AB} \right) 
 \right]
\nonumber \\
&&+ 
e^{- \frac{A}{G^{abs(L)}_0 (0 \vert 0)}}  
       e^{- \frac{B}{G^{abs(0)}_0 (L \vert L)}}
        \left( \frac{    {\hat G}^{abs(0)}_0 (L \vert x_0)  }{ {\hat G}^{abs(0)}_0 (L \vert L)} \right) 
 \nonumber \\
\times
&& \left[
      \frac{   c_0 \left(1- \frac{{\hat G}_0 (0 \vert L)}{{\hat G}_0 (0 \vert 0)} \right) }{ {\hat G}^{abs(0)}_0 (L \vert L)}    
             \frac{  \sqrt{B}}{ \sqrt{A} }  
              I_1 \left( 2 c_0  \sqrt{AB} \right) 
+   \frac{    {\hat G}_0 (0 \vert L)   \left( 1 - \frac{  {\hat G}_0 (L \vert 0)}{{\hat G}_0 (L \vert L)} \right)   }
  {\Delta_0{\hat G}^{abs(L)}_0 (0 \vert 0)} 
   I_0 \left( 2 c_0  \sqrt{AB} \right) 
\right]       
\label{calClimitfromlaplace}
\end{eqnarray}


\subsection{When $X(t)$ is recurrent : large deviations properties of the two intensive local times $(a,b)$ at large time   }

\label{subsec_rec}

When the diffusion process $X(t)$ is recurrent,  
it is more appropriate to introduce the two intensive local times of Eq. \ref{additiveIntensive}
and to analyze their large deviations properties 
thereby generalizing the previous works concerning the single intensive local time $a$ \cite{occupationsinai,us_LocalTime}.

\subsubsection{Rate function $I(a,b)$ for the two intensive local times $(a,b)$   }

Let us plug $A=ta $ and $B=tb $ into Eq. \ref{propagABalone} with its four contributions
\begin{eqnarray}
 \Pi_t(A=ta,B=tb \vert x_0)  && =   \delta(a) \delta(b) \frac{1}{t^2} S^{abs(0,L)}_t (  x_0)
+ \theta(a>0) \delta(b)\frac{1}{t} {\cal A}^{]-\infty,L[}_t (ta \vert x_0)
 + \delta(a)\theta(b>0) \frac{1}{t} {\cal B}^{]0,+\infty[}_t (tb\vert x_0)
\nonumber \\
&&+ \theta(a>0) \theta(b>0)  {\cal C}_t (ta,tb \vert x_0)
\label{propagABaloneintensive}
\end{eqnarray}
Since the leading behavior is the exponential decay with respect to the time $t$
of Eq. \ref{rateab} that involves the positive rate function $ I (a, b ) \geq 0 $ 
defined for $a \in [0,+\infty[$ and $b \in [0,+\infty[$,
the four contributions of Eq. \ref{propagABaloneintensive}
are then governed by the four different exponential factors
\begin{eqnarray}
 S^{abs(0,L)}_t (x_0) &&  \oppropto_{t \to +\infty} e^{- t I (a=0, b=0 )} 
\nonumber \\
 {\cal A}^{]-\infty,L[}_t (A=t a \vert x_0) &&  \oppropto_{t \to +\infty} e^{- t I (a, b=0 )} 
\nonumber \\
 {\cal B}^{]0,+\infty[}_t ( B=t b \vert x_0) &&  \oppropto_{t \to +\infty} e^{- t I (a=0, b )} 
\nonumber \\
  {\cal C}_t (A=t a,B=t b  \vert x_0) &&  \oppropto_{t \to +\infty} e^{- t I (a, b )} 
\label{rateabfour}
\end{eqnarray}

In the following subsections, the goal is to evaluate at leading order 
the various contributions including the prefactors in order to have the dependence with respect
to the initial position $x_0$
that will be needed later to construct conditioned processes.
It is more convenient to begin with the fourth contribution ${\cal C}_t (A=t a,B=t b  \vert x_0) $ 
since it is the only one that involves the full joint rate function $I(a,b)$ of the two variables $a$ and $b$.


\subsubsection { Evaluation of the fourth contribution ${\cal C}_t (A=t a,B=t b  \vert x_0)  $
at leading order for large time $t$}

The contribution ${\hat  {\cal C}}_s (A=t a,B=t b  \vert x_0)$
of Eq. \ref{laplaceCabcinvinteg} displays the asymptotic behavior given by Eq. \ref{laplaceCabcinvinteglargetot} for large time $t$
\begin{eqnarray}
 {\hat {\cal C}}_{s} (A=ta,B=tb \vert x_0)  && \opsimeq_{t \to +\infty}    \frac{1}{2 \sqrt{\pi t} }
      \left[    \frac{ \alpha_s^{]-\infty,L[}( x_0)     \sqrt{c_s} a^{\frac{1}{4}} }{    b^{\frac{3}{4}}}   
      +  \frac{ \beta_s^{]0,+\infty[}(x_0 )   \sqrt{c_s }b^{\frac{1}{4}} }{   a^{\frac{3}{4}}}  
      +   \frac{  \gamma_s( x_0)}{\sqrt{ c_s }  a^{\frac{1}{4}}b^{\frac{1}{4}}}   
      \right]  
      \nonumber \\
  &&    \times      e^{- t \left[ \frac{a}{{\hat G}^{abs(L)}_s (0 \vert 0)} + \frac{b}{{\hat G}^{abs(0)}_s (L \vert L)} - 2 c_s  \sqrt{ab} \right] } 
\label{laplaceCabcinvinteglargetotC}
\end{eqnarray}

The Laplace inversion of this asymptotic behavior
allows to compute
the fourth contribution ${\cal C}_t (A=t a,B=t b  \vert x_0)   $ at leading order for large $t$
\begin{eqnarray}
&& {\cal C}_t(A=t a,B=t b \vert x_0)  
= \int_{c-i\infty}^{c+i\infty} \frac{ds}{2 i \pi}  e^{s t}
{\hat  {\cal C}}_s (A=ta,B=tb \vert x_0)
\label{Ccontour}
 \\
&& \opsimeq_{t \to +\infty} 
 \int_{c-i\infty}^{c+i\infty} \frac{ds}{2 i \pi} 
  \frac{1}{2 \sqrt{\pi t} }
      \left[   \sqrt{c_s} \left( \frac{ \alpha_s^{]-\infty,L[}( x_0)     a^{\frac{1}{4}} }{    b^{\frac{3}{4}}}   
      +  \frac{ \beta_s^{]0,+\infty[}(x_0 )  b^{\frac{1}{4}} }{   a^{\frac{3}{4}}}  \right)
      +   \frac{  \gamma_s( x_0)}{\sqrt{ c_s }  a^{\frac{1}{4}}b^{\frac{1}{4}}}   
      \right]  
           e^{- t \left[ \frac{a}{{\hat G}^{abs(L)}_s (0 \vert 0)} + \frac{b}{{\hat G}^{abs(0)}_s (L \vert L)} - 2 c_s  \sqrt{ab} - s \right] }  
\nonumber
\end{eqnarray}

 For large $t$, the saddle-point evaluation of this integral over $s$ involves the solution $s_{a,b}$ of
\begin{eqnarray}
  0  =   \partial_s \left[\frac{a}{ {\hat G}^{abs(L)}_s (0 \vert 0)}+ \frac{b}{ {\hat G}^{abs(0)}_s (L \vert L)} -2 c_s \sqrt{ab} -s   \right] 
  \label{saddlesab}
\end{eqnarray}
One then needs to make the change of variable
\begin{eqnarray}
 s = s_{a,b} + i \omega
\label{complex}
\end{eqnarray}
and to use the Taylor expansion at second order in $\omega$ 
\begin{eqnarray}
 \left[  \frac{a}{ {\hat G}^{abs(L)}_s (0 \vert 0)}
 + \frac{b}{ {\hat G}^{abs(0)}_s (L \vert L)} -2 c_s \sqrt{ab} -s  \right]_{s=s_{a,b} + i \omega}  
 = I(a,b) + 0 + \frac{\omega^2}{2} K(a,b) + o\left( \omega^2 \right)
\label{taylorab}
\end{eqnarray}
that involves the two functions
\begin{eqnarray}
I(a,b) && = \left[  \frac{a}{ {\hat G}^{abs(L)}_s (0 \vert 0)}
 + \frac{b}{ {\hat G}^{abs(0)}_s (L \vert L)} -2 c_s \sqrt{ab} -s  \right]_{s=s_{a,b} }
\nonumber \\
K(a,b) && = -  \left[ \partial^2_s \left(   \frac{a}{ {\hat G}^{abs(L)}_s (0 \vert 0)}
 + \frac{b}{ {\hat G}^{abs(0)}_s (L \vert L)} -2 c_s \sqrt{ab} -s    \right) \right]_{s=s_{a,b}}
\label{iafromsaddle}
\end{eqnarray}
The final result for the leading order of Eq. \ref{Ccontour} based on the 
remaining Gaussian integral over $\omega$
reads 
\begin{eqnarray}
 {\cal C}_t(A=t a,B=t b \vert x_0)  
&& \opsimeq_{t \to +\infty} 
  \frac{e^{- t I (a, b )}}{2 \sqrt{\pi t} }
      \left[    \sqrt{c_s} \left( \frac{ \alpha_s^{]-\infty,L[}( x_0)     a^{\frac{1}{4}} }{    b^{\frac{3}{4}}}   
      +  \frac{ \beta_s^{]0,+\infty[}(x_0 )  b^{\frac{1}{4}} }{   a^{\frac{3}{4}}}  \right) 
      +   \frac{  \gamma_s( x_0)}{\sqrt{ c_s }  a^{\frac{1}{4}}b^{\frac{1}{4}}}   
      \right]_{s=s_{a,b}} 
 \int_{-\infty}^{+\infty} \frac{d\omega}{2  \pi} 
           e^{- t \frac{K(a,b)}{2}  \omega^2}  
           \nonumber \\
&& \opsimeq_{t \to +\infty} 
  \frac{e^{- t I (a, b )}}{ 2 \pi t \sqrt{  2  K(a,b)}  }
      \left[   \sqrt{c_s} \left( \frac{ \alpha_s^{]-\infty,L[}( x_0)     a^{\frac{1}{4}} }{    b^{\frac{3}{4}}}   
      +  \frac{ \beta_s^{]0,+\infty[}(x_0 )  b^{\frac{1}{4}} }{   a^{\frac{3}{4}}}  \right)
      +   \frac{  \gamma_s( x_0)}{\sqrt{ c_s }  a^{\frac{1}{4}}b^{\frac{1}{4}}}   
      \right]_{s=s_{a,b}}        
\label{Ccontoursaddle}
\end{eqnarray}


\subsubsection { Evaluation of the second contribution $ {\cal A}^{]-\infty,L[}_t (A=t a \vert x_0)  $ at leading order
for large time $t$}

The second contribution $ {\cal A}^{]-\infty,L[}_t (A=t a \vert x_0)  $
can be obtained from the Laplace inverse of ${\hat {\cal A}}^{]-\infty,L[}_s (A=t a \vert x_0) $
of Eq. \ref{calApinvinteg}
\begin{eqnarray}
 {\cal A}^{]-\infty,L[}_t (A=t a \vert x_0) 
 = \int_{c-i\infty}^{c+i\infty} \frac{ds}{2 i \pi}  e^{s t}  {\hat {\cal A}}^{]-\infty,L[}_s (A=t a \vert x_0)
 = \int_{c-i\infty}^{c+i\infty} \frac{ds}{2 i \pi} \alpha_s^{]-\infty,L[}(x_0)
  e^{ - t \left[ \frac{a}{ {\hat G}^{abs(L)}_s (0 \vert 0)} -s \right] }
\label{Acontour}
\end{eqnarray}

For large $t$, the saddle-point evaluation of this integral involves the solution $s_{a,0}$ of
\begin{eqnarray}
  0  =   \partial_s \left[  \frac{a}{ {\hat G}^{abs(L)}_s (0 \vert 0)} -s   \right] 
  \label{saddlesazero}
\end{eqnarray}
that corresponds to the special case $b=0$ in Eq. \ref{saddlesab}.
So the Taylor expansion of Eq. \ref{taylorab} for special case $b=0$
\begin{eqnarray}
 \left[  \frac{a}{ {\hat G}^{abs(L)}_s (0 \vert 0)} -s  \right]_{s=s_{a,0} + i \omega}  
 = I(a,0) + 0 + \frac{\omega^2}{2} K(a,0) + o\left( \omega^2 \right)
\label{taylora}
\end{eqnarray}
yields that Eq. \ref{Acontour}
reads the leading order
\begin{eqnarray}
 {\cal A}^{]-\infty,L[}_t (A=t a \vert x_0) 
&&  \opsimeq_{t \to + \infty} e^{-t I(a,0) } \alpha_{s_{a,0}}^{]-\infty,L[}(x_0)
  \int_{-\infty}^{+\infty} \frac{d\omega}{2  \pi} 
  e^{ - t  \frac{ K(a,0)}{2} \omega^2}
  \nonumber \\
 &&  \opsimeq_{t \to + \infty} e^{-t I(a,0) } \frac{ \alpha_{s_{a,0}}^{]-\infty,L[}(x_0) }{\sqrt{ 2 \pi K(a,0) t} } 
\label{Acontoursaddle}
\end{eqnarray}


\subsubsection { Evaluation of the third contribution $ {\cal B}^{]0,+\infty[}_t (B=t b \vert x_0)    $ 
at leading order for large time $t$}

The third contribution $ {\cal B}^{]0,+\infty[}_t (B=t b \vert x_0)    $
can be obtained from the Laplace inverse of $ {\hat {\cal B}}^{]0,+\infty[}_s (B=t b \vert x_0) $
of Eq. \ref{calBqinvinteg}
\begin{eqnarray}
 {\cal B}^{]0,+\infty[}_t (B=t b \vert x_0)  
 = \int_{c-i\infty}^{c+i\infty} \frac{ds}{2 i \pi}  e^{s t}
  {\hat {\cal B}}^{]0,+\infty[}_s (B=t b \vert x_0)
 = \int_{c-i\infty}^{c+i\infty} \frac{ds}{2 i \pi}  
 \beta_s^{]0,+\infty[}(x_0)
  e^{ - t \left[ \frac{b}{ {\hat G}^{abs(0)}_s (L \vert L)} -s \right]}
\label{Bcontour}
\end{eqnarray}

For large $t$, the saddle-point evaluation of this integral involves the solution $s_{0,b}$ of
\begin{eqnarray}
  0  =   \partial_s \left[ \frac{b}{ {\hat G}^{abs(0)}_s (L \vert L)}  -s   \right] 
  \label{saddleszerob}
\end{eqnarray}
that corresponds to the special case $a=0$ in Eq. \ref{saddlesab}.
So the Taylor expansion of Eq. \ref{taylorab} for special case $a=0$
\begin{eqnarray}
 \left[  \frac{b}{ {\hat G}^{abs(0)}_s (L \vert L)}  -s  \right]_{s=s_{0,b} + i \omega}  
 = I(0,b) + 0 + \frac{\omega^2}{2} K(0,b) + o\left( \omega^2 \right)
\label{taylorb}
\end{eqnarray}
yields that Eq. \ref{Bcontour}
reads the leading order
\begin{eqnarray}
 {\cal B}^{abs(0)}_t (B=t b \vert x_0)  
&&  \opsimeq_{t \to + \infty} e^{-t I(0,b) } \beta_{s_{0,b}}^{]0,+\infty[}(x_0)
  \int_{-\infty}^{+\infty} \frac{d\omega}{2  \pi} 
  e^{ - t  \frac{ K(0,b)}{2} \omega^2}
  \nonumber \\
 &&  \opsimeq_{t \to + \infty} e^{-t I(0,b) } \frac{ \beta_{s_{0,b}}^{]0,+\infty[} (x_0) }{\sqrt{ 2 \pi K(0,b) t} } 
\label{Bcontoursaddle}
\end{eqnarray}


\subsubsection { Conclusion}

Putting everything together, one obtains the asymptotic behavior at leading order for large time $t$
of Eq. \ref{propagABaloneintensive}
\begin{eqnarray}
 \Pi_t(A=ta,B=tb \vert x_0) && \opsimeq_{t \to + \infty}    \delta(a) \delta(b) \frac{1}{t^2} S^{abs(0,L)}_t (  x_0)
 \nonumber \\
&&+ \theta(a>0) \delta(b)  \frac{ \alpha_{s_{a,0}}^{]-\infty,L[}(x_0) e^{-t I(a,0) }}{\sqrt{ 2 \pi K(a,0) } t^{\frac{3}{2}} } 
 \nonumber \\
&& + \delta(a)\theta(b>0)  \frac{ \beta_{s_{0,b}}^{]0,+\infty[} (x_0)  e^{-t I(0,b) }}{\sqrt{ 2 \pi K(0,b) } t^{\frac{3}{2}}} 
 \nonumber \\
&&+ \theta(a>0) \theta(b>0) 
  \frac{e^{- t I (a, b )}}{ 2 \pi t \sqrt{ 2  K(a,b)}  }
      \left[   \sqrt{c_s} \left(  \frac{ \alpha_s^{]-\infty,L[}( x_0)     a^{\frac{1}{4}} }{    b^{\frac{3}{4}}}   
      +  \frac{ \beta_s^{]0,+\infty[}(x_0 )  b^{\frac{1}{4}} }{   a^{\frac{3}{4}}}  \right)
      +   \frac{  \gamma_s( x_0)}{\sqrt{ c_s }  a^{\frac{1}{4}}b^{\frac{1}{4}}}   
      \right]_{s=s_{a,b}}       
\label{propagABaloneintensivelarget}
\end{eqnarray}


\section{Statistics of the sum $\Sigma(t)=A(t)+B(t)$ of the two local times $A(t)$ and $B(t)$ }

\label{sec_sum}

In this section, we focus on the sum $\Sigma(t) $
of the two local times
\begin{eqnarray}
\Sigma(t) \equiv A(t) +B(t) && = \int_0^t d\tau \left[ \delta ( X(\tau) ) + \delta ( X(\tau)-L ) \right]
\label{defsumab}
\end{eqnarray}
whose statistical properties are simpler. 


\subsection{ Propagator $P_t(x,\Sigma \vert x_0) $ for the position $x$ and the sum $\Sigma$ of the two local times }

\subsubsection{ Singular contribution in $ \delta(\Sigma) $ and regular contribution in $\theta(\Sigma>0) $ }

The joint propagator $P_t(x,\Sigma \vert x_0) $ 
can be computed from the joint propagator $P_t(x,A,B \vert x_0) $ of Eq. \ref{four}
with its four contributions
\begin{eqnarray}
P_t(x,\Sigma \vert x_0) && = \int_0^{+\infty} dA \int_0^{+\infty} dB P_t(x,A,B \vert x_0) \delta(\Sigma-(A+B))
\nonumber \\
&& =  \delta(\Sigma) G^{abs(0,L)}_t (x \vert x_0)
+  \theta(\Sigma>0) \Upsilon_t(x,\Sigma \vert x_0)
 \label{fourforsigma}
\end{eqnarray}
where the regular contribution $\Upsilon_t(x,\Sigma \vert x_0) $ reads
\begin{eqnarray}
 \Upsilon_t(x,\Sigma \vert x_0) = {\cal A}^{abs(L)}_t (x,\Sigma \vert x_0) +  {\cal B}^{abs(0)}_t (x,\Sigma \vert x_0)
+ \int_0^{\Sigma} dA      {\cal C}_t (x,A,\Sigma-A \vert x_0)
 \label{upsilont}
\end{eqnarray}
or equivalently for the time Laplace transform
\begin{eqnarray}
{\hat P}_s(x,\Sigma \vert x_0)  \equiv \int_0^{+\infty} dt e^{-st } P_t(x,\Sigma \vert x_0)
 =  \delta(\Sigma) {\hat G}^{abs(0,L)}_s (x \vert x_0)
+  \theta(\Sigma>0) {\hat \Upsilon}_s(x,\Sigma \vert x_0)
 \label{timelaplacepropagatorsigma}
\end{eqnarray}
with 
\begin{eqnarray}
{\hat \Upsilon}_s(x,\Sigma \vert x_0) =   {\hat  {\cal A}}_s (x,\Sigma \vert x_0)
+ {\hat  {\cal B}}_s (x,\Sigma \vert x_0)
+\int_0^{\Sigma} dA    {\hat  {\cal C}}_s (x,A,\Sigma-A \vert x_0)
 \label{upsilons}
\end{eqnarray}

However, instead of computing the convolution of the last term,
it is simpler to 
consider the further Laplace transform with respect to $\Sigma$ of parameter $p$
as we now describe.

\subsubsection{ Explicit regular contribution ${\hat {\tilde \Upsilon}}_{s,p}(x \vert x_0) $ }

The further Laplace transform of Eq. \ref{timelaplacepropagatorsigma}
with respect to $\Sigma$ of parameter $p$
corresponds to ${\hat {\tilde P}}_{s,p,p} (x \vert x_0)   $ computed in Eq. \ref{dysonpoly}
for the special case $q=p$
 \begin{eqnarray}
&& \int_0^{+\infty} d \Sigma e^{- p \Sigma}{\hat P}_s(x,\Sigma \vert x_0) 
 = {\hat G}^{abs(0,L)}_s (x \vert x_0)
+  {\hat {\tilde \Upsilon}}_{s,p}(x \vert x_0)
 =\int_0^{+\infty} dA e^{-p A } \int_0^{+\infty} dB e^{-p B }{\hat P}_s(x,A,B \vert x_0)
\nonumber \\
&& = {\hat {\tilde P}}_{s,p,p} (x \vert x_0)   
= {\hat G}_s (x \vert x_0)
 -   \frac{p^2  \Delta_s \Omega_s (x \vert x_0)  
  +p  \left[ {\hat G}_s (x \vert 0) {\hat G}_s (0 \vert x_0)
   +  {\hat G}_s (x \vert L) {\hat G}_s (L \vert x_0) \right] }
{ \Delta_s p^2 + p  {\hat G}_s (0 \vert 0)   + p {\hat G}_s (L \vert L)  + 1 } 
\label{dysonpolyupscal}
\end{eqnarray}
Using Eq. \ref{pqinfty}, one obtains the explicit form of ${\hat {\tilde \Upsilon}}_{s,p}(x \vert x_0) $
 \begin{eqnarray}
   {\hat {\tilde \Upsilon}}_{s,p}(x \vert x_0)
&& =   \Omega_s (x \vert x_0)
 -   \frac{p^2  \Delta_s \Omega_s (x \vert x_0)  
  +p  \left[ {\hat G}_s (x \vert 0) {\hat G}_s (0 \vert x_0)
   +  {\hat G}_s (x \vert L) {\hat G}_s (L \vert x_0) \right] }
{ \Delta_s p^2 + p \left[ {\hat G}_s (0 \vert 0)   +  {\hat G}_s (L \vert L) \right] + 1 } 
\nonumber \\
&& =  
    \frac{ \Omega_s (x \vert x_0)
   +  p \Delta_s \left[  \alpha_s^{[00]}(x \vert x_0)+  \beta_s^{[LL]}(x \vert x_0) \right] }
{ \Delta_s p^2 + p \left[ {\hat G}_s (0 \vert 0)   +  {\hat G}_s (L \vert L) \right] + 1 } 
\label{dysonpolyups}
\end{eqnarray}
in terms of the functions $\alpha_s^{[00]}(x \vert x_0) $ and $\beta_s^{[LL]}(x \vert x_0) $ introduced in Eqs \ref{alpha}
and \ref{beta}.


\subsubsection{ Laplace inversion of ${\hat {\tilde \Upsilon}}_{s,p}(x \vert x_0) $ with respect to $p$
to obtain ${\hat  \Upsilon}_{s}(x,\Sigma \vert x_0) $}

The factorization of the denominator of Eq. \ref{dysonpolyups}
 \begin{eqnarray}
\Delta_s p^2 + p \left[ {\hat G}_s (0 \vert 0)   +  {\hat G}_s (L \vert L) \right] + 1 
= \Delta_s (p+\lambda_s^+)(p+\lambda_s^-)
\label{dysonpolyupsfactor}
\end{eqnarray}
involves the two positive functions $\lambda_s^{\pm} >0 $ given by
 \begin{eqnarray}
\lambda_s^{\pm} && \equiv \frac{  \left[ {\hat G}_s (0 \vert 0)   +  {\hat G}_s (L \vert L) \right] 
\pm \sqrt{ \left[ {\hat G}_s (0 \vert 0)   +  {\hat G}_s (L \vert L) \right]^2
- 4\Delta_s } }{ 2 \Delta_s}
\nonumber \\
&& = \frac{  \left[ {\hat G}_s (0 \vert 0)   +  {\hat G}_s (L \vert L) \right] 
\pm \sqrt{ \left[ {\hat G}_s (0 \vert 0)   -  {\hat G}_s (L \vert L) \right]^2
+ 4 {\hat G}_s (0 \vert L) {\hat G}_s (L \vert 0) } }{ 2 \Delta_s}
\nonumber \\
&& = \frac{1}{2} \left[\frac{1}{ {\hat G}^{abs(0)}_s (L \vert L) } +  \frac{1}{ {\hat G}^{abs(L)}_s (0 \vert 0) }    
\pm D_s \right]
\label{dysonpolyupsroots}
\end{eqnarray}
where we have introduced the following notation for the difference
 \begin{eqnarray}
D_s \equiv \lambda_s^+ -\lambda_s^- =
\sqrt{ \left(\frac{1}{ {\hat G}^{abs(0)}_s (L \vert L) } -  \frac{1}{ {\hat G}^{abs(L)}_s (0 \vert 0) }  \right)^2
+ 4 c_s^2 }
\label{dssquareroot}
\end{eqnarray}

So the partial fraction decomposition of Eq. \ref{dysonpolyups} with respect to the variable $p$
reduces to
 \begin{eqnarray}
   {\hat {\tilde \Upsilon}}_{s,p}(x \vert x_0)
 =  \frac{\Lambda_s^+(x \vert x_0) }{p+\lambda_s^+} +  \frac{\Lambda_s^-(x \vert x_0) }{p+\lambda_s^-}
\label{dysonpolyupsfrac}
\end{eqnarray}
where the identification with Eq. \ref{dysonpolyups} yields that
the two numerators $\Lambda_s^{\pm}(x \vert x_0) $ satisfy the following system using Eq. \ref{omegaabc}
 \begin{eqnarray}
\lambda_s^- \Lambda_s^+(x \vert x_0)  +\lambda_s^+ \Lambda_s^-(x \vert x_0) 
&& = \frac{\Omega_s (x \vert x_0)}{\Delta_s}  
=   \frac{ \alpha_s^{[00]}(x \vert x_0)  }{ {\hat G}^{abs(0)}_s (L \vert L) }   
   +     \frac{ \beta_s^{[LL]}(x \vert x_0)  }{ {\hat G}^{abs(L)}_s (0 \vert 0) }  
 +   \gamma_s(x \vert x_0)   
\nonumber \\
 \Lambda_s^+(x \vert x_0) + \Lambda_s^-(x \vert x_0)  
&& = \alpha_s^{[00]}(x \vert x_0)+  \beta_s^{[LL]}(x \vert x_0)
\label{dysonpolyupsfracueq}
\end{eqnarray}
and thus read
 \begin{eqnarray}
 \Lambda_s^+(x \vert x_0) && = \frac{
   \alpha_s^{[00]}(x \vert x_0) \left[ \lambda_s^+ -  \frac{ 1  }{ {\hat G}^{abs(0)}_s (L \vert L) } \right]
 +  \beta_s^{[LL]}(x \vert x_0) \left[ \lambda_s^+    -     \frac{ 1  }{ {\hat G}^{abs(L)}_s (0 \vert 0) }  \right]
 -   \gamma_s(x \vert x_0)   }
 {D_s}
 \nonumber \\
  \Lambda_s^-(x \vert x_0) && 
  = \frac{ 
 \alpha_s^{[00]}(x \vert x_0) \left[ \frac{ 1  }{ {\hat G}^{abs(0)}_s (L \vert L) } -  \lambda_s^- \right]
   +  \beta_s^{[LL]}(x \vert x_0)  \left[   \frac{ 1 }{ {\hat G}^{abs(L)}_s (0 \vert 0) }  -  \lambda_s^- \right]
 +   \gamma_s(x \vert x_0)   
}{D_s}
\label{dysonpolyupsfracusol}
\end{eqnarray}

The Laplace inversion with respect to $p$ of Eq. \ref{dysonpolyupsfrac} yields
the following two exponential functions with respect to $\Sigma$ of parameters $\lambda_s^{\pm}$
 \begin{eqnarray}
   {\hat  \Upsilon}_{s}(x,\Sigma \vert x_0)
    =  \Lambda_s^+(x \vert x_0) e^{- \lambda_s^+ \Sigma } +  \Lambda_s^-(x \vert x_0) e^{- \lambda_s^- \Sigma }
\label{upsilonres}
\end{eqnarray}


\subsection{ Distribution $\Pi_t(\Sigma \vert x_0)= \int dx P_t(x,\Sigma \vert x_0)$ of the sum $\Sigma$ at time $t$}

\label{subsec_sigma}

The distribution $\Pi_t(\Sigma \vert x_0)  $ of the sum $\Sigma$ at time $t$ when starting at position $x_0$ can be obtained from
the integration over the final position $x$ of
the $P_t(x,\Sigma \vert x_0) $ of Eq. \ref{fourforsigma}
\begin{eqnarray}
 \Pi_t(\Sigma \vert x_0)  \equiv \int_{-\infty}^{+\infty} dx  P_t(x,\Sigma \vert x_0)
 =    \delta(\Sigma) S^{abs(0,L)}_t ( x_0)
+  \theta(\Sigma>0) \Upsilon_t(\Sigma \vert x_0)
\label{pisigma}
\end{eqnarray}
where
\begin{eqnarray}
\Upsilon_t(\Sigma \vert x_0) \equiv \int_{-\infty}^{+\infty} dx   \Upsilon_t(x,\Sigma \vert x_0)
\label{upsiloninteg}
\end{eqnarray}
Its time Laplace transform ${\hat  \Upsilon}_{s}(\Sigma \vert x_0) $
can be obtained from the integration over $x$ of ${\hat  \Upsilon}_{s}(x,\Sigma \vert x_0) $ Eq. \ref{upsilonres}
 \begin{eqnarray}
{\hat  \Upsilon}_{s}(\Sigma \vert x_0)= \int_{-\infty}^{+\infty} dx  {\hat  \Upsilon}_{s}(x,\Sigma \vert x_0)
    =  \Lambda_s^+( x_0) e^{- \lambda_s^+ \Sigma } +  \Lambda_s^-( x_0) e^{- \lambda_s^- \Sigma }
\label{upsilonresinteg}
\end{eqnarray}
where the amplitudes $\Lambda_s^{\pm}( x_0) $ 
can be computed using the explicit expressions of 
the functions $\alpha_s^{]-\infty,L[}( x_0) $, $\beta_s^{]0,+\infty[}( x_0) $ 
and $\gamma_s( x_0)  $ of Eqs \ref{alphainteg}, \ref{betainteg} and \ref{gammainteg}
 \begin{eqnarray}
 \Lambda_s^+( x_0) && = \int_{-\infty}^{+\infty} dx \Lambda_s^+(x \vert x_0) 
  =\frac{
   \alpha_s^{]-\infty,L[}( x_0) \left[ \lambda_s^+ -  \frac{ 1  }{ {\hat G}^{abs(0)}_s (L \vert L) } \right]
 +  \beta_s^{]0,+\infty[}( x_0) \left[ \lambda_s^+    -     \frac{ 1  }{ {\hat G}^{abs(L)}_s (0 \vert 0) }  \right]
 -   \gamma_s( x_0)   }
 {D_s}
 \nonumber \\
 && =\lambda_s^+ \frac{\left(   {\hat G}_s (L \vert L) - {\hat G}_s (L \vert 0)  -\frac{1}{\lambda_s^+}
  \right)  \frac{{\hat G}^{abs(L)}_s (0 \vert x_0)}{{\hat G}^{abs(L)}_s (0 \vert 0)}
 +\left(  {\hat G}_s (0 \vert 0) - {\hat G}_s (0 \vert L) -\frac{1}{\lambda_s^+} \right)   \frac{{\hat G}^{abs(0)}_s (L \vert x_0)}{{\hat G}^{abs(0)}_s (L \vert L)}
  }{ s  \left[ \frac{1}{\lambda_s^-} - \frac{1}{\lambda_s^+} \right]   }
 \nonumber \\
 \Lambda_s^-( x_0) && = \int_{-\infty}^{+\infty} dx  \Lambda_s^-(x \vert x_0)  
 =\frac{ 
 \alpha_s^{]-\infty,L[}( x_0) \left[ \frac{ 1  }{ {\hat G}^{abs(0)}_s (L \vert L) } -  \lambda_s^- \right]
   +  \beta_s^{[LL]}( x_0)  \left[   \frac{ 1 }{ {\hat G}^{abs(L)}_s (0 \vert 0) }  -  \lambda_s^- \right]
 +   \gamma_s( x_0)   
}{D_s}
 \nonumber \\
 && =\lambda_s^- \frac{\left( \frac{1}{\lambda_s^-}-   \left[{\hat G}_s (L \vert L) - {\hat G}_s (L \vert 0) \right]  \right)  \frac{{\hat G}^{abs(L)}_s (0 \vert x_0)}{{\hat G}^{abs(L)}_s (0 \vert 0)}
 +\left( \frac{1}{\lambda_s^-} -   \left[{\hat G}_s (0 \vert 0) - {\hat G}_s (0 \vert L) \right]  \right) 
   \frac{{\hat G}^{abs(0)}_s (L \vert x_0)}{{\hat G}^{abs(0)}_s (L \vert L)}
  }{ s \left[ \frac{1}{\lambda_s^-} - \frac{1}{\lambda_s^+} \right]  }
\label{dysonpolyupsfracusolinteg}
\end{eqnarray}
so that the normalization of Eq. \ref{upsilonresinteg} over $\Sigma$
 \begin{eqnarray}
\int_0^{+\infty} d \Sigma {\hat  \Upsilon}_{s}(\Sigma \vert x_0)
    = \frac{ \Lambda_s^+( x_0) }{ \lambda_s^+ } + \frac{ \Lambda_s^-( x_0) }{ \lambda_s^-  }
    = \frac{1}{s} \left[\frac{{\hat G}^{abs(L)}_s (0 \vert x_0)}{{\hat G}^{abs(L)}_s (0 \vert 0)}
    + \frac{{\hat G}^{abs(0)}_s (L \vert x_0)}{{\hat G}^{abs(0)}_s (L \vert L)} \right] =  \frac{1}{s} -  {\hat S}^{abs(0,L)}_s (x_0)
\label{upsilonresintegnorma}
\end{eqnarray}
is complementary to $ {\hat S}^{abs(0,L)}_s (x_0) $ of Eq. \ref{laplacesurvival}
as it should.


\subsection{  Moments $m^{[k]}_t( x_0)= \langle \Sigma^k(t) \rangle_{x_0}$ of the sum $\Sigma(t)$ when starting at $x_0$  }

The moments of order $k >0 $ of $\Sigma$ 
at time $t$ can be computed from the distribution 
$ \Pi_t(\Sigma \vert x_0) $ of Eq. \ref{pisigma}
\begin{eqnarray}
m^{[k]}_t( x_0) = \langle \Sigma^k(t) \rangle_{x_0}  
= \int_{0}^{+\infty} d\Sigma \Sigma^k \Pi_t(\Sigma \vert x_0)
 = \int_{0}^{+\infty} d\Sigma \Sigma^k \Upsilon_t(\Sigma \vert x_0)
\label{mkdef}
\end{eqnarray}
Their time-Laplace transforms 
can be computed from the time Laplace transform ${\hat  \Upsilon}_{s}(\Sigma \vert x_0)  $
of Eq. \ref{upsiloninteg}
\begin{eqnarray}
 {\hat m}^{[k]}_s( x_0)  && \equiv \int_{0}^{+\infty} dt e^{-s t} m^{[k]}_t( x_0)
   =\int_{0}^{+\infty} d\Sigma \Sigma^k {\hat  \Upsilon}_{s}(\Sigma \vert x_0)
   = \int_{0}^{+\infty} d\Sigma \Sigma^k \left[ 
\Lambda_s^+( x_0) e^{- \lambda_s^+ \Sigma } +  \Lambda_s^-( x_0) e^{- \lambda_s^- \Sigma }
\right]
\nonumber \\
&&   = k! \left( \frac{\Lambda_s^+( x_0) } { (\lambda_s^+)^{k+1} } 
+ \frac{\Lambda_s^-( x_0) } { ( \lambda_s^-)^{k+1} }
\right)
\label{mklaplace}
\end{eqnarray}


\subsection{When $X(t)$ is transient : the sum 
$\Sigma(t) $ remain a finite random variable for $t \to +\infty$  }

As already explained in subsection \ref{subsec_trans}, the two local times $(A,B)$ 
remain finite random variables for $t \to +\infty$.
It is thus interesting to write the asymptotic distribution of the sum $\Sigma$
obtained from Eq. \ref{pisigma}
\begin{eqnarray}
 \Pi_{\infty}(\Sigma \vert x_0) = \lim_{t \to +\infty}  \Pi_t(\Sigma \vert x_0)
 =    \delta(\Sigma) S^{abs(0,L)}_{\infty} ( x_0)
+  \theta(\Sigma>0) \Upsilon_{\infty}(\Sigma \vert x_0)
\label{pisigmainfty}
\end{eqnarray}
where the singular contribution $S^{abs(0,L)}_{\infty} ( x_0) $ has been already given in Eq. \ref{survivallimitfromlaplace}, while the regular contribution $\Upsilon_{\infty}(\Sigma \vert x_0) $
 involving the infinite-time limit $t \to +\infty$ can be obtained 
from the Laplace transform ${\hat  \Upsilon}_{s}(\Sigma \vert x_0) $ of Eq. \ref{upsilonresinteg}
 by considering the limit $s \to 0$ of
\begin{eqnarray}
\Upsilon_{\infty}(\Sigma \vert x_0) && = \lim_{s \to 0} \left[ s {\hat  \Upsilon}_{s}(\Sigma \vert x_0)\right]
= \lim_{s \to 0} \left[ s  \Lambda_s^+( x_0) e^{- \lambda_s^+ \Sigma } +  s \Lambda_s^-( x_0) e^{- \lambda_s^- \Sigma } \right]
\nonumber \\
&& = \xi^+(x_0) e^{- \lambda_0^+ \Sigma } +   \xi^-(x_0)  e^{- \lambda_0^- \Sigma } 
\label{upsilonresinteginfty}
\end{eqnarray}
where the two amplitudes $\xi^{\pm}(x_0) $ can be obtained from Eq. \ref{dysonpolyupsfracusolinteg}
 \begin{eqnarray}
\xi^+(x_0) && = \lim_{s \to 0} \left[ s \Lambda_s^+( x_0) \right] 
 \nonumber \\ && 
= \lambda_0^+ \frac{\left(   {\hat G}_0 (L \vert L) - {\hat G}_0 (L \vert 0)  -\frac{1}{\lambda_0^+}
  \right)  \frac{{\hat G}^{abs(L)}_0 (0 \vert x_0)}{{\hat G}^{abs(L)}_0 (0 \vert 0)}
 +\left(  {\hat G}_0 (0 \vert 0) - {\hat G}_0 (0 \vert L) -\frac{1}{\lambda_0^+} \right)   \frac{{\hat G}^{abs(0)}_0 (L \vert x_0)}{{\hat G}^{abs(0)}_0 (L \vert L)}
  }{   \left[ \frac{1}{\lambda_0^-} - \frac{1}{\lambda_0^+} \right]   }
 \nonumber \\  
\xi^-(x_0) && = \lim_{s \to 0} \left[ s \Lambda_s^-( x_0) \right] 
 \nonumber \\ &&
 =\lambda_0^- \frac{\left( \frac{1}{\lambda_0^-}-   \left[{\hat G}_0 (L \vert L) - {\hat G}_0 (L \vert 0) \right]  \right)  \frac{{\hat G}^{abs(L)}_0 (0 \vert x_0)}{{\hat G}^{abs(L)}_0 (0 \vert 0)}
 +\left( \frac{1}{\lambda_0^-} -   \left[{\hat G}_0 (0 \vert 0) - {\hat G}_0 (0 \vert L) \right]  \right) 
   \frac{{\hat G}^{abs(0)}_0 (L \vert x_0)}{{\hat G}^{abs(0)}_0 (L \vert L)}
  }{  \left[ \frac{1}{\lambda_0^-} - \frac{1}{\lambda_0^+} \right]  }
\label{xipm}
\end{eqnarray}
so that the normalization of Eq. \ref{upsilonresinteginfty} over $\Sigma$
 \begin{eqnarray}
\int_0^{+\infty} d \Sigma \Upsilon_{\infty}(\Sigma \vert x_0)
    = \frac{ \xi^+(x_0) }{ \lambda_0^+ } + \frac{ \xi^-(x_0) }{ \lambda_0^-  }
    = \frac{{\hat G}^{abs(L)}_0 (0 \vert x_0)}{{\hat G}^{abs(L)}_0 (0 \vert 0)}
    + \frac{{\hat G}^{abs(0)}_0 (L \vert x_0)}{{\hat G}^{abs(0)}_0 (L \vert L)} 
     =  1 -  S^{abs(0,L)}_{\infty} ( x_0)
\label{upsilonresinteginftynorma}
\end{eqnarray}
is complementary to $S^{abs(0,L)}_{\infty} ( x_0) $ of Eq. \ref{survivallimitfromlaplace}
as it should.


\subsection{When $X(t)$ is recurrent : large deviations properties of the intensive sum 
$\sigma= \frac{\Sigma(t)}{t} $ at large time $t$  }

When $X(t)$ is recurrent, it is interesting to analyze the large deviations properties
of the intensive sum of the two local times $A(t)$ and $B(t)$
\begin{eqnarray}
\sigma= \frac{\Sigma(t)}{t} = \frac{A(t)+B(t)}{t} 
\label{sigmaIntensive}
\end{eqnarray}

\subsubsection{ Evaluation of the distribution $  \Pi_t(\Sigma=\sigma t  \vert x_0) $ at leading order in $t$ }

The distribution $\Pi_t(\Sigma=\sigma t \vert x_0) $ of $\Sigma=\sigma t$ 
for $\sigma \in]0,+\infty[$ can be evaluated from
the Laplace inversion of Eq. \ref{upsilonresinteg}
\begin{eqnarray}
 \Upsilon_t(\Sigma=\sigma t  \vert x_0) 
 && = \int_{c-i\infty}^{c+i\infty} \frac{ds}{2 i \pi}  e^{s t}
 {\hat  \Upsilon}_{s}(\Sigma \vert x_0)
    =  \int_{c-i\infty}^{c+i\infty} \frac{ds}{2 i \pi}  
     \left[ \Lambda_s^+( x_0) e^{- t \left[ \sigma \lambda_s^+   -s \right]} 
     +  \Lambda_s^-( x_0) e^{- t \left[ \sigma \lambda_s^- -s \right] } \right]
\label{sigmacontourcal}
\end{eqnarray}
 For large $t$, the dominant exponential is the second term involving $\lambda_s^-$ 
 (since $\lambda_s^-<\lambda_s^+$ in Eq. \ref{dysonpolyupsroots}),
 so that the saddle-point evaluation involves the solution $s_{\sigma}$ of
 \begin{eqnarray}
  0  =   \partial_s \left[\sigma \lambda_s^- -s   \right] = \sigma  \partial_s \lambda_s^- -1
  \label{saddlessigma}
\end{eqnarray}
The change of variable
\begin{eqnarray}
 s = s_{\sigma} + i \omega
\label{complexsigma}
\end{eqnarray}
and the Taylor expansion at second order in $\omega$ 
\begin{eqnarray}
 \left[ \sigma \lambda_s^- -s \right]_{s=s_{\sigma} + i \omega}  
 = J(\sigma) + 0 + \frac{\omega^2}{2} \chi(\sigma) + o\left( \omega^2 \right)
\label{taylorsigma}
\end{eqnarray}
that involves the two functions
\begin{eqnarray}
J(\sigma) && =\left[ \sigma \lambda_s^- -s \right]_{s=s_{\sigma}}  
\nonumber \\
\chi(\sigma) && = -  \sigma \left[ \partial^2_s  \lambda_s^-   \right]_{s=s_{\sigma}}
\label{jsigmafromsaddle}
\end{eqnarray}
leads to the final result for the leading order of Eq. \ref{sigmacontourcal} 
\begin{eqnarray}
 \Pi_t(\Sigma=\sigma t  \vert x_0) 
 && \opsimeq_{t \to +\infty} 
      \Lambda_{s_{\sigma}}^-( x_0) e^{- t J(\sigma)}
  \int_{-\infty}^{+\infty} \frac{d\omega}{2  \pi}  
  e^{ - t  \frac{\chi(\sigma)}{2}  \omega^2} 
  = \frac{\Lambda_{s_{\sigma}}^-( x_0) e^{- t J(\sigma)}}{\sqrt{2 \pi t \chi(\sigma) } }
\label{sigmacontour}
\end{eqnarray}


\subsubsection{ Link between the rate function $J(\sigma)$ of the intensive sum $\sigma=a+b$
and the joint rate function $I(a,b)$ }

The integration over $x$ of Eq. \ref{upsilont}
\begin{eqnarray}
 \Upsilon_t(\Sigma \vert x_0) = {\cal A}^{abs(L)}_t (\Sigma \vert x_0) +  {\cal B}^{abs(0)}_t (\Sigma \vert x_0)
+ \int_0^{\Sigma} dA      {\cal C}_t (A,\Sigma-A \vert x_0)
 \label{upsilontinteg}
\end{eqnarray}
reads for $\Sigma = \sigma t $
\begin{eqnarray}
 \Upsilon_t(\Sigma=\sigma t \vert x_0) = {\cal A}^{abs(L)}_t (t \sigma \vert x_0) 
 +  {\cal B}^{abs(0)}_t (t \sigma \vert x_0)
+ \int_0^{\sigma} t da       {\cal C}_t (t a,t (\sigma-a) \vert x_0)
 \label{upsilontintegintensive}
\end{eqnarray}
where the two first terms are governed by the rate functions 
$I (a=\sigma, b=0 ) $ and $I (a=0, b=\sigma ) $ of Eq. \ref{rateabfour},
while the convolution of the last term involves the rate function $I(a,b=\sigma-a)$
\begin{eqnarray}
 \int_0^{\sigma}  da       {\cal C}_t (t a,t (\sigma-a) \vert x_0) 
 \oppropto_{t \to +\infty}  \int_0^{\sigma}  da     e^{-t I(a,\sigma-a)}  \oppropto_{t \to +\infty}  e^{-t J(\sigma)}
 \label{convolution}
\end{eqnarray}
As a consequence, the rate function $J(\sigma)$ should correspond to the optimization of
the joint rate function $I(a,\sigma-a)$ over all possible values $a \in [0,\sigma]$
\begin{eqnarray}
 J(\sigma) = \min_{0 \leq a \leq \sigma} I(a,\sigma-a)
 \label{jfromi}
\end{eqnarray}
Since the rate function $I(a,b)$ is the minimum over $s$ of the function 
in the exponential in Eq. \ref{Ccontour}
\begin{eqnarray}
I(a,b) = \min_{s} \left[ \frac{a}{{\hat G}^{abs(L)}_s (0 \vert 0)} + \frac{b}{{\hat G}^{abs(0)}_s (L \vert L)} - 2 c_s  \sqrt{ab} - s\right]
 \label{iabmins}
\end{eqnarray}
one can exchange the two minimizations to rewrite Eq. \ref{jfromi}
as
\begin{eqnarray}
 J(\sigma) = \min_s  \min_{0 \leq a \leq \sigma} 
 \left[ \frac{a}{{\hat G}^{abs(L)}_s (0 \vert 0)} + \frac{\sigma-a}{{\hat G}^{abs(0)}_s (L \vert L)} - 2 c_s  \sqrt{a(\sigma-a)} - s\right]
 \label{jtwomin}
\end{eqnarray}
The minimization over $a$
\begin{eqnarray}
0 && = \partial_a 
 \left[ \frac{a}{{\hat G}^{abs(L)}_s (0 \vert 0)} + \frac{\sigma-a}{{\hat G}^{abs(0)}_s (L \vert L)} - 2 c_s  \sqrt{a(\sigma-a)} - s\right]
\nonumber \\
&& =  \frac{1}{{\hat G}^{abs(L)}_s (0 \vert 0)} - \frac{1}{{\hat G}^{abs(0)}_s (L \vert L)} -  c_s \frac{ \sigma-2 a}{ \sqrt{a(\sigma-a)} }
 \label{jtwominovera}
\end{eqnarray}
yields the explicit optimal value 
\begin{eqnarray}
a_{opt} && = 
\frac{\sigma}{2} \left( 1 + \frac{\frac{1}{{\hat G}^{abs(0)}_s (L \vert L)}-\frac{1}{{\hat G}^{abs(L)}_s (0 \vert 0)}}
{ \sqrt{ \left[ \frac{1}{{\hat G}^{abs(0)}_s (L \vert L)}-\frac{1}{{\hat G}^{abs(L)}_s (0 \vert 0)}\right]^2+4 c_s^2} } \right)
\nonumber \\
&& = \frac{\sigma}{2} \left( 1 + \frac{\frac{1}{{\hat G}^{abs(0)}_s (L \vert L)}-\frac{1}{{\hat G}^{abs(L)}_s (0 \vert 0)}}
{ D_s } \right)
 \label{optimals}
\end{eqnarray}
that can be plugged into Eq. \ref{jtwomin}
to obtain
\begin{eqnarray}
 J(\sigma) && = \min_s  
 \left[ \frac{a_{opt}}{{\hat G}^{abs(L)}_s (0 \vert 0)} + \frac{\sigma-a_{opt}}{{\hat G}^{abs(0)}_s (L \vert L)} - 2 c_s  \sqrt{a_{opt}(\sigma-a_{opt})} - s\right]
 \nonumber \\
= && \min_s    \left[ \frac{\sigma}{2} \left( \frac{1}{{\hat G}^{abs(0)}_s (L \vert L)}+\frac{1}{{\hat G}^{abs(L)}_s (0 \vert 0)}- D_s\right)  - s\right]
 \nonumber \\
= && \min_s    \left[ \sigma \lambda_s^-  - s\right]
 \label{jonemin}
\end{eqnarray}
so that one recovers the function $\lambda_s^-$ of Eq. \ref{dysonpolyupsroots},
as it should for consistency with the direct calculation of Eqs \ref{saddlessigma} and \ref{jsigmafromsaddle}
of the previous subsection.


\section{ Conditioned processes involving the two local times } 

\label{sec_doob}

In this section, the goal is to construct conditioned joint processes $[X^*(t),A^*(t),B^*(t)] $
satisfying certain constraints involving the two local times, thereby generalizing our previous work 
\cite{us_LocalTime} concerning the conditioning with respect to the single local time $A(t)$.


\subsection{ Conditioned Bridge towards the local times $[A^*_T,B^*_T] $ at the time horizon $T$ }

\label{subsec_bridgea}

If the conditioning is towards the local times $[A^*_T,B^*_T] $ at the time horizon $T$,
without any condition on the final position $x_T$,
the conditioned distribution of the position $x$ and the local times $(A,B)$ at some interior time $t \in ]0,T[ $
involves the unconditioned 
joint distribution $\Pi_{t_2-t_1}( A_2-A_1,B_2-B_1 \vert  x_1) $ 
of the two increments $[A_2-A_1,B_2-B_1]$ during the time interval $(t_2-t_1)$ described in section \ref{sec_ab}
\begin{eqnarray}
P^{[A_T^*,B_T^*]}_T( x,A,B,t) =\frac{\Pi_{T-t}( A_T^*-A,B_T^*-B \vert  x) P_t(x,A,B \vert x_0)}
{\Pi_T( A_T^*,B_T^* \vert  x_0)}  
\label{pstarbridgea}
\end{eqnarray}
The corresponding conditioned process $[X^*(t),A^*(t),B^*(t)]$ then satisfies the SDE of Eq. \ref{itoabstar}
where the conditioned drift $ \mu^{[A_T^*,B_T^*]}_T( x,A,B,t ) $ is given by Eq. \ref{mustarbridgepi}.
The decomposition of $\Pi_{T-t}( A_T^*-A,B_T^*-B \vert  x) $ into its four contributions of Eq. \ref{propagABalone}
\begin{eqnarray}
&& \Pi_{T-t}(A_T^*-A,B_T^*-B \vert x)  
  =   \delta(A_T^*-A) \delta(B_T^*-B) S^{abs(0,L)}_{T-t} (  x)
+ \theta(A_T^*>A) \delta(B_T^*-B) {\cal A}^{]-\infty,L[}_{T-t} (A_T^*-A \vert x)
\nonumber \\
&& + \delta(A_T^*-A)\theta(B_T^*>B)  {\cal B}^{]0,+\infty[}_{T-t} (B_T^*-B\vert x)
+ \theta(A_T^*>A) \theta(B_T^*>B)  {\cal C}_{T-t} (A_T^*-A,B_T^*-B \vert x)
\label{pifour}
\end{eqnarray}
yields that the conditioned dynamics can be decomposed into the following regimes  :

(1) for $[0 \leq A<A_T^*,0 \leq B < B_T^*]$ where the two local times $A$ and $B$ 
have not yet reached their conditioned final values $A_T^*$ and $B_T^*$, 
the conditioned drift of Eq. \ref{mustarbridgepi} involves the contribution ${\cal C}_{T-t} (A_T^*-A,B_T^*-B \vert x) $
\begin{eqnarray}
\mu^{[A_T^*,B_T^*]}_T( x,A<A_T^*,B< B_T^*,t )  = \mu(x) +  \partial_x    \ln  {\cal C}_{T-t} (A_T^*-A,B_T^*-B \vert x)
\label{mustarbridgepiC}
\end{eqnarray}

(2) the regime (1) ends when either $A$ or $B$ reaches its conditioned final value, 
so the next regime (2) contains two possibilities :

(2-A)  if $B$ has already reached its conditioned final value $B=B_T^*$ while $A$ has not yet 
reached its conditioned final value $A_T^*$,
then the conditioned drift of Eq. \ref{mustarbridgepi} involves the contribution $ {\cal A}^{]-\infty,L[}_{T-t} (A_T^*-A \vert x) $
\begin{eqnarray}
\mu^{[A_T^*,B_T^*]}_T( x,A<A_T^*,B= B_T^*,t )  = \mu(x) +  \partial_x    \ln   {\cal A}^{]-\infty,L[}_{T-t} (A_T^*-A \vert x)
\label{mustarbridgepiA}
\end{eqnarray}

(2-B)  if $A$ has already reached its conditioned final value $A=A_T^*$ while $B$ has not yet 
reached its conditioned final value $B_T^*$,
then the conditioned drift of Eq. \ref{mustarbridgepi} involves the contribution $ {\cal B}^{]0,+\infty[}_{T-t} (B_T^*-B\vert x)  $
\begin{eqnarray}
\mu^{[A_T^*,B_T^*]}_T( x,A=A_T^*,B< B_T^*,t )  = \mu(x) +  \partial_x    \ln   {\cal B}^{]0,+\infty[}_{T-t} (B_T^*-B\vert x)
\label{mustarbridgepiB}
\end{eqnarray}

(3) in the last regime 
$[A=A_T^*,B = B_T^*]$ where the two local times $A$ and $B$ 
have already reached their conditioned final values $A_T^*$ and $B_T^*$, 
the conditioned drift of Eq. \ref{mustarbridgepi} involves the contribution $S^{abs(0,L)}_{T-t} (  x) $,
since the process cannot visit $0$ or $L$ anymore
\begin{eqnarray}
\mu^{[A_T^*,B_T^*]}_T( x,A=A_T^*,B= B_T^*,t )  = \mu(x) +  \partial_x    \ln  S^{abs(0,L)}_{T-t} (  x)
\label{mustarbridgepiS}
\end{eqnarray}

It is now interesting to consider two possibilities in the limit of the infinite horizon $T \to +\infty$,
as described in the two next subsections.


\subsection{ Conditioning towards the finite local times $[A_{\infty},B_{\infty}]$ for the infinite horizon $T \to +\infty$  }

\label{subsec_doobinfinity}

If one wishes to impose the finite asymptotic local times $A_{\infty} <+\infty$ and $B_{\infty} <+\infty $ 
at the infinite horizon $T \to +\infty$,
one needs to analyze the limit of the infinite horizon $T \to +\infty$
for the conditioned drift of Eq \ref{mustarbridgepi} 
\begin{eqnarray}
\mu^{[A_{\infty}^*,B_{\infty}^*]}_{\infty}( x,A,B ) = \mu(x) +  \lim_{\tau \to +\infty} \partial_x    \ln \Pi_{\tau}( A_{\infty}^*-A,B_{\infty}^*-B \vert  x) 
\label{mustarbridgepitinfty} 
 \end{eqnarray}
with its different expressions for the various regimes of Eqs \ref{mustarbridgepiC}
\ref{mustarbridgepiA} \ref{mustarbridgepiB} \ref{mustarbridgepiS}
\begin{eqnarray}
\mu^{[A_{\infty}^*,B_{\infty}^*]}_{\infty}( x,A<A_{\infty}^*,B< B_{\infty}^*) 
&&= \mu(x) +  \lim_{\tau \to +\infty} \partial_x    \ln    {\cal C}_{\tau} (A_{\infty}^*-A,B_{\infty}^*-B \vert x)
\nonumber \\
\mu^{[A_{\infty}^*,B_{\infty}^*]}_{\infty}( x,A<A_{\infty}^*,B= B_{\infty}^*) 
&&= \mu(x) +  \lim_{\tau \to +\infty} \partial_x    \ln    {\cal A}^{]-\infty,L[}_{\tau} (A_{\infty}^*-A \vert x) 
\nonumber \\
\mu^{[A_{\infty}^*,B_{\infty}^*]}_{\infty}( x,A=A_{\infty}^*,B< B_{\infty}^*) 
&&= \mu(x) +  \lim_{\tau \to +\infty} \partial_x    \ln  {\cal B}^{]0,+\infty[}_{\tau} (B_{\infty}^*-B\vert x)
\nonumber \\
\mu^{[A_{\infty}^*,B_{\infty}^*]}_{\infty}( x,A=A_{\infty}^*,B= B_{\infty}^*) 
&&= \mu(x) +  \lim_{\tau \to +\infty} \partial_x    \ln S^{abs(0,L)}_{\tau} (  x)
\label{mustarbridgepitinftyregions} 
 \end{eqnarray}


\subsection{ Conditioning towards the intensive local times $a_* = \frac{A_T^*}{T}$ and $b_* = \frac{B_T^*}{T}$
for large time horizon $T \to + \infty$  }

\label{subsec_doobintensive}

If one wishes to impose instead the fixed intensive local times $a_*>0$ and $b_*>0$ for large time horizon $T \to + \infty $, 
one needs to plug the values
\begin{eqnarray}
 A_T^* && = T a_*
 \nonumber \\
  B_T^* && = T b_*
 \label{atstarintensive}
\end{eqnarray}
into the conditioned drift of Eq. \ref{mustarbridgepi}
\begin{eqnarray}
\mu^{[T a_*,T b_*]}_T( x,A,B,t )  = \mu(x) +  \partial_x    \ln \Pi_{T-t}( T a_*-A,T b_*-B \vert  x) 
\label{mustarbridgepiintensive}
\end{eqnarray}
and to use the large-time behavior of Eq. \ref{propagABaloneintensivelarget}
that involves the corresponding intensive local times $a_t$ and $b_t$ on the time interval $(T-t)$ 
\begin{eqnarray}
a_t && \equiv \frac{ T a_* - A}{ T-t} \opsimeq_{T \to +\infty } a_*
\nonumber \\
b_t && \equiv \frac{ T b_* - B}{ T-t} \opsimeq_{T \to +\infty } b_*
\label{aeffective}
\end{eqnarray}
that reduces to $a_*$ and $b_*$ at leading order when $T \to +\infty$.
Let us discuss two special cases before the general case $(a_*>0,b_*>0)$


\subsubsection{ Case $a_*>0$ and $b_*=0$ }

For $a_*>0$ and $b_*=0$, the asymptotic behavior of Eq. \ref{propagABaloneintensivelarget} 
\begin{eqnarray}
\Pi_{T-t}( T a_*-A,0 \vert  x)   \opsimeq_{T \to + \infty}  
        \frac{ \alpha_{s_{a_*,0}}^{]-\infty,L[}(x) e^{-t I(a_*,0) }}{\sqrt{ 2 \pi K(a_*,0) } (T-t)^{\frac{3}{2}} } 
\label{propagABaloneintensivelargetdooba}
\end{eqnarray}
can be plugged into Eq. \ref{mustarbridgepiintensive}
to obtain the limit $\mu^{[ a_*, b_*=0]}_{\infty}( x) $ of the conditioned drift 
using Eq. \ref{alphainteg} 
\begin{eqnarray}
\mu^{[ a_*, b_*=0]}_{\infty}( x) && \equiv \lim_{T \to +\infty} \mu^{[T a_*,T b_*]}_T( x,A,B,t ) 
= \mu(x) +  \partial_x    \ln \left[  \alpha_{s_{a_*,0}}^{]-\infty,L[}(x)   \right]
\nonumber \\
&& = \mu(x) +  \partial_x    \ln \left[  {\hat G}^{abs(L)}_{s_{a_*,0}} (0 \vert x)   \right]
\label{conditioneddriftastaralone}
\end{eqnarray}
where $s_{a_*,0}$ is the solution of Eq. \ref{saddlesazero}
\begin{eqnarray}
  0  =   \partial_s \left[  \frac{a_*}{ {\hat G}^{abs(L)}_s (0 \vert 0)} -s   \right] 
  = a_* \partial_s \left[  \frac{1}{ {\hat G}^{abs(L)}_s (0 \vert 0)}    \right] -1
  \label{saddlesazerostar}
\end{eqnarray}
So the conditioned drift of Eq. \ref{conditioneddriftastaralone}
can be also rewritten in the following parametric form of parameter $s$
\begin{eqnarray}
\mu^{[ a_*(s), b_*=0]}_{\infty}( x) 
&& = \mu(x) +  \partial_x    \ln \left[  {\hat G}^{abs(L)}_s (0 \vert x)   \right]
 \nonumber \\
a^*(s) && = \frac{1}{ \partial_s \left[  \frac{1}{ {\hat G}^{abs(L)}_s (0 \vert 0)}    \right] } 
\label{conditioneddriftastaraloneparametric}
\end{eqnarray}


\subsubsection{ Case $a_*=0$ and $b_*>0$ }

For $a_*=0$ and $b_*>0$, the asymptotic behavior of 
Eq. \ref{propagABaloneintensivelarget} 
\begin{eqnarray}
\Pi_{T-t}( 0,T b_*-B \vert  x)   \opsimeq_{T \to + \infty}  
       \frac{ \beta_{s_{0,b}}^{]0,+\infty[} (x)  e^{-t I(0,b) }}{\sqrt{ 2 \pi K(0,b) } t^{\frac{3}{2}}} 
\label{propagABaloneintensivelargetdoobb}
\end{eqnarray}
can be plugged into Eq. \ref{mustarbridgepiintensive}
to obtain the limit $\mu^{[ a_*, b_*=0]}_{\infty}( x) $ of the conditioned drift 
using Eq. \ref{betainteg} 
\begin{eqnarray}
\mu^{[ a_*=0, b_*]}_{\infty}( x) && \equiv \lim_{T \to +\infty} \mu^{[T a_*,T b_*]}_T( x,A,B,t ) 
= \mu(x) +  \partial_x    \ln \left[ \beta_{s_{0,b_*}}^{]0,+\infty[} (x)   \right]
\nonumber \\
&& = \mu(x) +  \partial_x    \ln \left[ {\hat G}^{abs(0)}_{s_{0,b_*}} (L \vert x)   \right]
\label{conditioneddriftbstaralone}
\end{eqnarray}
where $s_{0,b_*}$ is the solution of Eq. \ref{saddleszerob}
\begin{eqnarray}
  0  =   \partial_s \left[ \frac{b_*}{ {\hat G}^{abs(0)}_s (L \vert L)}  -s   \right] 
  = b_* \partial_s \left[ \frac{1}{ {\hat G}^{abs(0)}_s (L \vert L)}     \right] -1
  \label{saddlesbzerostar}
\end{eqnarray}
So the conditioned drift of Eq. \ref{conditioneddriftbstaralone}
can be also written in the following parametric form of parameter $s$
\begin{eqnarray}
\mu^{[ a_*=0, b_*(s)]}_{\infty}( x) &&  = \mu(x) +  \partial_x    \ln \left[ {\hat G}^{abs(0)}_s (L \vert x)   \right]
 \nonumber \\
 b_*(s) &&= \frac{1}{\partial_s \left[ \frac{1}{ {\hat G}^{abs(0)}_s (L \vert L)}     \right]}
\label{conditioneddriftbstaraloneparametric}
\end{eqnarray}


\subsubsection{ General case $a_*>0$ and $b_*>0$ }

For $a_*>0$ and $b_*>0$, the asymptotic behavior of
Eq. \ref{propagABaloneintensivelarget} 
\begin{eqnarray}
\Pi_{T-t}( T a_*-A,T b_*-B \vert  x)  && \opsimeq_{T \to + \infty}  
  \frac{e^{- (T-t) I (a_*, b_* )}}{ 2 \pi (T-t) \sqrt{  2  K(a_*,b_*)}   }
      \left[   \sqrt{c_s} \left(  \frac{ \alpha_s^{]-\infty,L[}( x)     a_*^{\frac{1}{4}} }{    b_*^{\frac{3}{4}}}   
      +  \frac{ \beta_s^{]0,+\infty[}(x )  b_*^{\frac{1}{4}} }{   a_*^{\frac{3}{4}}}  \right)
      +   \frac{  \gamma_s( x)}{\sqrt{ c_s }  a_*^{\frac{1}{4}} b_*^{\frac{1}{4}}}   
      \right]_{s=s_{a_*,b_*}}       
\label{propagABaloneintensivelargetdoob}
\end{eqnarray}
 can be plugged into Eq. \ref{mustarbridgepiintensive}
to obtain the limit $\mu^{[ a_*, b_*]}_{\infty}( x) $ of the conditioned drift 
\begin{eqnarray}
\mu^{[ a_*, b_*]}_{\infty}( x) && \equiv \lim_{T \to +\infty} \mu^{[T a_*,T b_*]}_T( x,A,B,t )  
= \mu(x) +  \partial_x    \ln  \left[   \sqrt{c_s} \left(  \frac{ \alpha_s^{]-\infty,L[}( x)     a_*^{\frac{1}{4}} }{    b_*^{\frac{3}{4}}}   
      +  \frac{ \beta_s^{]0,+\infty[}(x )  b_*^{\frac{1}{4}} }{   a_*^{\frac{3}{4}}}  \right)
      +   \frac{  \gamma_s( x)}{\sqrt{ c_s }  a_*^{\frac{1}{4}} b_*^{\frac{1}{4}}}   
      \right]_{s=s_{a_*,b_*}} 
      \nonumber \\
&&    = \mu(x) +  \partial_x    \ln  
\left[   c_s \left(  \sqrt{ \frac{a_*}{b_*} } \alpha_s^{]-\infty,L[}( x)      
      +   \sqrt{ \frac{b_*}{a_*} } \beta_s^{]0,+\infty[}( x)  \right)
      +     \gamma_s( x)     
      \right]_{s=s_{a_*,b_*}}        
\label{mustarbridgepiintensivefinal}
\end{eqnarray}

Using the explicit expressions of Eqs \ref{alphainteg} \ref{betainteg} \ref{gammainteg} and \ref{const},
this conditioned drift becomes
 \begin{eqnarray}
&& \mu^{[ a_*, b_*]}_{\infty}( x) = \mu(x) 
\nonumber \\
&& +  \partial_x    \ln  
\bigg[
\left(  \sqrt{{\hat G}_s (0 \vert L){\hat G}_s (L \vert 0) \frac{a_*}{b_*} } 
   \left[{\hat G}_s (L \vert L) - {\hat G}_s (L \vert 0) \right] 
     +    \left[{\hat G}_s (0 \vert 0)- {\hat G}_s (0 \vert L) \right] 
                  {\hat G}_s (L \vert 0)  \right)
                   {\hat G}_s (L \vert L)  
  {\hat G}^{abs(L)}_s (0 \vert x)
 \nonumber \\
 &&     +  
   \left( \sqrt{{\hat G}_s (0 \vert L){\hat G}_s (L \vert 0) \frac{b_*}{a_*} } 
    \left[{\hat G}_s (0 \vert 0)  - {\hat G}_s (0 \vert L) \right] 
              +       \left[{\hat G}_s (L \vert L) - {\hat G}_s (L \vert 0) \right]   
   {\hat G}_s (0 \vert L) 
   \right) {\hat G}_s (0 \vert 0)       {\hat G}^{abs(0)}_s (L \vert x) \bigg]_{s=s_{a_*,b_*}}  
\label{mustarbridgepiintensivefinalsimplif}
\end{eqnarray}
where $s_{a_*,b_*}$ is the solution of Eq. \ref{saddlesab}
\begin{eqnarray}
    0  =   \partial_s \left[\frac{a_*}{ {\hat G}^{abs(L)}_s (0 \vert 0)}
    + \frac{b_*}{ {\hat G}^{abs(0)}_s (L \vert L)} -2 c_s \sqrt{a_*b_*} -s   \right] 
\label{saddlesabstar}
\end{eqnarray}

The conditioned drift of Eq. \ref{mustarbridgepiintensivefinalsimplif}
is much simpler in the two external regions :

(i) for $x \in ]-\infty,0] $, the function ${\hat G}^{abs(0)}_s (L \vert x) $ vanishes,
so that Eq. \ref{mustarbridgepiintensivefinalsimplif} reduces to
 \begin{eqnarray}
 \mu^{[ a_*, b_*]}_{\infty}( x<0) = \mu(x) 
 +  \partial_x    \ln  \left[ {\hat G}^{abs(L)}_s (0 \vert x) \right]_{s=s_{a_*,b_*}} 
 \label{mustarbridgepiintensivefinalsimplifleft}
\end{eqnarray}

(ii) for $x \in [L,+\infty[$
the function ${\hat G}^{abs(L)}_s (0 \vert x) $ vanishes,
so that Eq. \ref{mustarbridgepiintensivefinalsimplif} reduces to
 \begin{eqnarray}
 \mu^{[ a_*, b_*]}_{\infty}( x>L) = \mu(x) 
 +  \partial_x    \ln  \left[ {\hat G}^{abs(0)}_s (L \vert x) \right]_{s=s_{a_*,b_*}} 
 \label{mustarbridgepiintensivefinalsimplifright}
\end{eqnarray}

Finally, if one replaces $(a_*,b_*)$ by the parametrization
\begin{eqnarray}
  a_* && = \frac{\sigma_*(1+r_*)}{2} 
  \nonumber \\
  b_* && = \frac{\sigma_*(1-r_*)}{2} 
\label{pbsigmar}
\end{eqnarray}
involving their sum $\sigma_*=a_*+b_* >0$ and the parameter $r_* \in ]-1,+1[$
Eq. \ref{saddlesabstar} becomes
\begin{eqnarray}
    1  = \sigma_*  \partial_s \left[\frac{1+r_*}{ 2{\hat G}^{abs(L)}_s (0 \vert 0)}
    + \frac{1-r_*}{ 2 {\hat G}^{abs(0)}_s (L \vert L)} - c_s \sqrt{1-r_*^2}    \right] 
\label{saddlesabstarsr}
\end{eqnarray}
then the conditioned drift of Eq. \ref{mustarbridgepiintensivefinalsimplif}
can be written in the following parametric form 
 \begin{eqnarray}
&& \sigma_*(s)    = \frac{1}{ \partial_s \left[\frac{1+r_*}{ 2{\hat G}^{abs(L)}_s (0 \vert 0)}
    + \frac{1-r_*}{ 2 {\hat G}^{abs(0)}_s (L \vert L)} - c_s \sqrt{1-r_*^2}    \right] }
 \nonumber \\
&& \mu^{[ \frac{\sigma_*(s)(1+r_*)}{2} , \frac{\sigma_*(s)(1-r_*)}{2} ]}_{\infty}( x) = \mu(x) 
\nonumber \\
&& +  \partial_x    \ln  
\bigg[
\left(  \sqrt{{\hat G}_s (0 \vert L){\hat G}_s (L \vert 0) \frac{1+r_*  }{1-r_*}  } 
   \left[{\hat G}_s (L \vert L) - {\hat G}_s (L \vert 0) \right] 
     +    \left[{\hat G}_s (0 \vert 0)- {\hat G}_s (0 \vert L) \right] 
                  {\hat G}_s (L \vert 0)  \right)
                   {\hat G}_s (L \vert L)  
  {\hat G}^{abs(L)}_s (0 \vert x)
 \nonumber \\
 &&     +  
   \left( \sqrt{{\hat G}_s (0 \vert L){\hat G}_s (L \vert 0) \frac{1-r_*  }{1+r_*}     } 
    \left[{\hat G}_s (0 \vert 0)  - {\hat G}_s (0 \vert L) \right] 
              +       \left[{\hat G}_s (L \vert L) - {\hat G}_s (L \vert 0) \right]   
   {\hat G}_s (0 \vert L) 
   \right) {\hat G}_s (0 \vert 0)       {\hat G}^{abs(0)}_s (L \vert x) \bigg] 
\label{mustarbridgepiintensivefinalsimplifparametric}
\end{eqnarray}


\subsection{ Conditioned Bridge towards the sum $\Sigma^*_T $ at the time horizon $T$ }

If the conditioning is towards the sum $\Sigma^*_T$ at time horizon $T$,
the conditioned distribution of the position $x$ and the sum $\Sigma$ at some interior time $t \in ]0,T[ $
involves the unconditioned distribution $\Pi_{t_2-t_1}( \Sigma_2-\Sigma_1, \vert  x_1) $ 
of the increment $(\Sigma_2-\Sigma_1)$ during the time interval $(t_2-t_1)$ described in subsection \ref{subsec_sigma}
\begin{eqnarray}
P^{[\Sigma_T^*]}_T( x,\Sigma,t) =\frac{\Pi_{T-t}( \Sigma_T^*-\Sigma \vert  x) P_t(x,\Sigma \vert x_0)}
{\Pi_T( \Sigma_T^* \vert  x_0)}  
\label{pstarbridgesigma}
\end{eqnarray}
The corresponding conditioned process $[X^*(t),\Sigma^*(t)]$ satisfies the following SDE system \begin{eqnarray}
dX^*(t) && =  \mu^{[\Sigma_T^*]}_T( X^*(t),\Sigma^*(t),t ) dt + dW(t)
\nonumber \\
d\Sigma^*(t) && = \left[  \delta ( X^*(t) ) + \delta ( X^*(t)-L ) \right]dt
\label{itosigma}
\end{eqnarray}
where the conditioned drift $ \mu^{[\Sigma_T^*]}_T( x,\Sigma,t ) $ reads
\begin{eqnarray}
\mu^{[\Sigma_T^*]}_T( x,\Sigma,t ) = \mu(x) +  \partial_x    \ln \Pi_{T-t}( \Sigma_T^*-\Sigma \vert  x)
\label{mustarbridgepisigma}
\end{eqnarray}
The decomposition of $\Pi_{T-t}( \Sigma_T^*-\Sigma \vert  x) $ into its 
singular and regular contributions of Eq. \ref{pisigma}
\begin{eqnarray}
\Pi_{T-t}( \Sigma_T^*-\Sigma \vert  x)
 =    \delta(\Sigma_T^*-\Sigma) S^{abs(0,L)}_{T-t} ( x)
+  \theta(\Sigma_T^*>\Sigma) \Upsilon_{T-t}(\Sigma_T^*-\Sigma \vert x)
\label{pisigmastar}
\end{eqnarray}
yields that the conditioned dynamics can be decomposed into the two regimes  :

(i) for $0 \leq \Sigma<\Sigma_T^*$ where the sum $\Sigma $ 
has not yet reached its conditioned final value $\Sigma_T^*$, 
the conditioned drift of Eq. \ref{mustarbridgepisigma}
 involves the regular contribution $\Upsilon_{T-t}(\Sigma_T^*-\Sigma \vert x) $
\begin{eqnarray}
\mu^{[\Sigma_T^*]}_T( x,\Sigma<\Sigma_T^*,t ) = \mu(x) +  \partial_x    \ln \Upsilon_{T-t}(\Sigma_T^*-\Sigma \vert x)
\label{mustarbridgepisigmaregular}
\end{eqnarray}

(ii) for $ \Sigma=\Sigma_T^*$ where the sum $\Sigma $ 
has already reached its conditioned final value $\Sigma_T^*$, 
the conditioned drift of Eq. \ref{mustarbridgepisigma}
 involves the survival probability $S^{abs(0,L)}_{T-t} ( x) $,
 since the process cannot visit $0$ or $L$ anymore
\begin{eqnarray}
\mu^{[\Sigma_T^*]}_T( x,\Sigma<\Sigma_T^*,t ) = \mu(x) +  \partial_x    \ln S^{abs(0,L)}_{T-t} ( x)
\label{mustarbridgepisigmasingular}
\end{eqnarray}

It is now interesting to consider two possibilities in the limit of the infinite horizon $T \to +\infty$,
as described in the two next subsections.


\subsection{ Conditioning towards the finite sum $\Sigma_{\infty}^*$ for the infinite horizon $T \to +\infty$  }

If one wishes to impose the finite asymptotic sum $\Sigma_{\infty}^* $ 
at the infinite horizon $T \to +\infty$,
one needs to analyze the limit of the infinite horizon $T \to +\infty$
for the conditioned drift of Eq \ref{mustarbridgepisigma} 
\begin{eqnarray}
\mu^{[\Sigma_{\infty}^*]}_{\infty}( x,\Sigma ) 
= \mu(x) +  \lim_{\tau \to +\infty} \partial_x    \ln \Pi_{\tau}( \Sigma_{\infty}^*-\Sigma \vert  x) 
\label{mustarbridgepisigmainfty} 
 \end{eqnarray}
with its two different expressions for the two regimes of Eqs \ref{mustarbridgepisigmaregular}
\ref{mustarbridgepisigmasingular} 
\begin{eqnarray}
\mu^{[\Sigma_{\infty}^*]}_{\infty}( x,\Sigma<\Sigma_{\infty}^*) 
&&= \mu(x) +  \lim_{\tau \to +\infty} \partial_x    \ln   \Upsilon_{\tau}(\Sigma_T^*-\Sigma \vert x)
\nonumber \\
\mu^{[\Sigma_{\infty}^*]}_{\infty}( x,\Sigma=\Sigma_{\infty}^*) 
&&= \mu(x) +  \lim_{\tau \to +\infty} \partial_x    \ln S^{abs(0,L)}_{\tau} (  x)
\label{mustarbridgepitinftyregions2} 
 \end{eqnarray}


\subsection{ Conditioning towards the intensive sum $\sigma_* = \frac{\Sigma_T^*}{T}$ 
for large time horizon $T \to + \infty$  }

If one wishes to impose the fixed intensive sum $\sigma_*>0$ for large time horizon $T \to + \infty $, 
one needs to plug the values
\begin{eqnarray}
 \Sigma_T^* && = T \sigma_*
 \label{sigmatstarintensive}
\end{eqnarray}
into the conditioned drift of Eq. \ref{mustarbridgepisigmaregular}
\begin{eqnarray}
\mu^{[T \sigma_*]}_T( x,\Sigma,t )  = \mu(x) +  \partial_x    \ln \Pi_{T-t}( T \sigma_*-\Sigma, \vert  x) 
\label{mustarbridgepiintensivesigma}
\end{eqnarray}
and to use the large-time behavior of Eq. \ref{sigmacontour}
that involves the corresponding intensive sum $\sigma_t$ on the time interval $(T-t)$ 
\begin{eqnarray}
\sigma_t && \equiv \frac{ T \sigma_* - \Sigma}{ T-t} \opsimeq_{T \to +\infty } \sigma_*
\label{sigmaeffective}
\end{eqnarray}
that reduces to $\sigma_*$ at leading order when $T \to +\infty$,
so that one can plug
\begin{eqnarray}
\Pi_{T-t}( T \sigma_*-\Sigma, \vert  x) 
  \opsimeq_{(T-t) \to + \infty} \frac{\Lambda_{s_{\sigma_*}}^-( x) e^{- (T-t) J(\sigma_*)}}{\sqrt{2 \pi (T-t) \chi(\sigma_*) } }
\label{sigmacontoursatr}
\end{eqnarray}
into Eq. \ref{mustarbridgepiintensivesigma}
to obtain the limit $\mu^{[\sigma_*]}_{\infty}( x) $ of the conditioned drift of Eq. \ref{mustarbridgepiintensivesigma} using Eq. \ref{dysonpolyupsfracusolinteg}
\begin{eqnarray}
&& \mu_{\infty}^{[\sigma_*]}  \equiv \lim_{T \to +\infty}
\mu^{[T \sigma_*]}_T( x,\Sigma,t )   \opsimeq_{(T-t) \to + \infty}
= \mu(x) +  \partial_x    \ln \Lambda_{s_{\sigma_*}}^-( x) 
\nonumber \\
&& =  \mu(x) +  \partial_x    \ln \left[ 
\left( 1- \lambda_s^-  \left[{\hat G}_s (L \vert L) - {\hat G}_s (L \vert 0) \right]  \right)  \frac{{\hat G}^{abs(L)}_s (0 \vert x)}{{\hat G}^{abs(L)}_s (0 \vert 0)}
 +\left( 1- \lambda_s^-  \left[{\hat G}_s (0 \vert 0) - {\hat G}_s (0 \vert L) \right]  \right)   \frac{{\hat G}^{abs(0)}_s (L \vert x)}{{\hat G}^{abs(0)}_s (L \vert L)}
 \right]_{s=s_{\sigma_*}}
\label{mustarbridgepiintensivesigmares}
\end{eqnarray}
where $s_{\sigma_*}$ is the solution of Eq. \ref{saddlessigma} 
 \begin{eqnarray}
  0  =   \sigma_*  \partial_s \lambda_s^- -1 
  \label{saddlessigmastar}
\end{eqnarray}

So the conditioned drift of Eq. \ref{mustarbridgepiintensivefinalsimplif}
can be written in the following parametric form of parameter $s$
\begin{eqnarray}
&& \mu_{\infty}^{\left[ \sigma_*(s)=\frac{1}{  \partial_s \lambda_s^- }\right]}  =  \mu(x) 
 \nonumber \\ &&
 +  \partial_x    \ln
 \left[ 
\left( 1- \lambda_s^-  \left[{\hat G}_s (L \vert L) - {\hat G}_s (L \vert 0) \right]  \right)  \frac{{\hat G}^{abs(L)}_s (0 \vert x)}{{\hat G}^{abs(L)}_s (0 \vert 0)}
 +\left( 1- \lambda_s^-  \left[{\hat G}_s (0 \vert 0) - {\hat G}_s (0 \vert L) \right]  \right)   \frac{{\hat G}^{abs(0)}_s (L \vert x)}{{\hat G}^{abs(0)}_s (L \vert L)}
 \right]_{s=s_{\sigma_*}} 
  \label{mustarbridgepiintensivesigmaresexpliparametric}
\end{eqnarray}

Again this conditioned drift 
is much simpler in the two external regions :

(i) for $x \in ]-\infty,0] $, the function ${\hat G}^{abs(0)}_s (L \vert x) $ vanishes,
so that Eq. \ref{mustarbridgepiintensivesigmaresexpliparametric} reduces to
 \begin{eqnarray}
 \mu^{\left[ \sigma_*(s)=\frac{1}{  \partial_s \lambda_s^- }\right]}_{\infty}( x<0) = \mu(x) 
 +  \partial_x    \ln  \left[ {\hat G}^{abs(L)}_s (0 \vert x) \right]_{s=s_{\sigma_*}} 
 \label{mustarbridgepiintensivesigmaresexplileft}
\end{eqnarray}

(ii) for $x \in [L,+\infty[$
the function ${\hat G}^{abs(L)}_s (0 \vert x) $ vanishes,
so that Eq. \ref{mustarbridgepiintensivesigmaresexpliparametric} reduces to
 \begin{eqnarray}
 \mu^{\left[ \sigma_*(s)=\frac{1}{  \partial_s \lambda_s^- }\right]}_{\infty}( x>L) = \mu(x) 
 +  \partial_x    \ln  \left[ {\hat G}^{abs(0)}_s (L \vert x) \right]_{s=s_{\sigma_*}} 
 \label{mustarbridgepiintensivesigmaresexpliright}
\end{eqnarray}


\section{ Application to the uniform drift $\mu(x) = \mu$ on the whole line $]-\infty,+\infty[$ }

\label{sec_brown}

In this section, the general framework described in the previous section is applied
to the simplest example of the uniform drift $\mu(x) = \mu$ on the whole line $]-\infty,+\infty[$ :

(i) for $\mu \ne 0$, the unconditioned diffusion $X(t)$ is transient,
and the two local times $A(t)$ and $B(t)$ remain finite random variables for $t \to + \infty$
as described in subsection \ref{subsec_trans}.

(ii) for $\mu=0$, the unconditioned diffusion $X(t)$ is recurrent, but does not converge towards an equilibrium. The large deviations properties of the intensive local times $a=\frac{A(t)}{t}$ 
and $b=\frac{B(t)}{t}$ for large times are 
governed by some rate function $I(a,b)$, as described in subsection \ref{subsec_rec}.
Here, the typical values where the rate function is vanishing and minimum are $a^{typ}=0=b^{typ}$ 
\begin{eqnarray}
0=I(a=0,b=0) = \partial_a I (a=0,b=0) = \partial_b I (a=0,b=0)
\label{iabmin00}
\end{eqnarray}


\subsection{ Useful propagators and notations  }

The unconditioned propagator $G(x,t \vert x_0,t_0) $  is Gaussian
\begin{eqnarray}
G(x,t \vert x_0,t_0) = \frac{1}{\sqrt{2 \pi (t-t_0)}}  e^{- \frac{[x-x_0-\mu(t-t_0) ]^2}{2(t-t_0)}} 
=\frac{1}{\sqrt{2 \pi (t-t_0)}}  
e^{- \frac{(x-x_0)^2}{2(t-t_0)} 
+\mu (x-x_0)
- \frac{\mu^2 }{2}(t-t_0)
}
\label{free}
\end{eqnarray}
and its time Laplace transform reads
\begin{eqnarray}
{\hat G}_s (x \vert x_0) && \equiv \int_{t_0}^{+\infty} dt e^{-s (t-t_0) } G(  x,t \vert   x_0,t_0) 
 = \frac{ e^{\mu (x-x_0) - \sqrt{\mu^2+2 s } \vert x-x_0 \vert } } { \sqrt{\mu^2+2 s }} 
 =  \frac{ e^{\mu (x-x_0) - \kappa_s \vert x-x_0 \vert } } { \kappa_s} 
\label{laplacefree}
\end{eqnarray}
where we have introduced the notation
\begin{eqnarray}
\kappa_s \equiv \sqrt{\mu^2+2 s } 
\label{kappas}
\end{eqnarray}
in order to see more clearly the structures of the other propagators of Eq. \ref{Gabsbonly} 
 \begin{eqnarray}
  {\hat G}_s^{abs(L)} (x \vert x_0) 
  =  {\hat G}_s (x \vert x_0) 
  - \frac{{\hat G}_s (x \vert L) {\hat G}_s (L \vert x_0)}{{\hat G}_s (L \vert L)}
  =  \frac{ e^{\mu (x-x_0) } } { \kappa_s} 
  \left[     e^{ - \kappa_s \vert x-x_0 \vert }
    -   e^{ - \kappa_s (\vert x-L \vert + \vert L-x_0 \vert ) }   
   \right]
\label{Gabsbonlyfree}
\end{eqnarray}
and of Eq. \ref{Gabsaonly} 
\begin{eqnarray}
{\hat G}^{abs(0)}_s (x \vert x_0) 
 = {\hat G}_s (x \vert x_0)
 - \frac{ {\hat G}_s (x \vert 0){\hat G}_s (0 \vert x_0) }{ {\hat G}_s (0 \vert 0) }  
  =  \frac{ e^{\mu (x-x_0) } } { \kappa_s} 
  \left[ e^{ - \kappa_s \vert x-x_0 \vert}
  -  e^{ - \kappa_s (\vert x \vert + \vert x_0 \vert )} \right]
 \label{Gabsaonlyfree}
\end{eqnarray}
with the special cases
    \begin{eqnarray}
  {\hat G}_s^{abs(L)} (0 \vert 0)   = \frac{ 1 } { \kappa_s} 
  \left[ 1  -  e^{ - 2 \kappa_s L} \right]
=  {\hat G}^{abs(0)}_s (L \vert L) 
        \label{GOLabs}
\end{eqnarray}
These various propagators allow to evaluate the useful notations of Eq. \ref{nota} \ref{const} \ref{dssquareroot} 
    \begin{eqnarray}
   \Delta_s && = {\hat G}_s (0 \vert 0)  {\hat G}_s (L \vert L)  -  {\hat G}_s (L \vert 0) {\hat G}_s (0 \vert L) 
   =  \frac{1 } { \kappa_s^2 } 
  \left[   1    -   e^{ -  2\kappa_s L }   
   \right]
 \nonumber  \\
  c_s && =    \frac{ \sqrt{ {\hat G}_s (0 \vert L){\hat G}_s (L \vert 0) }}{\Delta_s} 
  = \frac{\kappa_s}{2 \sinh(\kappa_s L ) }
 \nonumber \\
  D_s && =
\sqrt{ \left(\frac{1}{ {\hat G}^{abs(0)}_s (L \vert L) } -  \frac{1}{ {\hat G}^{abs(L)}_s (0 \vert 0) }  \right)^2
+ 4 c_s^2 } = \frac{\kappa_s}{ \sinh(\kappa_s L ) }
  \label{notauseful}
\end{eqnarray}
as well as Eq. \ref{dysonpolyupsrootsfree} 
 \begin{eqnarray}
\lambda_s^{\pm}   
= \frac{1}{2} \left[\frac{1}{ {\hat G}^{abs(0)}_s (L \vert L) } +  \frac{1}{ {\hat G}^{abs(L)}_s (0 \vert 0) }    
\pm D_s \right]
= \kappa_s \frac{ 1 \pm   e^{ -  \kappa_s L} } {  1  -  e^{ - 2 \kappa_s L} }
 =  \frac{  \kappa_s } {  1 \mp   e^{ -  \kappa_s L}}
\label{dysonpolyupsrootsfree}
\end{eqnarray}



\subsection{ Conditioning towards the intensive sum $\sigma_* = \frac{\Sigma_T^*}{T}$ 
for large time horizon $T \to + \infty$  }

The conditioned drift of Eq. \ref{mustarbridgepiintensivesigmaresexpliparametric}
yields
\begin{eqnarray}
 \mu_{\infty}^{\left[ \sigma_*(s)\right]}  =  \mu +  \partial_x    \ln  
\bigg[
\left(  1+e^{\mu L}      \right)   {\hat G}^{abs(L)}_s (0 \vert x)
     +     \left(  1+e^{-\mu L}      \right)     {\hat G}^{abs(0)}_s (L \vert x) \bigg]
  \label{mustarbridgepiintensivesigmaresexpliparametricfree}
\end{eqnarray}
with the parametrization 
\begin{eqnarray}
\sigma_*(s)=\frac{1}{  \partial_s \lambda_s^- }
= \frac{1}{  \partial_s \left[  \frac{  \kappa_s } {  1 +   e^{ -  \kappa_s L}}\right] }
= \kappa_s \frac{( 1 +   e^{ -  \kappa_s L})^2}{1 + (1+\kappa_s L)  e^{ -  \kappa_s L} }
\label{sigmastarsparafree}
\end{eqnarray}
In the three regions, the conditioned drift of Eq. \ref{mustarbridgepiintensivesigmaresexpliparametricfree}
reduces to
\begin{eqnarray}
 \mu_{\infty}^{\left[ \sigma_*(s)\right]} = \left\lbrace
  \begin{array}{lll}
  \mu  +  \partial_x    \ln  \left[ {\hat G}^{abs(L)}_s (0 \vert x) \right] 
    =   \kappa_s  
    &~~\mathrm{for~~} x \in ]-\infty,0[
    \\
 \mu  +  \partial_x    \ln  
\left[e^{\kappa_s x} +  e^{\kappa_s (L-x)} \right]
 =   
\kappa_s \frac{ e^{\kappa_s x} -  e^{\kappa_s (L-x)}}{ e^{\kappa_s x} +  e^{\kappa_s (L-x)}} 
=
\kappa_s \frac{ \sinh \left[  \kappa_s \left(x-\frac{L}{2}\right) \right] } 
{\cosh \left[  \kappa_s \left(x-\frac{L}{2}\right) \right]   }
    &~~\mathrm{for~~}   x \in ]0,L[
      \\
       \mu +  \partial_x    \ln  \left[ {\hat G}^{abs(0)}_s (L \vert x) \right]
= - \kappa_s 
    &~~\mathrm{for~~}   x \in ]L,+\infty[
  \end{array}
\right.
\ \ 
\label{mustarbridgepiintensivesigmaresexpliparametricfree3regions}
\end{eqnarray}


\subsection{ Conditioning towards the intensive local times $(a_*,b_*)$ for the infinite horizon $T \to +\infty$ }


\subsubsection{ Special case $a_*>0$ and $b_*=0$ }

The parametric form of Eq. \ref{conditioneddriftastaraloneparametric}
involves the conditioned drift
\begin{eqnarray}
\mu^{[ a_*(s), b_*=0]}_{\infty}( x) 
&& = \mu +  \partial_x    \ln \left[  {\hat G}^{abs(L)}_s (0 \vert x)   \right]
= \mu +  \partial_x    \ln \left[  
 \frac{ e^{ - \mu x } } { \kappa_s} 
  \left(     e^{ - \kappa_s \vert x \vert }
    -   e^{ - \kappa_s (L + \vert L-x \vert ) }   
   \right)
 \right]
 \nonumber \\
 && =   \partial_x    \ln 
  \left(     e^{ - \kappa_s \vert x \vert }
    -   e^{ - \kappa_s (L + \vert L-x \vert ) }      \right)
\label{conditioneddriftastaraloneparametricfree}
\end{eqnarray}
and the parametrization
\begin{eqnarray}
a^*(s) && = \frac{1}{ \partial_s \left[  \frac{1}{ {\hat G}^{abs(L)}_s (0 \vert 0)}    \right] } 
=  \frac{1}{ \partial_s \left[  \frac{\kappa_s}{ 1  -  e^{ - 2 \kappa_s L}}    \right]  } 
=  \frac{\kappa_s}{ \partial_{\kappa_s} \left[  \frac{\kappa_s}{ 1  -  e^{ - 2 \kappa_s L}}    \right]  }
= \kappa_s\left[ \frac{  1  -  e^{ - 2 \kappa_s L}  }
{1- \frac{2 L \kappa_s }{e^{2 \kappa_s L} -1 }}
   \right]
\label{conditioneddriftastaraloneparametricfreea}
\end{eqnarray}
As a consequence, the parametric form of parameter $s$ 
actually only involves the parameter $\kappa =\kappa_s$ of Eq. \ref{kappas}
with $\kappa \in ]0,+\infty[$,
where the initial drift $\mu$ does not appear anymore.
In addition, the conditioned drift does not exist in the region $x \geq L$
(where it is impossible to obtain $b_*=0$ and $a_*>0$),
and can be simplified in the two remaining regions
\begin{eqnarray}
\mu^{[ a_*(s),b_*=0]}(x) = \left\lbrace
  \begin{array}{lll}
   \partial_x    \ln 
  \left(     e^{  \kappa_s x }    -   e^{  \kappa_s (x-2L  ) }      \right)
  =
    \kappa_s  
    &~~\mathrm{for~~} x \in ]-\infty,0[
    \\
  \partial_x    \ln 
  \left(     e^{ - \kappa_s x }   -   e^{  \kappa_s (x-2 L  ) }      \right)
  =  - \kappa_s \frac{e^{  \kappa_s (L-x) }   +   e^{  - \kappa_s (L-x  ) } }
  {e^{  \kappa_s (L-x)  }   -   e^{ - \kappa_s (L-x  ) } }
  =  - \kappa_s \frac{ \cosh[  \kappa_s (L-x) ] } {  \sinh[  \kappa_s (L-x) ] }
    &~~\mathrm{for~~}   x \in ]0,L[
  \end{array}
\right.
\ \ 
\label{conditioneddriftastaraloneparametricfreekappa2regions}
\end{eqnarray}


\subsubsection{ Special case $a_*=0$ and $b_*>0$ }

The parametric form of Eq. \ref{conditioneddriftbstaraloneparametric}
involves the conditioned drift
\begin{eqnarray}
\mu^{[ a_*=0, b_*(s)]}_{\infty}( x) &&  = \mu +  \partial_x    \ln \left[ {\hat G}^{abs(0)}_s (L \vert x)   \right]
=  \mu +  \partial_x    \ln \left[
\frac{ e^{\mu (L-x) } } { \kappa_s} 
  \left( e^{ - \kappa_s \vert L-x \vert}
  -  e^{ - \kappa_s (L + \vert x \vert )} \right)
    \right]
    \nonumber \\
    && =   \partial_x    \ln 
  \left( e^{ - \kappa_s \vert L-x \vert}  -  e^{ - \kappa_s (L + \vert x \vert )} \right)
\label{conditioneddriftbstaraloneparametricfree}
\end{eqnarray}
and the parametrization
\begin{eqnarray}
 b_*(s) &&= \frac{1}{\partial_s \left[ \frac{1}{ {\hat G}^{abs(0)}_s (L \vert L)}     \right]}
 = \frac{1}{\partial_s \left[  \frac{\kappa_s}{ 1  -  e^{ - 2 \kappa_s L}}     \right]}
 = \kappa_s\left[ \frac{  1  -  e^{ - 2 \kappa_s L}  }
{1- \frac{2 L \kappa_s }{e^{2 \kappa_s L} -1 }}
   \right]
\label{conditioneddriftbstaraloneparametricfreeb}
\end{eqnarray}
the parametric form of parameter $s$ 
actually only involves the parameter $\kappa =\kappa_s$ of Eq. \ref{kappas}
with $\kappa \in ]0,+\infty[$,
where the initial drift $\mu$ does not appear anymore.
In addition, the conditioned drift does not exist in the region $x \leq 0$
(where it is impossible to obtain $a_*=0$ and $b_*>0$),
and can be simplified in the two remaining regions
\begin{eqnarray}
\mu^{[ a_*=0, b_*(s)]}_{\infty}( x) 
  = 
 \left\lbrace
  \begin{array}{lll}
   \partial_x    \ln 
  \left(    e^{ - \kappa_s (L-x)}  -  e^{ - \kappa_s (L + x )}         \right)
  = \kappa_s \frac{ e^{  \kappa_s x}  +  e^{ - \kappa_s  x }  }
  { e^{  \kappa_s x}  -  e^{ - \kappa_s x}  } 
  = \kappa_s \frac{ \cosh(  \kappa_s x)  }  { \sinh(\kappa_s x )  }
    &~~\mathrm{for~~} x \in ]0,L[
    \\
  \partial_x    \ln 
  \left(   e^{ - \kappa_s (x-L) }  -  e^{ - \kappa_s (L + x )}        \right)
  =  -  \kappa_s  
    &~~\mathrm{for~~}   x \in ]L,+\infty[
  \end{array}
\right.
\label{conditioneddriftbstaraloneparametricfreekappa2regions}
\end{eqnarray}


\subsubsection{ General case $a_*>0$ and $b_*>0$ }

The conditioned drift of Eq. \ref{mustarbridgepiintensivefinalsimplifparametric}
reads
 \begin{eqnarray}
 \mu^{[ \frac{\sigma_*(s)(1+r_*)}{2} , \frac{\sigma_*(s)(1-r_*)}{2} ]}_{\infty}( x) 
= \mu
&& +  \partial_x    \ln  
\bigg[
\left(  \sqrt{ \frac{1+r_*}{1-r_*} }  \left[1 - e^{(\mu  - \kappa_s )L }  \right] 
     +    \left[e^{\mu L} - e^{  - \kappa_s L }  \right]                  \right)
                   {\hat G}^{abs(L)}_s (0 \vert x)
 \nonumber \\
 &&     +  
   \left( \sqrt{ \frac{1-r_*}{1+r_*} } \left[1 - e^{(-\mu  - \kappa_s )L }  \right]
                 +      \left[e^{-\mu L} - e^{  - \kappa_s L }  \right]  
   \right)     {\hat G}^{abs(0)}_s (L \vert x) \bigg]
\label{mustarbridgepiintensivefinalsimplifparametricfree}
\end{eqnarray}
with the parametrization
 \begin{eqnarray}
 \sigma_*(s)    = \frac{1}{ \partial_s \left[\frac{\kappa_s}{ 1  -  e^{ - 2 \kappa_s L}} 
 -  \sqrt{1-r_*^2} \frac{\kappa_s}{2 \sinh(\kappa_s L ) }   \right] }
 = \frac{\kappa_s} { \partial_{\kappa_s} \left[\frac{\kappa_s ( 1- \sqrt{1-r_*^2} e^{- \kappa_s L} ) }{ 1  -  e^{ - 2 \kappa_s L}}   \right] }
\label{mustarbridgepiintensivefinalsimplifparametricfreesigma}
\end{eqnarray}

The conditioned drift can be simplified in the three regions :

(i) in the left region $x \in ]-\infty,0] $ where the function ${\hat G}^{abs(0)}_s (L \vert x) $ vanishes,
the conditioned drift of Eq. \ref{mustarbridgepiintensivefinalsimplifparametricfree}
 reduces to
 \begin{eqnarray}
 \mu^{[ \frac{\sigma_*(s)(1+r_*)}{2} , \frac{\sigma_*(s)(1-r_*)}{2} ]}_{\infty}( x<0)  
&& = \mu  +  \partial_x    \ln  \left[ {\hat G}^{abs(L)}_s (0 \vert x) \right] 
 =  \mu +  \partial_x    \ln \left[  
 \frac{ e^{ - \mu x } } { \kappa_s} 
  \left(     e^{  \kappa_s  x  }
    -   e^{  \kappa_s (x-2L  ) }   
   \right)
 \right] 
 \nonumber \\
 && = \kappa_s
 \label{mustarbridgepiintensivefinalsimplifleftfree}
\end{eqnarray}

(ii) in the right region $x \in [L,+\infty[$ where
the function ${\hat G}^{abs(L)}_s (0 \vert x) $ vanishes,
the conditioned drift of Eq. \ref{mustarbridgepiintensivefinalsimplifparametricfree}
 reduces to
 \begin{eqnarray}
 \mu^{[ \frac{\sigma_*(s)(1+r_*)}{2} , \frac{\sigma_*(s)(1-r_*)}{2} ]}_{\infty}( x>L) 
&&   = \mu +  \partial_x    \ln  \left[ {\hat G}^{abs(0)}_s (L \vert x) \right]
  =  \mu +  \partial_x    \ln \left[
\frac{ e^{\mu (L-x) } } { \kappa_s} 
  \left( e^{ - \kappa_s (x-L)}
  -  e^{ - \kappa_s (L +  x  )} \right)
    \right]
\nonumber \\
&& = -\kappa_s
 \label{mustarbridgepiintensivefinalsimplifrightfree}
\end{eqnarray}

(iii) in the middle region $x \in ]0,L[$, the conditioned drift of Eq. \ref{mustarbridgepiintensivefinalsimplifparametricfree}
reads
 \begin{eqnarray}
 \mu^{[ \frac{\sigma_*(s)(1+r_*)}{2} , \frac{\sigma_*(s)(1-r_*)}{2} ]}_{\infty}( x\in ]0,L[) 
&& = \mu
\nonumber \\
&& +  \partial_x    \ln  
\bigg[
\left(  \sqrt{ \frac{1+r_*}{1-r_*} }  \left[1 - e^{(\mu  - \kappa_s )L }  \right] 
     +    \left[e^{\mu L} - e^{  - \kappa_s L }  \right]                  \right)
   \frac{ e^{ - \mu x } } { \kappa_s} 
  \left(     e^{ - \kappa_s x }
    -   e^{  \kappa_s (x-2L ) }   
   \right)                   
 \nonumber \\
 &&     +  
   \left( \sqrt{ \frac{1-r_*}{1+r_*} } \left[1 - e^{(-\mu  - \kappa_s )L }  \right]
                 +      \left[e^{-\mu L} - e^{  - \kappa_s L }  \right]  
   \right)    
    \frac{ e^{\mu (L-x) } } { \kappa_s} 
  \left( e^{  \kappa_s (x-L)}  -  e^{ - \kappa_s (x+L )} \right)   
    \bigg]
    \nonumber \\ 
&& = \partial_x    \ln  
\left[ K_s^+(r_*) e^{\kappa_s x} + K_s^-(r_*) e^{\kappa_s (L-x)}\right]
= \kappa_s \frac{K_s^+(r_*) e^{\kappa_s x} - K_s^-(r_*) e^{\kappa_s (L-x)}}{K_s^+(r_*) e^{\kappa_s x} + K_s^-(r_*) e^{\kappa_s (L-x)}}
\label{mustarbridgepiintensivefinalsimplifparametricfreemiddle}
\end{eqnarray}
where we have introduced the two constants
 \begin{eqnarray}
K_s^+(r_*)  && =
 \sqrt{ \frac{1-r_*}{1+r_*} } \left[e^{\mu L }  - e^{  - \kappa_s L }  \right]
 -     e^{ -  \kappa_s L  }     \sqrt{ \frac{1+r_*}{1-r_*} }  \left[1 - e^{(\mu  - \kappa_s )L }  \right]        
                 +      1 - 2 e^{ (\mu - \kappa_s) L }      +     e^{ - 2 \kappa_s L  }                     
 \nonumber \\
K_s^-(r_*)  &&    = 
 \sqrt{ \frac{1+r_*}{1-r_*} }  \left[1 - e^{(\mu  - \kappa_s )L }  \right] 
  -  e^{ -  \kappa_s L  } 
   \sqrt{ \frac{1-r_*}{1+r_*} } \left[e^{\mu L }  - e^{  - \kappa_s L }  \right]        
     +   e^{\mu L} - 2 e^{  - \kappa_s L }        +        e^{ (\mu - 2\kappa_s) L }                                          
\label{kspm}
\end{eqnarray}
even if it is only their ratio that determines the conditioned drift of 
Eq. \ref{mustarbridgepiintensivefinalsimplifparametricfreemiddle}.


\subsection{ Case $\mu>0$ : Conditioning towards the finite sum $\Sigma_{\infty}^*$ for the infinite horizon $T \to +\infty$ }


\subsubsection{Unconditioned distribution $\Pi_{\infty}( \Sigma \vert  x_0)$ of the sum $\Sigma$ for $t=+\infty$ 
when starting at $x_0$}

For the transient case $\mu>0$, the unconditioned distribution $ \Pi_{\infty}(\Sigma \vert x_0) $ 
is given by Eq. \ref{pisigmainfty}
\begin{eqnarray}
 \Pi_{\infty}(\Sigma \vert x_0) 
 =    \delta(\Sigma) S^{abs(0,L)}_{\infty} ( x_0)
+  \theta(\Sigma>0) \Upsilon_{\infty}(\Sigma \vert x_0)
\label{pisigmainftyfree}
\end{eqnarray}
The forever-survival probability $S^{abs(0,L)}_{\infty} ( x_0) $ of Eq. \ref{survivallimitfromlaplace} 
\begin{eqnarray}
S^{abs(0,L)}_{\infty} (  x_0) && =   1  - \frac{ {\hat G}^{abs(L)}_0 (0 \vert x_0)}{{\hat G}^{abs(L)}_0 (0 \vert 0)} 
  - \frac{{\hat G}^{abs(0)}_0 (L \vert x_0)}{{\hat G}^{abs(0)}_0 (L \vert L)}
  \nonumber \\
  && = 1 - e^{\mu (L-x_0- \vert L-x_0 \vert )} 
 =   \left\lbrace
  \begin{array}{lll}
  0
    &~~\mathrm{for~~}   x_0 \in ]-\infty,L]
     \\
    1-e^{-2 \mu (x_0-L)} 
    &~~\mathrm{for~~} x_0 \in]L,+\infty[
  \end{array}
\right.
\label{survivallimitfromlaplacefree}
\end{eqnarray}
is non-vanishing only in the region $x_0>L$, where the particle can escape towards $(+\infty)$ without touching $L$.

The regular contribution of Eq. \ref{upsilonresinteginfty}
\begin{eqnarray}
\Upsilon_{\infty}(\Sigma \vert x_0)  = \xi^+(x_0) e^{- \lambda_0^+ \Sigma } +   \xi^-(x_0)  e^{- \lambda_0^- \Sigma } 
\label{upsilonresinteginftyfree}
\end{eqnarray}
with Eq. \ref{dysonpolyupsrootsfree}
 \begin{eqnarray}
\lambda_0^{\pm}   =  \frac{  \mu } {  1 \mp   e^{ -  \mu L}}
\label{dysonpolyupsrootsfreeszero}
\end{eqnarray}
and Eq. \ref{xipm}
 \begin{eqnarray}
\xi^+(x_0) && = \frac{\mu e^{ \mu (L-x_0)}  }{2 (1-e^{-2 \mu L})} 
\left[ - e^{-\mu \vert x_0 \vert} + e^{ -\mu (L+\vert L-x_0 \vert)} 
+ e^{ -\mu \vert L-x_0 \vert } - e^{ -\mu (L+\vert x_0 \vert)} 
\right]
\nonumber \\
&& =  
 \left\lbrace
  \begin{array}{lll}
  \frac{\mu e^{ \mu (L-x_0)}  }{2 (1-e^{-2 \mu L})} 
\left[ - e^{\mu  x_0 } + e^{ -\mu (2 L-x_0 )} 
\right] = -  \frac{\mu}{2} e^{ \mu L}
    &~~\mathrm{for~~} x_0 \in ]-\infty,0[
    \\ 
 \frac{\mu e^{ \mu (L-x_0)}  }{2 (1-e^{-2 \mu L})} 
\left[ - e^{-\mu  x_0 } + e^{ -\mu (2L-x_0)} 
+ e^{ -\mu (L-x_0) } - e^{ -\mu (L+ x_0 )} 
\right] = \frac{ \mu \left[ 1- e^{\mu( L-2 x_0)} \right]}{2 (1-e^{- \mu L})} 
    &~~\mathrm{for~~}   x_0 \in ]0,L[
     \\ 
  \frac{\mu e^{ \mu (L-x_0)}  }{2 (1-e^{-2 \mu L})} 
\left[ e^{ -\mu (x_0-L) } - e^{ -\mu (L+ x_0 )} 
\right] = \frac{\mu}{2} e^{-2 \mu (x_0-L)}
    &~~\mathrm{for~~} x_0 \in]L,+\infty[
  \end{array}
\right.
 \nonumber \\  
\xi^-(x_0) && = \frac{\mu e^{ \mu (L-x_0)}  }{2 (1-e^{-2 \mu L})} 
\left[  e^{-\mu \vert x_0 \vert} - e^{ -\mu (L+\vert L-x_0 \vert)} 
+ e^{ -\mu \vert L-x_0 \vert } - e^{ -\mu (L+\vert x_0 \vert)} 
\right]
\nonumber \\
&& =  
 \left\lbrace
  \begin{array}{lll}
  \frac{\mu e^{ \mu (L-x_0)}  }{2 (1-e^{-2 \mu L})} 
\left[  e^{\mu  x_0 } - e^{ -\mu (2 L-x_0 )} 
\right]  = \frac{\mu}{2} e^{ \mu L}
    &~~\mathrm{for~~} x_0 \in ]-\infty,0[
    \\ 
 \frac{\mu e^{ \mu (L-x_0)}  }{2 (1-e^{-2 \mu L})} 
\left[  e^{-\mu  x_0 } - e^{ -\mu (2L-x_0)} 
+ e^{ -\mu (L-x_0) } - e^{ -\mu (L+ x_0 )} 
\right]= \frac{ \mu \left[ 1+ e^{\mu( L-2 x_0)} \right]}{2 (1+e^{- \mu L})} 
    &~~\mathrm{for~~}   x_0 \in ]0,L[
     \\ 
  \frac{\mu e^{ \mu (L-x_0)}  }{2 (1-e^{-2 \mu L})} 
\left[   e^{ -\mu (x_0-L) } - e^{ -\mu (L+ x_0 )} 
\right] = \frac{\mu}{2} e^{-2 \mu (x_0-L)}
    &~~\mathrm{for~~} x_0 \in]L,+\infty[
  \end{array}
\right.
\label{xipmfree}
\end{eqnarray}

In summary, the regular contribution of Eq. \ref{upsilonresinteginftyfree}
reads in the three regions
\begin{eqnarray}
\Upsilon_{\infty}(\Sigma \vert x_0)  
=  
 \left\lbrace
  \begin{array}{lll}
    \frac{\mu}{2} e^{ \mu L} \left( - e^{- \lambda_0^+ \Sigma } 
    + e^{- \lambda_0^- \Sigma }\right)
    &~~\mathrm{for~~} x_0 \in ]-\infty,0[
    \\ 
 \frac{ \left[ 1- e^{\mu( L-2 x_0)} \right]}{2 } \lambda_0^+ e^{- \lambda_0^+ \Sigma } 
 + \frac{ \left[ 1+ e^{\mu( L-2 x_0)} \right]}{2 } \lambda_0^- e^{- \lambda_0^- \Sigma } 
    &~~\mathrm{for~~}   x_0 \in ]0,L[
     \\ 
 \frac{\mu}{2} e^{-2 \mu (x_0-L)} \left(  e^{- \lambda_0^+ \Sigma } 
    + e^{- \lambda_0^- \Sigma }\right)
    &~~\mathrm{for~~} x_0 \in]L,+\infty[
  \end{array}
\right.
\label{upsilonresinteginftyfree3regions}
\end{eqnarray}


\subsubsection{ Conditioned drift to obtain the finite sum $\Sigma_{\infty}^*$ for the infinite horizon $T \to +\infty$  }

\begin{figure}[h]
\centering
\includegraphics[width=6.in,height=6.2in]{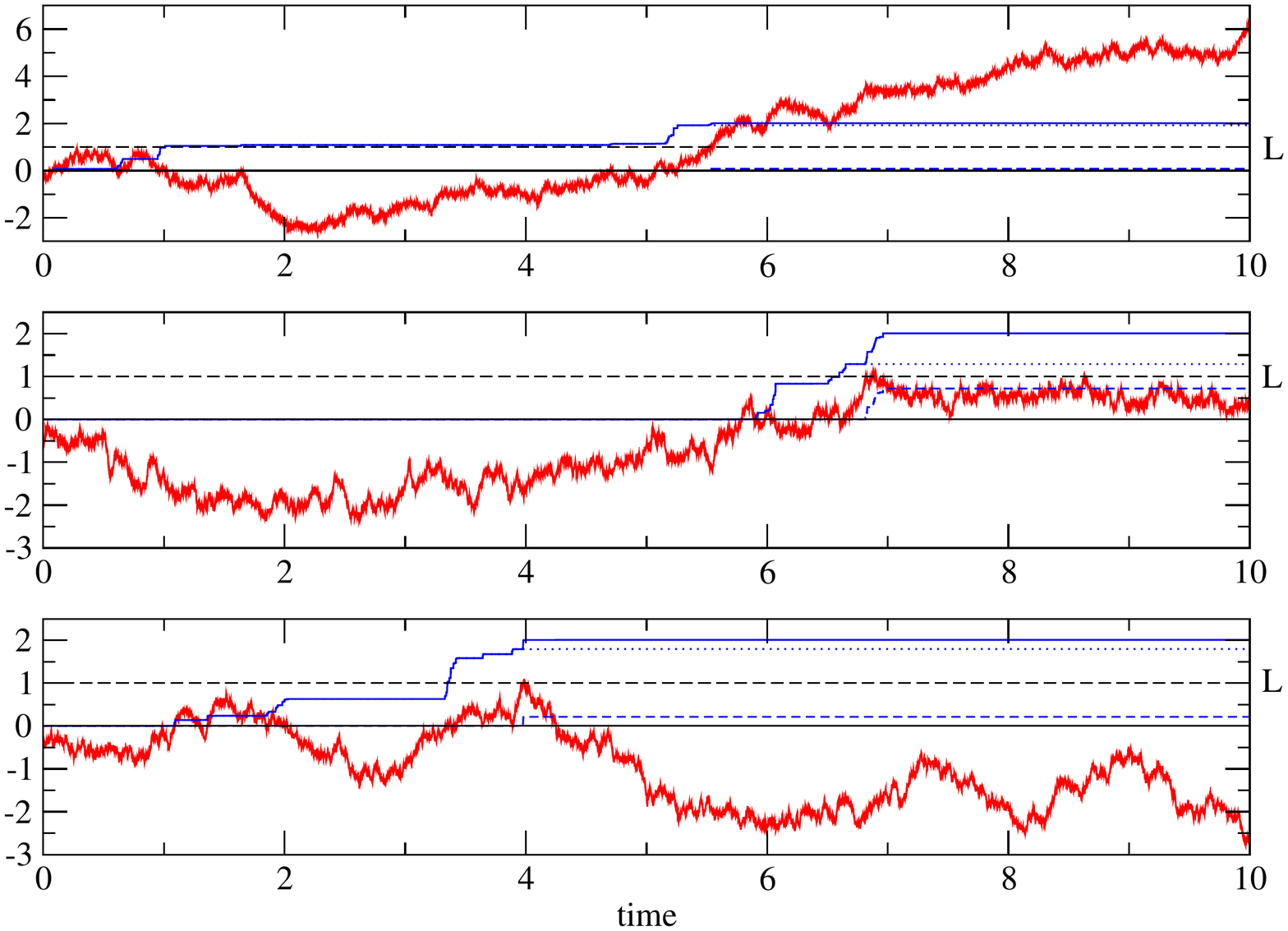}
\setlength{\abovecaptionskip}{15pt}  
\caption{Red lines: Examples of realizations of diffusions conditioned to have a finite sum of local times $\Sigma_{\infty}^* = 2$ at the infinite time horizon (see the conditioned drift of Eqs. \ref{mustarbridgepitinftyregions2freei}, \ref{mustarbridgepitinftyregions2freeii} and \ref{mustarbridgepitinftyregions2taboo}). For each trajectory, the associated local time $A(t)$ at $x=0$ (blue dotted line), the associated local time $B(t)$ at $x=L=1$ (blue dashed line) as well as their sum $\Sigma(t)=A(t)+B(t)$ (blue line) are shown as a function of the time $t $. All processes start at $x_0=-0.5$, the constant drift $\mu$ is equal to $0.5$ and the time step used in the discretization is $dt = 10^{-4}$. Once the desired sum of local times $\Sigma_{\infty}^* = 2$ is reached, there are three possibilities (i)  Top figure:  the process lives in the region $x\in ]L=1,+\infty[$
(ii)  Middle figure: the process lives in the interval $x\in ]0,L=1[$
(iii) Bottom figure: the process lives in the region $x\in ]-\infty,0[$ }
\label{fig3}
\end{figure}

If one takes directly the limit $\tau=+\infty$ in
 the conditioned drift of Eq. \ref{mustarbridgepitinftyregions2},
 one obtains 
 for $\Sigma<\Sigma_{\infty}^* $ using Eq. \ref{upsilonresinteginftyfree3regions}
\begin{eqnarray}
&&\mu^{[\Sigma_{\infty}^*]}_{\infty}( x,\Sigma<\Sigma_{\infty}^*) 
= \mu +   \partial_x    \ln   \Upsilon_{\infty}(\Sigma_T^*-\Sigma \vert x)
\nonumber \\
&& =  
 \left\lbrace
  \begin{array}{lll}
    \mu
    &~~\mathrm{for~~} x \in ]-\infty,0[
    \\ 
\mu +   \partial_x    \ln \left[  \left[ 1- e^{\mu( L-2 x)} \right] \lambda_0^+ e^{- \lambda_0^+ (\Sigma_T^*-\Sigma) } 
 +  \left[ 1+ e^{\mu( L-2 x)} \right] \lambda_0^- e^{- \lambda_0^- (\Sigma_T^*-\Sigma) } \right]
    &~~\mathrm{for~~}   x \in ]0,L[
     \\ 
-\mu
    &~~\mathrm{for~~} x \in]L,+\infty[
  \end{array}
\right.
\label{mustarbridgepitinftyregions2freei} 
 \end{eqnarray}
and for $\Sigma=\Sigma_{\infty}^* $ using Eq. \ref{survivallimitfromlaplacefree} in the region $x \in]L,+\infty[ $
\begin{eqnarray}
\mu^{[\Sigma_{\infty}^*]}_{\infty}( x,\Sigma=\Sigma_{\infty}^*) 
= \mu +  \partial_x    \ln S^{abs(0,L)}_{\infty} (  x)
 = \mu +  \partial_x    \ln \left[  1-e^{-2 \mu (x-L)} \right] = \mu \coth[\mu (x-L) ]
 ~~ \mathrm{for~~} x \in]L,+\infty[
\label{mustarbridgepitinftyregions2freeii} 
 \end{eqnarray}
while for the two other regions where $S^{abs(0,L)}_{\infty} (  x) $ vanishes in Eq. \ref{survivallimitfromlaplacefree},
one should use the asymptotic behavior of $S^{abs(0,L)}_{\tau} (  x) $ for large time $\tau$ to obtain 
that the conditioned drift of Eq. \ref{mustarbridgepitinftyregions2}
corresponds to the taboo process on $]-\infty,0[$ and on the interval $]0,L[$ respectively. We postpone these calculations in the appendix \ref{app_taboo} and state the results
\begin{eqnarray}
\mu^{[\Sigma_{\infty}^*]}_{\infty}( x,\Sigma=\Sigma_{\infty}^*) 
&&= \mu +  \lim_{\tau \to +\infty} \partial_x    \ln S^{abs(0,L)}_{\tau} (  x)
 =  
 \left\lbrace
  \begin{array}{lll}
\frac{1}{x}
    &~~\mathrm{for~~} x \in ]-\infty,0[
    \\ 
    \frac{\pi}{L} \cot \left(  \frac{\pi x}{L} \right)
    &~~\mathrm{for~~}   x \in ]0,L[
  \end{array}
\right.
\label{mustarbridgepitinftyregions2taboo} 
 \end{eqnarray}
Figure \ref{fig3} shows some realizations of this conditioned process with a drift $\mu = 1/2$.



\subsection{ Case $\mu>0$ : Conditioning towards the finite local times $(A_{\infty}^*,B_{\infty}^*)$ for the infinite horizon $T \to +\infty$ }

\subsubsection{Unconditioned distribution $\Pi_{\infty}( A,B\vert  x_0)$ of the two local times $(A,B)$ for $t=+\infty$ 
when starting at $x_0$}

For the transient case $\mu>0$, the unconditioned distribution $ \Pi_{\infty}(A, B \vert x_0) $ 
of Eq. \ref{piAlimit}
\begin{eqnarray}
\Pi_{\infty}( A, B \vert  x_0) 
&&   =   \delta(A) \delta(B) S^{abs(0,L)}_{\infty} (  x_0)
+ \theta(A>0) \delta(B) {\cal A}^{]-\infty,L[}_{\infty} (A \vert x_0)
 + \delta(A)\theta(B>0)  {\cal B}^{]0,+\infty[}_{\infty} (B \vert x_0)
\nonumber \\
&&+ \theta(A>0) \theta(B>0)  {\cal C}_{\infty} (A,B \vert x_0)
\label{piAlimitfree}
\end{eqnarray}
involves the four contributions :

(1) the first contribution
$ S^{abs(0,L)}_{\infty} (  x_0)$ has been given in Eq. \ref{survivallimitfromlaplacefree}

(2) the second contribution $ {\cal A}^{]-\infty,L[}_{\infty} (A \vert x_0) $ of Eq. \ref{calAlimitfromlaplace} vanishes
\begin{eqnarray}
{\cal A}^{]-\infty,L[}_{\infty} (A \vert x_0)   =0
\label{calAlimitfromlaplacefree}
\end{eqnarray}

(3) the third contribution ${\cal B}^{]0,+\infty[}_{\infty} (B \vert x_0) $ of Eq. \ref{calBlimitfromlaplace} reads
\begin{eqnarray}
{\cal B}^{]0,+\infty[}_{\infty} (B \vert x_0) 
&& = e^{\mu (L-x_0)} \left[ e^{- \mu \vert L-x_0\vert} - e^{- \mu (L+\vert x_0\vert) }\right] 
 \frac{\mu}{1-e^{-2 \mu L}} e^{ - \frac{\mu}{1-e^{-2 \mu L}} B }
 \nonumber \\
 && 
 = \left\lbrace
  \begin{array}{lll}
   0
    &~~\mathrm{for~~} x_0 \in ]-\infty,0[
    \\ 
 \left[ 1-e^{-2 \mu x_0} \right] \frac{\mu}{1-e^{-2 \mu L}} e^{ - \frac{\mu}{1-e^{-2 \mu L}} B }
    &~~\mathrm{for~~}   x_0 \in ]0,L[
     \\ 
e^{-2 \mu (x_0-L) } \mu e^{ - \frac{\mu}{1-e^{-2 \mu L}} B }
    &~~\mathrm{for~~} x_0 \in]L,+\infty[
  \end{array}
\right.
\label{calBlimitfromlaplacefree}
\end{eqnarray}
with the normalization over $B \in ]0,+\infty[$
\begin{eqnarray}
\int_0^{+\infty} dB {\cal B}^{]0,+\infty[}_{\infty} (B \vert x_0) 
 = \left\lbrace
  \begin{array}{lll}
   0
    &~~\mathrm{for~~} x_0 \in ]-\infty,0[
    \\ 
 1-e^{-2 \mu x_0}  
    &~~\mathrm{for~~}   x_0 \in ]0,L[
     \\ 
e^{-2 \mu (x_0-L) } -e^{-2 \mu x_0} 
    &~~\mathrm{for~~} x_0 \in]L,+\infty[
  \end{array}
\right.
\label{calBlimitfromlaplacefreenorma}
\end{eqnarray}

(4) the fourth contribution of Eq. \ref{calClimitfromlaplace} reads
\begin{eqnarray}
\label{calClimitfromlaplacefree}
&& {\cal C}_{\infty} (A,B \vert x_0) 
   =  \frac{\mu^2}{(1-e^{-2 \mu L})^2} e^{ - \frac{\mu}{1-e^{-2 \mu L}} (A+B) } e^{-\mu x_0} 
 \\
&&   \times
\left[      
\left( e^{-\mu \vert x_0 \vert} - e^{-\mu (L + \vert L-x_0 \vert)} \right)
I_0 \left( \frac{\mu}{ \sinh(\mu L ) }   \sqrt{AB} \right) 
+   \left(e^{- \mu \vert L-x_0 \vert } - e^{-\mu (L + \vert x_0 \vert)} \right)
         \frac{  \sqrt{B}}{ \sqrt{A} }       I_1 \left(  \frac{\mu}{ \sinh(\mu L ) }  \sqrt{AB} \right) 
\right]  
\nonumber \\
 && 
 = \left\lbrace
  \begin{array}{lll}
  \frac{\mu^2  e^{ - \frac{\mu}{1-e^{-2 \mu L}} (A+B) }}{1-e^{-2 \mu L}}  
I_0 \left( \frac{\mu}{ \sinh(\mu L ) }   \sqrt{AB} \right)    
    &~~\mathrm{for~~} x_0 \in ]-\infty,0[
    \\ 
    \frac{\mu^2  e^{ - \frac{\mu}{1-e^{-2 \mu L}} (A+B) }}{(1-e^{-2 \mu L})^2}       
    \left[      
\left( e^{-2 \mu x_0} - e^{-2 \mu L } \right)
I_0 \left(  \frac{\mu}{ \sinh(\mu L ) }   \sqrt{AB} \right) 
+  e^{- \mu L} \left( 1 - e^{- 2 \mu x_0} \right)
          \frac{  \sqrt{B}}{ \sqrt{A} }      I_1 \left(  \frac{\mu}{ \sinh(\mu L ) }  \sqrt{AB} \right) 
\right] 
    &~~\mathrm{for~~}   x_0 \in ]0,L[
     \\ 
     \frac{\mu^2  e^{ - \frac{\mu}{1-e^{-2 \mu L}} (A+B) }}{1-e^{-2 \mu L}}  e^{\mu L - 2\mu x_0} 
          \frac{  \sqrt{B}}{ \sqrt{A} }      I_1 \left(  \frac{\mu}{ \sinh(\mu L ) }  \sqrt{AB} \right) 
    &~~\mathrm{for~~} x_0 \in]L,+\infty[
  \end{array}
\right.   
\nonumber
\end{eqnarray}
with the normalization over $(A \in ]0,+\infty[;B \in ]0,+\infty[)$
using Eqs \ref{bessel}
and \ref{besselderi}
\begin{eqnarray}
 \int_0^{+\infty} dA \int_0^{+\infty} dB {\cal C}_{\infty} (A,B \vert x_0)  
 = \left\lbrace
  \begin{array}{lll}
  1
    &~~\mathrm{for~~} x_0 \in ]-\infty,0[
    \\ 
    e^{-2 \mu x_0} 
    &~~\mathrm{for~~}   x_0 \in ]0,L[
     \\ 
    e^{-2 \mu x_0} 
    &~~\mathrm{for~~} x_0 \in]L,+\infty[
  \end{array}
\right.   
\label{calClimitfromlaplacefreenorma}
\end{eqnarray}


\subsubsection{Conditioned drift to obtain the finite local times $(A_{\infty}^*,B_{\infty}^*)$ for the infinite horizon $T \to +\infty$}

\begin{figure}[h]
\centering
\includegraphics[width=6.in,height=6.2in]{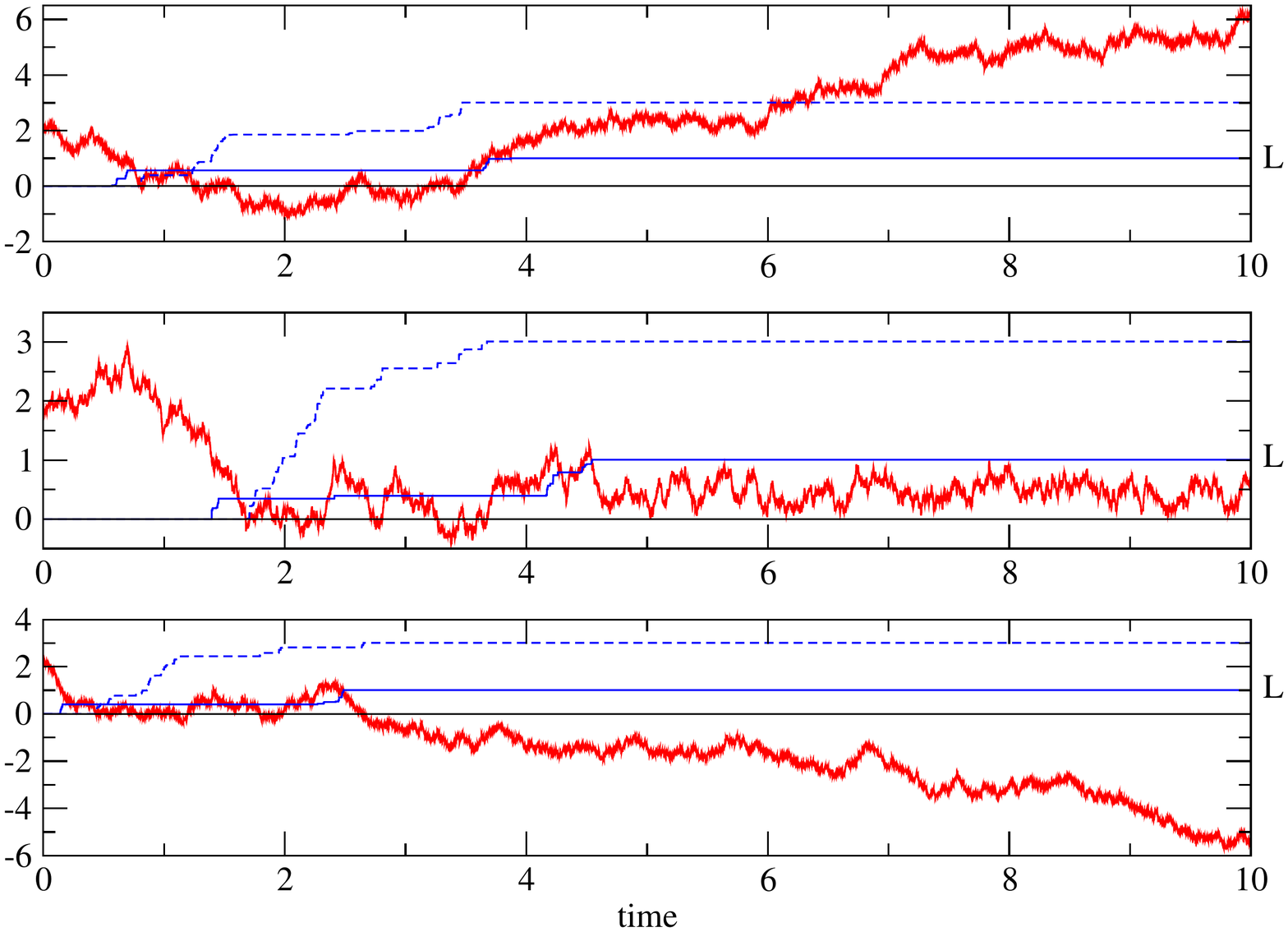}
\setlength{\abovecaptionskip}{15pt}  
\caption{Red lines: Examples of realizations of diffusions conditioned to have finite local times $(A_{\infty}^* = 3,B_{\infty}^* = 1)$ at the infinite time horizon (see the conditioned drift of Eqs. \ref{mustarbridgepitinftyregionsCfree}, \ref{mustarbridgepitinftyregionsBfree} and \ref{mustarbridgepitinftyregionsAfree}). For each trajectory, the associated local time $A(t)$ at $x=0$ (blue line) and the associated local time $B(t)$ at $x=L=1$ (blue dashed line) are shown as a function of the time $t$. All processes start at $x_0 = 2$, the constant drift $\mu$ is equal to $0.5$ and the time step used in the discretization is $dt = 10^{-4}$. Once both the desired local times $A_{\infty}^* = 3$ and $B_{\infty}^* = 1$ are reached, 
there are three possibilities (i)  Top figure:  the process lives in the region $x\in ]L=1,+\infty[$
(ii)  Middle figure: the process lives in the interval $x\in ]0,L=1[$
(iii) Bottom figure: the process lives in the region $x\in ]-\infty,0[$
}
\label{fig4}
\end{figure}

If one takes directly the limit $\tau=+\infty$ in
 the conditioned drift of Eq. \ref{mustarbridgepitinftyregions},
 one obtains 
 for $(A<A_{\infty}^*,B< B_{\infty}^*)$ using Eq. \ref{calClimitfromlaplacefree}
\begin{eqnarray}
&& \mu^{[A_{\infty}^*,B_{\infty}^*]}_{\infty}( x,A<A_{\infty}^*,B< B_{\infty}^*) 
= \mu +   \partial_x    \ln    {\cal C}_{\infty} (A_{\infty}^*-A,B_{\infty}^*-B \vert x)
\label{mustarbridgepitinftyregionsCfree}  \\
&& =  
 \left\lbrace
  \begin{array}{lll}
    \mu 
    &~~\mathrm{for~~} x \in ]-\infty,0[
    \\ 
   \mu \frac{\sqrt{B_{\infty}^*-B} \cosh(\mu x)  I_1 \left(  \frac{\mu}{ \sinh(\mu L ) }  \sqrt{(A_{\infty}^*-A)(B_{\infty}^*-B)} \right) -  \sqrt{A_{\infty}^*-A} \cosh(\mu (L - x)) I_0 \left(  \frac{\mu}{ \sinh(\mu L ) }   \sqrt{(A_{\infty}^*-A)(B_{\infty}^*-B)} \right)}
    {\sqrt{B_{\infty}^*-B} \sinh(\mu x)  I_1 \left(  \frac{\mu}{ \sinh(\mu L ) }  \sqrt{(A_{\infty}^*-A)(B_{\infty}^*-B)} \right) +  \sqrt{A_{\infty}^*-A} \sinh(\mu (L - x)) I_0 \left(  \frac{\mu}{ \sinh(\mu L ) }   \sqrt{(A_{\infty}^*-A)(B_{\infty}^*-B)} \right)}
    &~~\mathrm{for~~}   x \in ]0,L[
     \\ 
-\mu
    &~~\mathrm{for~~} x \in]L,+\infty[
  \end{array}
\right.
\nonumber
 \end{eqnarray}

 and for $(A=A_{\infty}^*,B< B_{\infty}^*)$ using Eq. \ref{calBlimitfromlaplacefree} 
\begin{eqnarray}
 \mu^{[A_{\infty}^*,B_{\infty}^*]}_{\infty}( x,A=A_{\infty}^*,B< B_{\infty}^*) 
&& = \mu +   \partial_x    \ln  {\cal B}^{]0,+\infty[}_{\infty} (B_{\infty}^*-B\vert x)
\nonumber \\
&& = \left\lbrace
  \begin{array}{lll}
  \mu +   \partial_x    \ln \left[ 1-e^{-2 \mu x} \right] = \mu \coth(\mu x)
    &~~\mathrm{for~~}   x \in ]0,L[
     \\ 
- \mu 
    &~~\mathrm{for~~} x \in]L,+\infty[
  \end{array}
\right.
\label{mustarbridgepitinftyregionsBfree} 
 \end{eqnarray}
For  $(A<A_{\infty}^*,B = B_{\infty}^*)$ where  $ {\cal A}^{]-\infty,L[}_{\infty} (A \vert x) $ vanished in Eq. \eqref{calAlimitfromlaplacefree}, one should use the asymptotic behavior of  $ {\cal A}^{]-\infty,L[}_{\tau} (A \vert x) $ for large time $\tau$ to obtain that the conditioned drift has a behavior similar to the previous case, as expected 
\begin{eqnarray}
 \mu^{[A_{\infty}^*,B_{\infty}^*]}_{\infty}( x,A<A_{\infty}^*,B= B_{\infty}^*) 
&& = \mu + \lim_{\tau \to +\infty}  \partial_x    \ln   {\cal A}^{]-\infty,L[}_{\tau} (A_{\infty}^*-A\vert x)
\nonumber \\
&& = \left\lbrace
  \begin{array}{lll}
  \mu  
    &~~\mathrm{for~~}   x \in ]-\infty,0[
     \\ 
- \mu \coth(\mu (L - x)) 
    &~~\mathrm{for~~} x \in ]0,L[
  \end{array}
\right.
\label{mustarbridgepitinftyregionsAfree} 
 \end{eqnarray} 
At last, for $(A=A_{\infty}^*,B=B_{\infty}^*)$, the conditioned drift involving the survival probability $S^{abs(0,L)}_{\tau} (  x) $ has already been discussed above in Eqs \ref{mustarbridgepitinftyregions2freeii} 
and \ref{mustarbridgepitinftyregions2taboo}.
Figure \ref{fig4} shows some realizations of this conditioned process with a drift $\mu = 1/2$.

\section{ Conclusions } 

\label{sec_conclusion}

In this paper, we have considered a diffusion process $X(t)$ of drift $\mu(x)$ and of diffusion coefficient $D=1/2$ in order to study the joint statistics of the two local times $A(t)= \int_{0}^{t} d\tau \delta(X(\tau))  $ and $B(t)= \int_{0}^{t} d\tau \delta(X(\tau)-L) $ at positions $x=0$ and $x=L$, as well as the simpler statistics of their sum $ \Sigma(t)=A(t)+B(t)$. We have discussed the asymptotic behavior for large time $t \to + \infty$ : (i) when the diffusion process $X(t)$ is transient, the two local times $[A(t);B(t)]$ remain finite random variables $[A^*(\infty),B^*(\infty)]$ and we have analyzed their limiting joint distribution ; (ii) when the diffusion process $X(t)$ is recurrent, we have described the large deviations properties of the two intensive local times $a = \frac{A(t)}{t}$ and $b = \frac{B(t)}{t}$ and of the intensive sum $\sigma = \frac{\Sigma(t)}{t}=a+b$. We have then used these properties to construct various conditioned processes $[X^*(t),A^*(t),B^*(t)]$ satisfying certain constraints involving the two local times, thereby generalizing our previous work \cite{us_LocalTime} concerning the conditioning with respect to the single local time $A(t)$. In particular for the infinite time horizon $T \to +\infty$, we have considered the conditioning towards the finite asymptotic values $[A^*(\infty),B^*(\infty)]$ or $\Sigma^*(\infty) $, as well as the conditioning towards the intensive values $[a^*,b^*] $ or $\sigma^*$, that we have compared in Appendix \ref{sec_canonical} with the appropriate 'canonical conditioning' based on the generating function of the local times in the regime of large deviations. 
Finally, we have applied this general construction to the simplest case where the unconditioned diffusion is the Brownian motion of uniform drift $\mu$, while the comparison with the canonical conditioning is described in Appendix \ref{sec_canonicalbrown}.


\appendix


\section{ Canonical conditioned process $X_{p,q}^*(t)$ of parameters $(p,q)$ } 

\label{sec_canonical}

In this Appendix, we describe the properties of the canonical conditioning of parameters $(p,q)$,
in order to compare with the other conditioned processes described in section \ref{sec_doob} of the main text.

\subsection{ Canonical conditioned process $X_{p,q}^*(t)$ of parameters $(p,q)$ based on the Laplace transform
${\tilde P}_{t,p,q} (x  \vert x_0)$ }

The canonical conditioning is based 
on the Laplace transform ${\tilde P}_{t,p,q} (x  \vert x_0)$ of Eq. \ref{laplacefour}
 where the Laplace parameters $(p,q)$ conjugated to the two local times $A$ and $B$ are fixed.
For the bridge conditioned to end at the position $x_T^*$ at the time horizon $T$,
the conditioned distribution for the position $x$ at an interior time $t \in [0,T]$ reads
\begin{eqnarray}
P^{[x_T^*;p,q]}_T(x,t) =  \frac{{\tilde P}_{T-t,p,q}(x_T \vert x) {\tilde P}_{t,p,q}(x \vert x_0)}{ {\tilde P}_{T,p,q}(x_T \vert x_0)}
\label{markovcondk}
\end{eqnarray}
The corresponding Ito dynamics for the conditioned process $X_{p,q}^*(t) $ of parameters $(p,q)$
\begin{eqnarray}
dX_{p,q}^*(t)  =  \mu_{p,q}^*( X_{p,q}^*(t),t ) dt + dB(t)
\label{Doobp}
\end{eqnarray}
involves the conditioned drift
\begin{eqnarray}
  \mu^{[x_T^*;p,q]}_T( x,t )  = \mu(x) + \partial_x \ln {\tilde P}_{T-t,p,q}(x_T \vert x)
\label{mustarp}
\end{eqnarray}


\subsection{ Properties of the $(p,q)$-deformed propagator ${\tilde P}_{t,p,q} (x  \vert x_0) $ }

The forward dynamics of the $(p,q)$-deformed propagator $  {\tilde P}_{t,p,q} (x  \vert x_0)$
is given by 
the Feynman-Kac formula of Eq. \ref{feynmankac}
\begin{eqnarray}
\partial_t  {\tilde P}_{t,p,q} (x  \vert x_0)
&& = -  p \delta(x)  {\tilde P}_{t,p,q} (x  \vert x_0)
-  q \delta(x-L)  {\tilde P}_{t,p,q} (x  \vert x_0)
 -    \partial_x \left[ \mu(x)  {\tilde P}_{t,p,q} (x  \vert x_0) \right] +  \frac{1}{2} \partial_x^2   {\tilde P}_{t,p,q} (x  \vert x_0) 
\label{forward}
\end{eqnarray}


\subsubsection{ Similarity transformation towards an hermitian quantum Hamiltonian $ H_{p,q} $ }

The potential $U(x)$ defined via the following integration of the drift $\mu(y) $
\begin{eqnarray}
U(x) \equiv - 2 \int_0^{x} dy \mu(y) 
\label{potentialU}
\end{eqnarray}
can be used to make the similarity transformation 
\begin{eqnarray}
{\tilde P}_{t,p,q} (x  \vert x_0)  =e^{- \frac{  U(x) }{2}}\psi_{t,p,q} (x  \vert x_0)e^{ \frac{  U(x_0) }{2}}
= e^{ \int_{x_0}^{x} dy \mu(y)} \psi_{t,p,q} (x  \vert x_0)
\label{defpsip}
\end{eqnarray}
in order to transform the dynamics of Eq. \ref{forward}
for ${\tilde P}_{t,p,q} (x  \vert x_0) $ 
 into the Euclidean Schr\"odinger Equation for 
$\psi_{t,p,q} (x  \vert x_0) $
\begin{eqnarray}
- \partial_t \psi_{t,p,q} (x  \vert x_0) = H_{p,q} \psi_{t,p,q} (x  \vert x_0)
\label{schrodingerp}
\end{eqnarray}
where the hermitian quantum Hamiltonian $H_{p,q} $ involves 
two additional delta impurities of amplitudes $p$ and $q$ at positions $x=0$ and $x=L$ respectively
\begin{eqnarray}
 H_{p,q} =  H_{0,0} + p \delta(x)+q \delta(x-L)
\label{hamiltonianpq}
\end{eqnarray}
with respect to the supersymmetric Hamiltonian
\begin{eqnarray}
 H_{0,0} =   \frac{1}{2}\left(  \partial_x + \mu(x) \right) \left( - \partial_x + \mu(x) \right) = -  \frac{1}{2} \partial_x^2 +V(x)
\label{hamiltonian}
\end{eqnarray}
that 
involves the quantum potential
\begin{eqnarray}
 V(x) \equiv \frac{ \mu^2(x)}{2} + \frac{\mu'(x)}{2} 
\label{susy}
\end{eqnarray}


\subsubsection{ Physical meaning of the normalizable zero-energy ground state of the supersymmetric Hamiltonian $H_{0,0}$ when it exists }

Let us recall the well-known discussion :

(i) if the following integral involving the potential $U(x)$ of Eq. \ref{potentialU}
converges
\begin{eqnarray}
   \int_{-\infty}^{+\infty} dx e^{-  U(x) } <+\infty
\label{qsusyannihilatenorma}
\end{eqnarray}
then the quantum Hamiltonian $H_{0,0}$ has the following normalizable ground state at zero-energy $E=0$
\begin{eqnarray}
 \phi^{GS}_{0,0}(  x) = \frac{ e^{- \frac{ U(x)}{2} }}{ \sqrt{ \int_{-\infty}^{+\infty} dy e^{-U(y) } } } 
\label{gs}
\end{eqnarray}
The propagator $G_t(x \vert x_0)= {\tilde P}_{t,p=0,q=0} (x  \vert x_0) $ 
obtained from the similarity transformation of Eq. \ref{defpsip}
\begin{eqnarray}
G(x,t \vert x_0,t_0) &&= e^{- \frac{  U(x) }{2}} \psi_{t,p=0,q=0}(x \vert x_0)e^{ \frac{  U(x_0) }{2}}
 \opsimeq_{(t-t_0) \to +\infty} e^{- \frac{  U(x) }{2}} \phi^{GS}_{0,0}(  x)
  \phi^{GS}_{0,0}(  x_0)e^{ \frac{  U(x_0) }{2}}
\nonumber \\
&& \opsimeq_{(t-t_0) \to +\infty}  \frac{ e^{-  U(x) }}{  \int_{-\infty}^{+\infty} dy e^{-U(y) }  } \equiv G_{eq}(x)
\label{boltzmann}
\end{eqnarray}
converges towards the Boltzmann equilibrium $G_{eq}(x) $ 
in the potential $U(x)$.

(ii) if the integral of Eq. \ref{qsusyannihilatenorma} diverges
\begin{eqnarray}
   \int_{-\infty}^{+\infty} dx e^{-  U(x) } =+\infty
\label{dvnoeq}
\end{eqnarray}
then the quantum Hamiltonian $H_{0,0}$ has no bound state, and the process $X(t)$
does not converge towards an equilibrium, but it can be either transient or recurrent.


\subsubsection{ Propagator ${\tilde P}_{t,p,q} (x  \vert x_0) $ for large time $t$ when 
the quantum Hamiltonian $H_{p,q}$ has a normalizable ground-state }

When the quantum Hamiltonian $H_{p,q}$ has 
a normalizable ground-state $\phi_{p,q}^{GS}(  x) $ of energy $E_{p,q}$
\begin{eqnarray}
 H_{p,q} \phi_{p,q}^{GS}(  x) =E_{p,q} \phi_{p,q}^{GS}(  x) 
\label{eigenhamiltonianp}
\end{eqnarray}
the ground state can be chosen real and positive $\phi_{p,q}^{GS}(  x) \geq 0$
with the normalization
\begin{eqnarray}
\langle  \phi_{p,q}^{GS} \vert  \phi_p^{GS} \rangle = \int_{-\infty}^{+\infty} dx \left[  \phi_{p,q}^{GS}(  x) \right]^2=1 
\label{quantumnorma}
\end{eqnarray}
This ground-state $\phi_{p,q}^{GS}(  x) $ and its energy $E_{p,q}$
determine the leading asymptotic behavior
of the quantum propagator 
\begin{eqnarray}
 \psi_{t,p,q} (  x \vert   x_0) \opsimeq_{t \to + \infty}  e^{- t E_{p,q}}
 \phi_{p,q}^{GS}(  x) \phi_{p,q}^{GS}(  x_0) 
\label{psipGS}
\end{eqnarray}
The corresponding asymptotic behavior of the propagator ${\tilde P}_{t,p,q} (x  \vert x_0) $ 
is then given by 
the similarity transformation of Eq. \ref{defpsip} 
\begin{eqnarray}
{\tilde P}_{t,p,q} (x  \vert x_0) && =e^{- \frac{  U(x) }{2}}  \psi_{t,p,q} (  x \vert   x_0)e^{ \frac{  U(x_0) }{2}}
\nonumber \\
&& \opsimeq_{t \to + \infty}  e^{-t E_{p,q}}
\left[ e^{- \frac{  U(x) }{2}} \phi_{p,q}^{GS}(  x) \right] \left[e^{ \frac{  U(x_0) }{2}} \phi_{p,q}^{GS}(  x_0) \right]
\label{ppGS}
\end{eqnarray}

For the further time Laplace transform $ {\hat {\tilde P}}_{s,p,q} (x \vert x_0)$
of Eq. \ref{laplacetriple}, the asymptotic behavior of Eq. \ref{ppGS} for large $t$
means that  $ {\hat {\tilde P}}_{s,p,q} (x \vert x_0)$ exists for $s \in ]-E_{p,q},+\infty[$
with the following pole singularity for $s \to (-E_{p,q})^+ $
\begin{eqnarray}
{\hat {\tilde P}}_{s,p,q} (x \vert x_0) = \int_{0}^{+\infty} dt e^{- s t } {\tilde P}_{t,p,q} (x  \vert x_0)
\opsimeq_{s \to (-E_{p,q})^+}
 \frac{ \left[ e^{- \frac{  U(x) }{2}} \phi_{p,q}^{GS}(  x) \right] \left[e^{ \frac{  U(x_0) }{2}} \phi_{p,q}^{GS}(  x_0) \right] } {s+E_{p,q} } 
\ \ \ \ 
\label{laplacetriplepole}
\end{eqnarray}
The comparison with the denominator of $ {\hat {\tilde P}}_{s,p,q} (x \vert x_0)$ in Eq. \ref{dysonpoly}
shows that $s=-E_{p,q} $ can be found as the solution of
 \begin{eqnarray}
0  =1 + p  {\hat G}_s (0 \vert 0)   + q {\hat G}_s (L \vert L)  +  p q \left[ {\hat G}_s (0 \vert 0)  {\hat G}_s (L \vert L)  -  {\hat G}_s (L \vert 0) {\hat G}_s (0 \vert L) \right]
\label{epqfroms}
\end{eqnarray}


\subsection{ Canonical conditioning for large horizon $T $ when the Hamiltonian $H_{p,q}$ has a normalizable ground-state} 

\label{sec_canonicalGS}

When the quantum Hamiltonian $H_{p,q}$ has 
a normalizable ground-state $\phi_{p,q}^{GS}(  x) $,
the asymptotic behavior of Eq. \ref{ppGS}
can be plugged into the the three propagators of Eq. \ref{markovcondk}
to obtain that the conditioned density at any interior time $0 \ll t \ll T$
\begin{eqnarray}
P^{[x_T^*;p,q]}_T(x,t) && \opsimeq_{ 0 \ll t \ll T}  
 \frac{ e^{-E_{p,q} (T-t) } \left[ e^{- \frac{  U(x_T) }{2}} \phi_{p,q}^{GS}(  x_T) \right] \left[e^{ \frac{  U(x) }{2}} \phi_{p,q}^{GS}(  x) \right]
  e^{-E_{p,q}t} \left[ e^{- \frac{  U(x) }{2}} \phi_{p,q}^{GS}(  x) \right] \left[e^{ \frac{  U(x_0) }{2}} \phi_{p,q}^{GS}(  x_0) \right]}
 {e^{-E_{p,q}T} \left[ e^{- \frac{  U(x_T) }{2}} \phi_{p,q}^{GS}(  x_T) \right] \left[e^{ \frac{  U(x_0) }{2}} \phi_{p,q}^{GS}(  x_0) \right] }
\nonumber \\
&& =\left[  \phi_{p,q}^{GS}(  x) \right]^2 \equiv P^*_{p,q}(x) 
\label{pkinterior}
\end{eqnarray}
does not depend on the interior time $t$ anymore.

The corresponding conditioned drift of Eq. \ref{mustarp} is also independent of the interior time $t$
and reduces to
\begin{eqnarray}
  \mu^{[x_T^*;p,q]}_T( x,t ) && \opsimeq_{ 0 \ll t \ll T} 
    \mu(x) + \partial_x \ln \left(  e^{-E_{p,q} (T-t) } \left[ e^{- \frac{  U(x_T) }{2}} \phi_{p,q}^{GS}(  x_T) \right] \left[e^{ \frac{  U(x) }{2}} \phi_{p,q}^{GS}(  x) \right]\right)
  \nonumber \\
  &&   = \partial_x \ln \left[  \phi_{p,q}^{GS}(  x)\right]   \equiv  \mu_{p,q}^*( x )
\label{mustarpinterior}
\end{eqnarray}
where we have used the derivative 
$U'(x) = - 2  \mu(x) $
of the potential $U(x)$ of Eq. \ref{potentialU}.

\subsubsection{ Physical meaning of this canonical conditioning when the unconditioned process $X(t)$ is recurrent }

When the unconditioned process $X(t)$ is recurrent,
the ground-state energy $E_{p,q}$ of the Hamiltonian $H_{p,q}$
that governs the leading exponential behavior of Eq. \ref{ppGS}
\begin{eqnarray}
 {\tilde P}_{T,p,q} (x  \vert x_0)  \oppropto_{T \to + \infty}  e^{- T E_{p,q}}
\label{laplaceAtlarge}
\end{eqnarray}
is directly related to the rate function $I(a,b)$ that governs the large deviations properties of
the intensive local times $(a,b)$ of Eq. \ref{rateab} :
indeed, the saddle-point evaluation of the generating function
of the two intensive local times $(a,b)$ 
\begin{eqnarray}
  \langle e^{-p T a-q T b } \rangle_{x_0} && =\int_0^{+\infty} da  \int_0^{+\infty} db e^{-p T a-q T b} \Pi(ta,tb \vert x_0) 
\nonumber \\
&& \opsimeq_{T \to +\infty} 
 \int_0^{+\infty} da \int_0^{+\infty} db e^{ -T \left[ p a +qb+ I ( a,b ) \right] }\opsimeq_{T \to +\infty} e^{ -T E_{p,q} }
\label{level1gen}
\end{eqnarray} 
yields that the ground-state
energy $E_{p,q}$ corresponds to the two-dimensional Legendre transform of the rate function $ I ( a,b ) $
 \begin{eqnarray}
 p a +qb+ I ( a,b ) && =  E_{p,q}
 \nonumber \\
 p+ \partial_a I(a,b) && =0
  \nonumber \\
 q+ \partial_b I(a,b) && =0
\label{legendre}
\end{eqnarray} 
while the reciprocal Legendre transform reads
 \begin{eqnarray}
I(a,b) && =E_{p,q}-pa -qb
 \nonumber \\
a && = \partial_p E_{p,q}
 \nonumber \\
b && = \partial_q E_{p,q}
\label{legendrereci}
\end{eqnarray} 
As a consequence, the canonical conditioning of parameters $(p,q)$ 
can be considered as asymptotically equivalent for large $T$
to the microcanonical conditioning 
towards 
the two intensive local times $a^*_{p,q}= \partial_p E_{p,q}$ 
and $b^*_{p,q}= \partial_q E_{p,q} $ corresponding to the Legendre values of Eq. \ref{legendrereci}.

These two relations have a very simple interpretation via
the first-order perturbation theory for the energy $E_{p,q}$ of the ground state $\phi^{GS}_{p,q}(x)$
 in quantum mechanics when the parameter $p$ is changed into $(p+\epsilon)$
 or when the parameter $q$ is changed into $(q+\eta)$
\begin{eqnarray}
a^*_{p,q} && = \partial_p E_{p,q} = \lim_{\epsilon \to 0} \left(\frac{E_{p+\epsilon,q}-E_{p,q}}{\epsilon} \right)
= \langle \phi^{GS}_{p,q} \vert \delta(x) \vert \phi^{GS}_{p,q} \rangle = 
\left[\phi^{GS}_{p,q}(x=0) \right]^2 = P^*_{p,q}(x=0) 
\nonumber \\
b^*_{p,q} && = \partial_q E_{p,q} = \lim_{\eta \to 0} \left(\frac{E_{p,q+\eta}-E_{p,q}}{\eta} \right)
= \langle \phi^{GS}_{p,q} \vert \delta(x-L) \vert \phi^{GS}_{p,q} \rangle = 
\left[\phi^{GS}_{p,q}(x=L) \right]^2 = P^*_{p,q}(x=L) 
\label{quantumperturbationfirstorder}
\end{eqnarray}


\subsubsection{ Emergence of a normalizable ground-state for $H_{p,q}$ when $H_{p=0,q=0}$ has no ground state  }

\label{subsec_emergence}

When the quantum Hamiltonian $H_{p=0,q=0}$ has no bound state,
the Hamiltonian $H_{p,q}$ of Eq. \ref{hamiltonianpq} can nevertheless have a normalizable bound state. 
Indeed, for the Laplace transform $ {\hat {\tilde P}}_{s,p,q} (x \vert x_0)$ of Eq. \ref{laplacetriple},
the result of Eq. \ref{dysonpoly}
shows that a new singularity can appear in ${\hat {\tilde P}}_{s,p,q} (x \vert x_0)$
with respect to ${\hat G}_s (x \vert x_0) $ when the variable $s=-E_{p,q}$ makes 
Eq. \ref{epqfroms} vanish
\begin{eqnarray}
0&& = 1+ p  {\hat G}_s (0 \vert 0)   + q {\hat G}_s (L \vert L) +  p q 
\left[ {\hat G}_s (0 \vert 0)  {\hat G}_s (L \vert L)  -  {\hat G}_s (L \vert 0) {\hat G}_s (0 \vert L) \right]
\nonumber \\
&& = \left( 1+p {\hat G}_s (0 \vert 0)\right) \left(1+ q {\hat G}_s (L \vert L) \right)
 -pq  {\hat G}_s (0 \vert L){\hat G}_s (L \vert 0)
\label{spole}
\end{eqnarray}

Let us mention the two limiting cases for the distance $L$ between the two delta impurities :

(i) if $L=0$, Eq. \ref{spole} becomes
\begin{eqnarray}
0 = 1+ (p+q)  {\hat G}_s (0 \vert 0) 
\label{spolecoinciding}
\end{eqnarray}
and corresponds to the case of a single delta impurity of amplitude $(p+q)$ at the origin.
So when $H_{0,0}$ has no bound state,
a normalizable ground state state will emerge for $H_{p,q}$ 
if the global amplitude is strictly negative $(p+q)<0$.

(ii) if $L \to +\infty$, where it is expected that the vanishing limit ${\hat G}_s (0 \vert L){\hat G}_s (L \vert 0)  \to 0$,
 Eq. \ref{spole} becomes
\begin{eqnarray}
0  = \left( 1+p {\hat G}_s (0 \vert 0)\right) \left(1+ q {\hat G}_s (L \vert L) \right)
\label{spoleinfinity}
\end{eqnarray}
and corresponds to two independent delta impurities of amplitude $p$ and of amplitude $q$
that become separated by the distance $L \to +\infty$.
Therefore, when $H_{0,0}$ has no bound state, 
a normalizable ground state state will emerge for $H_{p,q}$
 if at least one of the two amplitudes $p$ or $q$
is strictly negative.
If the two amplitudes are strictly negative $p<0$ and $q<0$, there will be two bound states,
so the ground state will correspond to the lower energy.

Let us now discuss three special cases for the parameters $(p,q)$ :

(a) In the limit $q \to + \infty$ that amounts to impose the vanishing local time $B=0$ at position $x=L$, 
 Eq. \ref{spole} for for $s=-E_{p,q=+\infty}$
 becomes
\begin{eqnarray}
- p =  \frac{1}{  {\hat G}_s (0 \vert 0) -
 \frac{  {\hat G}_s (0 \vert L){\hat G}_s (L \vert 0) }{{\hat G}_s (L \vert L) }} 
 = \frac{1}{ {\hat G}^{abs(L)}_s (0 \vert 0) }
\label{spoleqinfty}
\end{eqnarray}
 in agreement with the pole of Eq. \ref{calA}.

(b) In the limit $p \to + \infty$ that amounts to impose the vanishing local time $A=0$ at position $x=L$, 
 Eq. \ref{spole} becomes for $s=-E_{p=+\infty,q}$
\begin{eqnarray}
- q =  \frac{1}{  {\hat G}_s (L \vert L) -
 \frac{  {\hat G}_s (L \vert 0){\hat G}_s (0 \vert L) }{{\hat G}_s (0 \vert 0) }} 
 = \frac{1}{ {\hat G}^{abs(0)}_s (L \vert L) }
\label{spolepinfty}
\end{eqnarray}
 in agreement with the pole of Eq. \ref{calB}.
 
 (c) In the special case of equal amplitudes $p=q$,  Eq. \ref{spole} becomes 
 \begin{eqnarray}
0 && = \left( 1+p {\hat G}_s (0 \vert 0)\right) \left(1+ p {\hat G}_s (L \vert L) \right)
 - p^2  {\hat G}_s (0 \vert L){\hat G}_s (L \vert 0)
 \nonumber \\
 && = 1+ p \left[ {\hat G}_s (0 \vert 0)+{\hat G}_s (L \vert L) \right] + p^2 \Delta_s
\label{spoleequal}
\end{eqnarray}
in agreement with Eq. \ref{dysonpolyupsroots} discussed during the analysis of the sum $\Sigma=A+B$.
The two roots $p^{\pm}=-\lambda_s^{\pm}$ correspond to two bound states,
while the energy $E_{p,p} =-s $ of the ground state corresponds to $p=-\lambda_s^{-}$
that governs the rate function $J(\sigma)$ as discussed in Eq. \ref{jsigmafromsaddle}
of the main text.

 
 \section{ Canonical conditioning of parameters $(p,q)$ for the case of uniform drift $\mu(x)=\mu $ }
 
 \label{sec_canonicalbrown}
 
 In this Appendix, the canonical conditioning described in the previous Appendix
 is applied to the case of uniform drift $\mu(x)=\mu$, 
in order to compare with the other conditioned processes described in section \ref{sec_brown} of the main text.
 
 The quantum Hamiltonian of Eq. \ref{hamiltonianpq} reduces to
\begin{eqnarray}
 H_{p,q} =  -  \frac{1}{2} \partial_x^2 + \frac{\mu^2}{2} + p \delta(x)+q \delta(x-L)
\label{hamiltonianpqfree}
\end{eqnarray}
The Hamiltonian $H_{p=0,q=0}$ has no bound state, 
but we will be interested in the regions of parameters $(p,q)$
where $H_{p,q}$ has a normalizable ground-state wave function $\phi^{GS}_{p,q}(x)$ 
(see subsection \ref{subsec_emergence})
in order to apply the framework described in subsection \ref{sec_canonicalGS}.

\subsection{ Analysis the energy $E_{p,q}$ of the normalizable ground state of $H_{p,q}$ }

Using ${\tilde G}_s (x \vert x_0)  $ of Eq. \ref{laplacefree}, Eq. \ref{spole} for $s=-E_{p,q}$ reads
\begin{eqnarray}
0&& = \left( 1+p {\hat G}_s (0 \vert 0)\right) \left(1+ q {\hat G}_s (L \vert L) \right)
 -pq  {\hat G}_s (0 \vert L){\hat G}_s (L \vert 0)
\nonumber \\
&& = \left( 1+\frac{p}{\kappa_s } \right) \left(1+\frac{q}{\kappa_s }  \right)
 -pq   \frac{ e^{ - 2\kappa_s L } } { \kappa_s^2 } 
\label{spolefree}
\end{eqnarray}
in terms of the parameter of Eq. \ref{kappas}
\begin{eqnarray}
\kappa_s \equiv  \sqrt{\mu^2+2 s } = \sqrt{\mu^2- 2 E_{p,q} }
\label{kappa}
\end{eqnarray}
that now parametrizes the ground-state energy 
\begin{eqnarray}
E_{p,q} = -s = \frac{\mu^2 -\kappa^2}{2}
\label{ekappa}
\end{eqnarray}
In summary, once the three parameters $(p,q,L)$ are given,
there will be a normalizable ground state for $H_{p,q}$
Eq. \ref{spolefree} that can be rewritten as
\begin{eqnarray}
0 = \left( 1+\frac{\kappa}{p}  \right) \left( 1+ \frac{\kappa}{q}  \right) -  e^{ - 2 \kappa L } 
\label{eqkappa}
\end{eqnarray}
that has a positive solution $\kappa_{p,q}>0$.


\subsection{ Analysis of the normalizable ground state wave function $\phi^{GS}_{p,q}(x)$ of $H_{p,q}$}

The eigenvalue equation for the ground state wave function $\phi^{GS}_{p,q}(x)$ of energy $E_{p,q}$
\begin{eqnarray}
 E_{p,q} \phi^{GS}_{p,q}(x) && = H_{p,q} \phi^{GS}_{p,q}(x)
 \nonumber \\
 && =   -  \frac{1}{2} \partial_x^2 \phi^{GS}_{p,q}(x)
 + \frac{\mu^2}{2}\phi^{GS}_{p,q}(x) + p \delta(x)\phi^{GS}_{p,q}(0)+q \delta(x-L)\phi^{GS}_{p,q}(L)
\label{hamiltonianpqeigenv}
\end{eqnarray}
can be rewritten using the parametrization $E_{p,q} = \frac{\mu^2 -\kappa_{p,q}^2}{2}$ of
Eq. \ref{ekappa} as
\begin{eqnarray}
0 =  -   \partial_x^2 \phi^{GS}_{p,q}(x)
 + \kappa_{p,q}^2\phi^{GS}_{p,q}(x) + 2 p \delta(x)\phi^{GS}_{p,q}(0)+2 q \delta(x-L)\phi^{GS}_{p,q}(L)
\label{hamiltonianpqeigen}
\end{eqnarray}
The continuous solution can be decomposed into the three following regions :

(i) for the middle region $x \in [0,L]$, the wave function can be written as the linear combination 
involving two constants $(K^+,K^-)$
\begin{eqnarray}
 \phi^{GS}_{p,q}(x) =K^+ e^{ \kappa_{p,q} x} +K^-  e^{ \kappa_{p,q} (L-x)} \ \ {\rm for } \ \ x \in [0,L]
\label{phiinter}
\end{eqnarray}

(ii) for the left region $x \in ]-\infty,0]$, the wave function should be normalizable at $x \to (-\infty)$ and 
should be continuous with Eq. \ref{phiinter} at $x=0$
\begin{eqnarray}
 \phi^{GS}_{p,q}(x) =\phi^{GS}_{p,q}(0) e^{ \kappa_{p,q} x} = (K^++K^- e^{\kappa_{p,q} L}) e^{ \kappa_{p,q} x} \ \ {\rm for } \ \ x \in ]-\infty,0]
\label{phiminus}
\end{eqnarray}

(iii) for the right region $x \in [L,+\infty[$, the wave function should be normalizable at $x \to (+\infty)$ and 
should be continuous with Eq. \ref{phiinter} at $x=L$
\begin{eqnarray}
 \phi^{GS}_{p,q}(x) =\phi^{GS}_{p,q}(L) e^{ -\kappa_{p,q} (x-L) } 
 = (K^+ e^{ \kappa_{p,q} L} +K^-  ) e^{ -\kappa_{p,q} (x-L) } \ \ {\rm for } \ \ x \in [L,+\infty[
\label{phiplus}
\end{eqnarray}

Now one needs to take into account the delta functions of Eq. \ref{hamiltonianpqeigen}
that impose the following discontinuities of the derivative of the wave function at $x=0$ and at $x=L$
 \begin{eqnarray}
2 p \phi^{GS}_{p,q}(0) 
&& = \!  \frac{d \phi^{GS}_{p,q}(x)}{dx}  \vert_{x=0^+} \!
- \! \frac{d \phi^{GS}_{p,q}(x)}{dx}  \vert_{x=0^-} \!
= \! \kappa_{p,q} (K^+  -K^- e^{\kappa_{p,q} L}  ) \!
- \! \kappa_{p,q} (K^++K^- e^{\kappa_{p,q} L})  \!
= \! -2 \kappa_{p,q} K^- e^{\kappa_{p,q} L} 
 \\
2 q \phi^{GS}_{p,q}(L) 
 && = \! \frac{d \phi^{GS}_{p,q}(x)}{dx}  \vert_{x=L^+} \!
- \! \frac{d \phi^{GS}_{p,q}(x)}{dx}  \vert_{x=L^-} \!
= \! -\kappa_{p,q} (K^+ e^{ \kappa_{p,q} L} +K^-  ) \!
- \! \kappa_{p,q} (K^+ e^{ \kappa_{p,q} L} -K^-  ) \!
= \! - 2 \kappa_{p,q} K^+ e^{ \kappa_{p,q} L} \nonumber
\label{matchingderi}
\end{eqnarray}
leading to the following homogeneous system for the two constants $(K^+,K^- )$
 \begin{eqnarray}
0 && =  p (K^++K^- e^{\kappa_{p,q} L})  +  \kappa_{p,q} K^- e^{\kappa_{p,q} L} = p K^+ + (p+\kappa_{p,q})e^{\kappa_{p,q} L} K^- 
\nonumber \\
0 && =  q (K^+ e^{ \kappa_{p,q} L} +K^-  ) +  \kappa_{p,q} K^+ e^{ \kappa_{p,q} L}
=  (q+\kappa_{p,q}) e^{ \kappa_{p,q} L} K^+ + q   K^- 
\label{matchingderisys}
\end{eqnarray}
In order to have a non-vanishing solution, the determinant of this system should vanish
 \begin{eqnarray}
0  =   
\begin{vmatrix} 
p &  (p+\kappa_{p,q})e^{\kappa_{p,q} L} \\
(q+\kappa_{p,q})e^{ \kappa_{p,q} L} & q  
 \end{vmatrix}  
 = pq  - (p+\kappa_{p,q})(q+\kappa_{p,q}) e^{2\kappa_{p,q} L}
\label{detvanish}
\end{eqnarray}
i.e. one recovers Eq. \ref{eqkappa} that determines the energy $E_{p,q}$, as it should for consistency.
Therefore, Eq. \ref{matchingderisys} allows to compute the ratio 
 \begin{eqnarray}
 \frac{K^-}{K^+}  = - \frac{e^{-\kappa_{p,q} L}}{1+\frac{\kappa_{p,q}}{p}}   = - \left( 1+\frac{\kappa_{p,q}}{q}\right) e^{\kappa_{p,q} L }
\label{K2fromK1}
\end{eqnarray}
Finally, the global normalization of the wave function $\phi^{GS}_{p,q}(x) $
\begin{eqnarray}
 1  = \int_{-\infty}^{+\infty} dx \left[  \phi^{GS}_{p,q}(x)\right]^2
\label{normawavefunction}
\end{eqnarray}
can be computed, but is not needed to evaluate the conditioned drift below.


\subsection{ Analysis of the conditioned drift $  \mu_{p,q}^*( x )$}

The conditioned drift of Eq. \ref{mustarpinterior} 
can be computed from the wave function $\phi^{GS}_{p,q}(x) $ 
given by Eqs \ref{phiinter}, 
\ref{phiminus}
and \ref{phiplus}
in the three regions
\begin{eqnarray}
 \mu_{p,q}^*( x ) && = \partial_x \ln \left[  \phi_{p,q}^{GS}(  x)\right]   
  = 
 \left\lbrace
  \begin{array}{lll}
    \kappa_{p,q}  
    &~~\mathrm{for~~} x \in ]-\infty,0[
    \\
  \kappa_{p,q}  \frac{K^+ e^{ \kappa_{p,q} x} -K^- e^{- \kappa_{p,q} (x-L)}}
  {K^+ e^{ \kappa_{p,q} x} +K^-  e^{- \kappa_{p,q} (x-L)}}  
    &~~\mathrm{for~~}   x \in ]0,L[
     \\
    -\kappa_{p,q} 
    &~~\mathrm{for~~} x \in]L,+\infty[
  \end{array}
\right.
\label{mustarpqfree}
\end{eqnarray}
that should be compared to Eqs 
\ref{mustarbridgepiintensivefinalsimplifleftfree},
\ref{mustarbridgepiintensivefinalsimplifrightfree} and 
  \ref{mustarbridgepiintensivefinalsimplifparametricfreemiddle}
  of the main text.

 
\subsubsection{ Special case $q \to + \infty$ for $p \in ]-\infty,0[$ corresponding to corresponding to the conditioning
towards $(a_*>0,b_*=0)$ }

In the limit $q \to + \infty$ for $p \in ]-\infty,0[$, the ground-state $\phi^{GS}_{p,q=\infty}(x) $ 
should vanish at $x=L$ and thus in the whole region $x \geq L$ as a consequence of Eq. \ref{phiplus}, 
while the ratio of Eq. \ref{K2fromK1}
reduces to
 \begin{eqnarray}
 \frac{K^-}{K^+}    \opsimeq_{q \to + \infty} - e^{\kappa_{p,\infty} L }
\label{K2fromK1qinfty}
\end{eqnarray}
where $\kappa_{p,\infty} >0  $ is the solution of Eq. \ref{eqkappa}
\begin{eqnarray}
0 = \left( 1+\frac{\kappa_{p,\infty}}{p}  \right)  -  e^{ - 2 \kappa_{p,\infty} L } 
\label{eqkappaqinfty}
\end{eqnarray}

The conditioned drift of Eq. \ref{mustarpqfree} reduces to
\begin{eqnarray}
 \mu_{p,q=+\infty}^*( x ) 
  = 
 \left\lbrace
  \begin{array}{lll}
    \kappa_{p,\infty}  
    &~~\mathrm{for~~} x \in ]-\infty,0[
    \\
  \kappa_{p,\infty}  \frac{ e^{ \kappa_{p,\infty} x} + e^{- \kappa_{p,\infty} (x-2L)}}
  { e^{ \kappa_{p,\infty} x} -  e^{- \kappa_{p,\infty} (x-2L)}}  
   =  - \kappa_{p,\infty} \frac{ \cosh[  \kappa_{p,\infty} (L-x) ] } {  \sinh[  \kappa_{p,\infty} (L-x) ] }
    &~~\mathrm{for~~}   x \in ]0,L[
  \end{array}
\right.
\label{mustarpqinftyfreeqinfty}
\end{eqnarray}
that should be compared to Eq. \ref{conditioneddriftastaraloneparametricfreekappa2regions}
of the main text.


\subsubsection{ Special case $p \to + \infty$ for $q \in ]-\infty,0[$ corresponding to the conditioning
towards $(a_*=0,b_*>0)$ }
 
In the limit $p \to + \infty$ for $q \in ]-\infty,0[$, the ground-state $\phi^{GS}_{p=\infty,q}(x) $ 
should vanish at $x=0$ and thus in the whole region $x \leq 0$ as a consequence of Eq. \ref{phiminus}, 
while the ratio of Eq. \ref{K2fromK1}
reduces to
 \begin{eqnarray}
 \frac{K^-}{K^+}    \opsimeq_{p \to + \infty} - e^{-\kappa_{\infty,q} L }
\label{K2fromK1pinfty}
\end{eqnarray}
where $\kappa_{\infty,q} >0  $ is the solution of Eq. \ref{eqkappa}
\begin{eqnarray}
0 =  \left( 1+ \frac{\kappa_{\infty,q}}{q}  \right) -  e^{ - 2 \kappa_{\infty,q} L } 
\label{eqkappapinfy}
\end{eqnarray}

The conditioned drift of Eq. \ref{mustarpqfree} reduces to
\begin{eqnarray}
 \mu_{p=+\infty,q}^*( x ) 
  = 
  \left\lbrace
  \begin{array}{lll}
  \kappa_{\infty,q}  \frac{ e^{ \kappa_{\infty,q} x} + e^{- \kappa_{\infty,q} x}}
  { e^{ \kappa_{\infty,q} x} -  e^{- \kappa_{\infty,q} x}}  
  =  \kappa_{\infty,q}\frac{ \cosh(  \kappa_{\infty,q} x)  }  { \sinh(\kappa_{\infty,q} x )  }
    &~~\mathrm{for~~}   x \in ]0,L[
     \\
    -\kappa_{\infty,q} 
    &~~\mathrm{for~~} x \in]L,+\infty[
  \end{array}
\right.
\label{mustarpqinftyfreepinfty}
\end{eqnarray}
that should be compared to Eq. \ref{conditioneddriftastaraloneparametricfreekappa2regions}
of the main text.


\subsubsection{ Special case $p=q \in ]-\infty,0[$ corresponding to the conditioning towards $\sigma_*>0$ }
 
In the case of equal negative amplitude $p=q \in ]-\infty,0[$,
Eq. \ref{eqkappa}
\begin{eqnarray}
0 = \left( 1+\frac{\kappa}{p}  \right)^2 -  e^{ - 2 \kappa L } 
\label{eqkappapqequal}
\end{eqnarray}
has two solutions 
\begin{eqnarray}
 \left( 1+\frac{\kappa}{p}  \right) =  \pm   e^{ - 2 \kappa L } 
\label{eqkappapqequaltwosol}
\end{eqnarray}
corresponding to two possible bound states,
while the corresponding ratio of Eq. \ref{K2fromK1}
takes the two possible values
 \begin{eqnarray}
 \frac{K^-}{K^+}  = - \frac{e^{-\kappa L}}{1+\frac{\kappa}{p}}   = \mp 1
\label{K2fromK1pqequal}
\end{eqnarray}
The positive ground-state wave function $\phi_{p,p}(x)$ corresponds to the ratio $  \frac{K^-}{K^+}=+1$
with $\kappa_{p,p} $ solution of
\begin{eqnarray}
 \left( 1+\frac{\kappa_{p,p}}{p}  \right) =  -  e^{ -  \kappa_{p,p} L } 
\label{eqkappapqequalgs}
\end{eqnarray}
corresponding to
\begin{eqnarray}
p =  - \frac{\kappa_{p,p}} {1+  e^{ -  \kappa_{p,p} L } } = - \lambda^-_s
\label{eqkappapqequalgslambdamoins}
\end{eqnarray}
where one recognizes the form $\lambda^-_s$ of Eq. \ref{dysonpolyupsrootsfree}.
The other bound state associated to the negative ratio $  \frac{K^-}{K^+}=-1$
corresponds to the first excited state characterized by
\begin{eqnarray}
p =  - \frac{\kappa_{p,p}} {1-  e^{ -  \kappa_{p,p} L } } = - \lambda^+_s
\label{eqkappapqequalgsexcited}
\end{eqnarray}
where one recognizes the form of $\lambda^+_s$ of Eq. \ref{dysonpolyupsrootsfree}.

Plugging the ratio $  \frac{K^-}{K^+}=+1$ for the ground state 
into the conditioned drift of Eq. \ref{mustarpqfree} yields
 \begin{eqnarray}
 \mu_{p,p}^*( x ) 
  = 
 \left\lbrace
  \begin{array}{lll}
    \kappa_{p,p}  
    &~~\mathrm{for~~} x \in ]-\infty,0[
    \\
  \kappa_{p,p}  \frac{ e^{ \kappa_{p,p} x} - e^{- \kappa_{p,p} (x-L)}}
  { e^{ \kappa_{p,p} x} +  e^{- \kappa_{p,p} (x-L)}}  
  = \kappa_{p,p} \frac{ \sinh \left[  \kappa_{p,p} \left(x-\frac{L}{2}\right) \right] } 
{\cosh \left[  \kappa_{p,p} \left(x-\frac{L}{2}\right) \right]   }
    &~~\mathrm{for~~}   x \in ]0,L[
     \\
    -\kappa_{p,p} 
    &~~\mathrm{for~~} x \in]L,+\infty[
  \end{array}
\right.
\label{mustarpqfreeequal}
\end{eqnarray}
that should be compared to Eq. \ref{mustarbridgepiintensivesigmaresexpliparametricfree3regions}
of the main text.


\section{ Explicit Laplace inversion with respect to $p$ and $q$ of the contribution 
${\hat {\tilde {\cal C}}}_{s,p,q} (x \vert x_0)  $ of Eq. \ref{calCdef}   }

\label{app_laplace}

In this Appendix, the goal is to compute the Laplace inversion with respect to $p$ and $q$ of the contribution 
$ {\hat {\tilde {\cal C}}}_{s,p,q} (x \vert x_0)   $ of Eq. \ref{calCdef}.

\subsection{ Partial fraction decomposition of ${\hat {\tilde {\cal C}}}_{s,p,q} (x \vert x_0) $   } 

To see more clearly the structure of ${\hat {\tilde {\cal C}}}_{s,p,q} (x \vert x_0)  $ of Eq. \ref{calCdef}
with respect to $(p,q)$, it is convenient to introduce the shifted variables
 \begin{eqnarray}
P && \equiv p + \frac{{\hat G}_s (L \vert L)}{\Delta_s} = p + \frac{1}{ {\hat G}^{abs(L)}_s (0 \vert 0) }
\nonumber \\
Q && \equiv q + \frac{{\hat G}_s (0 \vert 0)}{\Delta_s} = q + \frac{1}{ {\hat G}^{abs(0)}_s (L \vert L) }
\label{bigPQ}
\end{eqnarray}
and the notation
 \begin{eqnarray}
c_s  \equiv  \frac{ \sqrt{ {\hat G}_s (0 \vert L){\hat G}_s (L \vert 0) }}{\Delta_s} 
= \    \frac{ \sqrt{ {\hat G}_s (0 \vert L){\hat G}_s (L \vert 0)}}
{ {\hat G}_s (0 \vert 0)  {\hat G}_s (L \vert L)  -  {\hat G}_s (L \vert 0) {\hat G}_s (0 \vert L) } 
\label{const}
\end{eqnarray}
in order to rewrite Eq. \ref{calCdef}
as 
 \begin{eqnarray}
 {\hat {\tilde {\cal C}}}_{s,p,q} (x \vert x_0) && =
    \frac{  \gamma_s(x \vert x_0) + P  \beta_s^{[LL]}(x \vert x_0) +Q  \alpha_s^{[00]}(x \vert x_0) } {PQ-c_s^2 } 
-  \frac{     \alpha_s^{[00]}(x \vert x_0) }{ P   } 
-   \frac{      \beta_s^{[LL]}(x \vert x_0)      }{  Q   } 
\nonumber \\
&& =     \frac{ \gamma_s(x \vert x_0)     } { PQ - c_s^2 } 
+ \alpha_s^{[00]}(x \vert x_0) 
\left[  \frac{  Q     }{ PQ - c_s^2 } -\frac{1}{P} \right] 
+ \beta_s^{[LL]}(x \vert x_0) 
\left[ \frac{ P      } { PQ - c_s^2 } - \frac{1}{Q} \right] 
\label{calCbig}
\end{eqnarray}
The amplitude of the first fraction 
  \begin{eqnarray}
        \gamma_s(x \vert x_0) 
        && =       
         \frac{{\hat G}_s (L \vert L) {\hat G}_s (x \vert 0) {\hat G}_s (0 \vert x_0)
  + {\hat G}_s (0 \vert 0){\hat G}_s (x \vert L) {\hat G}_s (L \vert x_0)
  - \left[ {\hat G}_s (0 \vert 0){\hat G}_s (L \vert L) +{\hat G}_s (0 \vert L){\hat G}_s (L \vert 0) 
         \right]  \Omega_s (x \vert x_0)
  }{\Delta^2 _s } 
   \nonumber \\
   && =    \gamma_s^{[L0]}(x \vert x_0) +     \gamma_s^{[0L]}(x \vert x_0) 
\label{gamma2calcul}
\end{eqnarray}
can be decomposed into the two terms $ \gamma_s^{[L0]}(x \vert x_0) $ and $\gamma_s^{[0L]}(x \vert x_0) $ given in Eq. \ref{gamma}
of the main text.


\subsection{ Laplace inversion of $ {\hat {\tilde {\cal C}}}_{s,p,q} (x \vert x_0)  $ with respect to $p$ and $q$} \

The Laplace inversion of the three terms of Eq. \ref{calCbig}
involves the modified Bessel function $I_0(z)$ of Eq. \ref{I0serie} and 
 its derivative $I_0'(z) =  I_1 (z) $ of Eq. \ref{I1serie}.
 The Laplace inversion of the first term of Eq. \ref{calCbig}
is based on the identity (already used in Eq. 39 of \cite{grebenkov_2local})
 \begin{eqnarray}
 \frac{1}{PQ - c^2 } = \int_0^{+\infty} dx e^{-Px}  \int_0^{+\infty} dy e^{-Qy}  I_0 \left( 2 c \sqrt{xy} \right) 
\label{bessel}
\end{eqnarray}
that can be checked by plugging the series representation of $I_0(z)$ of Eq. \ref{I0serie}
 into the right hand-side of Eq. \ref{bessel}
 \begin{eqnarray}
&&  \int_0^{+\infty} dx e^{-Px}  \int_0^{+\infty} dy e^{-Qy}  I_0 \left( 2c \sqrt{xy} \right) 
 =  \int_0^{+\infty} dx e^{-Px}  \int_0^{+\infty} dy e^{-Qy} 
 \sum_{k=0}^{+\infty} \frac{\left(  c \sqrt{xy} \right)^{2k}}{( k!)^2}
 \nonumber \\
 && = \sum_{k=0}^{+\infty} \frac{c^{2k}}{( k!)^2}  
 \int_0^{+\infty} dx x^{k} e^{-Px}  \int_0^{+\infty} dy y^{k} e^{-Qy}  
 = \sum_{k=0}^{+\infty} \frac{ c^{2k}}{( k!)^2}  
  \frac{ (k!)^2}{ (PQ)^{k+1}}
  \nonumber \\
 && = \frac{1}{PQ} \sum_{k=0}^{+\infty}\left( \frac{c^2 }{ PQ} \right)^k =  \frac{1}{PQ} \left( \frac{1}{1-\frac{c^2 }{ PQ} } \right) =  \frac{1}{PQ - c^2 }
 \label{laplacebesselserie}
\end{eqnarray}

The multiplication of Eq. \ref{bessel} by $P$ yields via integration by parts
 \begin{eqnarray}
 \frac{P}{PQ - c^2 } && =
 \int_0^{+\infty} dx (P e^{-Px})  \int_0^{+\infty} dy e^{-Qy}  I_0 \left( 2 c \sqrt{xy} \right) 
 = \int_0^{+\infty} dy e^{-Qy}
 \int_0^{+\infty} dx (- \partial_x e^{-Px} )   I_0 \left( 2c \sqrt{xy} \right) 
 \nonumber \\
 && = - \int_0^{+\infty} dy e^{-Qy}
\left[ e^{-Px}    I_0 \left( 2c \sqrt{xy} \right) \right]_{x=0}^{x=+\infty} 
+  \int_0^{+\infty} dy e^{-Qy}
 \int_0^{+\infty} dx   e^{-Px}  \left(  \partial_x   I_0 \left( 2c \sqrt{xy} \right) \right)
  \nonumber \\
 && =  \int_0^{+\infty} dy e^{-Qy}
+  \int_0^{+\infty} dy e^{-Qy}
 \int_0^{+\infty} dx   e^{-Px}  \frac{ c \sqrt{y}}{ \sqrt{x} }   I_0' \left( 2c \sqrt{xy} \right) 
   \nonumber \\
 && =  \frac{1}{Q}
+  \int_0^{+\infty} dy e^{-Qy}
 \int_0^{+\infty} dx   e^{-Px}  \frac{ c \sqrt{y}}{ \sqrt{x} }   I_1 \left( 2c \sqrt{xy} \right) 
 \label{besseldericalcul}
\end{eqnarray}
so one obtains the identity (already used in Eq. 47 of \cite{grebenkov_2local})
 \begin{eqnarray}
 \frac{P}{PQ - c^2 }   -  \frac{1}{Q}
= \int_0^{+\infty} dy e^{-Qy}
 \int_0^{+\infty} dx   e^{-Px}  \frac{ c \sqrt{y}}{ \sqrt{x} }   I_1 \left( 2c \sqrt{xy} \right) 
 \label{besselderi}
\end{eqnarray}
and its analog (already used in Eq. 48 of \cite{grebenkov_2local})
 \begin{eqnarray}
 \frac{Q}{PQ - c^2 }   -  \frac{1}{P}
=  \int_0^{+\infty} dy e^{-Qy}
 \int_0^{+\infty} dx   e^{-Px}  \frac{ c \sqrt{x}}{ \sqrt{y} }   I_1 \left( 2c \sqrt{xy} \right) 
 \label{besselderi2}
\end{eqnarray}
that will allow to write the Laplace inversions of the second and third terms of Eq. \ref{calCbig}.

If one replaces the shifted variables $(P,Q)$ of Eq. \ref{bigPQ} in terms of the original Laplace variables $(p,q)$,
the three terms of the fourth contribution $   {\cal C}_{s,p,q} (x \vert x_0)   $ of Eq. \ref{calCbig}
\begin{eqnarray}
  {\hat {\tilde {\cal C}}}_{s,p,q} (x \vert x_0) && =   {\hat {\tilde {\cal C}}}^{[\alpha]}_{s,p,q} (x \vert x_0) 
  +   {\hat {\tilde {\cal C}}}^{[\beta]}_{s,p,q} (x \vert x_0) 
 + {\hat {\tilde {\cal C}}}^{[\gamma]}_{s,p,q} (x \vert x_0) 
\label{calCthree}
\end{eqnarray}
read
\begin{eqnarray}
  {\hat {\tilde {\cal C}}}^{[\alpha]}_{s,p,q} (x \vert x_0)    && 
 = \alpha_s^{[00]}(x \vert x_0)  \left[  \frac{  Q     }{ PQ - c_s^2 } -\frac{1}{P} \right] 
\nonumber \\
&&  =\alpha_s^{[00]}(x \vert x_0)
  \left[  \frac{  q + \frac{1}{ {\hat G}^{abs(0)}_s (L \vert L) }     }
  { \left( p + \frac{1}{ {\hat G}^{abs(L)}_s (0 \vert 0) }\right) \left( q + \frac{1}{ {\hat G}^{abs(0)}_s (L \vert L) }\right) 
   -  c_s^2  } 
   -\frac{1}{p + \frac{1}{ {\hat G}^{abs(L)}_s (0 \vert 0) }} \right] 
  \nonumber \\
 {\hat {\tilde {\cal C}}}^{[\beta]}_{s,p,q} (x \vert x_0) 
 && = \beta_s^{[LL]}(x \vert x_0) 
\left[ \frac{ P      } { PQ -  c_s^2 } - \frac{1}{Q} \right]
\nonumber \\
&& = 
\beta_s^{[LL]}(x \vert x_0) 
\left[ \frac{ p + \frac{1}{ {\hat G}^{abs(L)}_s (0 \vert 0) }      }
 {  \left( p + \frac{1}{ {\hat G}^{abs(L)}_s (0 \vert 0) }\right) \left( q + \frac{1}{ {\hat G}^{abs(0)}_s (L \vert L) }\right)
 - c_s^2  } 
 - \frac{1}{q + \frac{1}{ {\hat G}^{abs(0)}_s (L \vert L) }} \right]
\nonumber \\
 {\hat {\tilde {\cal C}}}^{[\gamma]}_{s,p,q} (x \vert x_0)  && = 
     \frac{ \gamma_s(x \vert x_0)      } { PQ -  c_s^2  } 
   =    \frac{ \gamma_s(x \vert x_0)  } 
  {  \left[ p + \frac{1}{ {\hat G}^{abs(L)}_s (0 \vert 0) }\right] \left[ q + \frac{1}{ {\hat G}^{abs(0)}_s (L \vert L) }\right] 
  -  c_s^2 }   
\label{calC}
\end{eqnarray}

So using Eqs \ref{bessel}, \ref{besselderi} and \ref{besselderi2},
the Laplace inversion with respect to $p$ and $q$ of Eq. \ref{calCthree}
leads to Eqs \ref{laplaceCabcinvthree} \ref{laplaceCabcinv} given in the main text.

\subsection{ Rewriting $\Omega_s (x \vert x_0)  $ 
 in terms of $ \alpha_s^{[00]}(x \vert x_0) $, $\beta_s^{[LL]}(x \vert x_0) $ and $ \gamma_s(x \vert x_0) $  }
 
 For another computation of the main text around Eq. \ref{dysonpolyupsfracueq}, it is useful to rewrite $\Omega_s (x \vert x_0) $
 in terms of the three functions $ \alpha_s^{[00]}(x \vert x_0) $, $\beta_s^{[LL]}(x \vert x_0) $ and $ \gamma_s(x \vert x_0) $ by considering the special case $p=0=q$ of Eq. \ref{calCdef}
  \begin{eqnarray}
 {\hat {\tilde {\cal C}}}_{s,p=0,q=0} (x \vert x_0)  
  =  \Omega_s (x \vert x_0)
-    \alpha_s^{[00]}(x \vert x_0)  {\hat G}^{abs(L)}_s (0 \vert 0)     
-       \beta_s^{[LL]}(x \vert x_0)   {\hat G}^{abs(0)}_s (L \vert L) 
\label{calCdef00}
\end{eqnarray}
 and of Eq. \ref{calCthree}
  \begin{eqnarray}
   {\hat {\tilde {\cal C}}}_{s,p=0,q=0} (x \vert x_0) 
   && =   {\hat {\tilde {\cal C}}}^{[\alpha]}_{s,p=0,q=0} (x \vert x_0) 
  +   {\hat {\tilde {\cal C}}}^{[\beta]}_{s,p=0,q=0} (x \vert x_0) 
 + {\hat {\tilde {\cal C}}}^{[\gamma]}_{s,p=0,q=0} (x \vert x_0) 
\label{calC00three}
\end{eqnarray}
 with Eq. \ref{calC}
 \begin{eqnarray}
  {\hat {\tilde {\cal C}}}^{[\alpha]}_{s,p=0,q=0} (x \vert x_0) 
  &&   =\alpha_s^{[00]}(x \vert x_0)
  \left[  \frac{   \frac{1}{ {\hat G}^{abs(0)}_s (L \vert L) }     }
  {  \frac{1}{ G^{abs(L)}_s (0 \vert 0) {\hat G}^{abs(0)}_s (L \vert L) }  -  c_s^2  } 
   - {\hat G}^{abs(L)}_s (0 \vert 0)  \right] 
   = \alpha_s^{[00]}(x \vert x_0)
  \left[    \frac{\Delta_s}{ {\hat G}^{abs(0)}_s (L \vert L) }  
   - {\hat G}^{abs(L)}_s (0 \vert 0)  \right] 
  \nonumber \\
 {\hat {\tilde {\cal C}}}^{[\beta]}_{s,p=0,q=0} (x \vert x_0) 
&& = 
\beta_s^{[LL]}(x \vert x_0) 
\left[ \frac{  \frac{1}{ {\hat G}^{abs(L)}_s (0 \vert 0) }      }
 {   \frac{1}{ G^{abs(L)}_s (0 \vert 0) {\hat G}^{abs(0)}_s (L \vert L) }  -  c_s^2  } 
 -  {\hat G}^{abs(0)}_s (L \vert L)  \right]
 = \beta_s^{[LL]}(x \vert x_0) 
\left[   \frac{\Delta_s}{ {\hat G}^{abs(L)}_s (0 \vert 0) }   
 -  {\hat G}^{abs(0)}_s (L \vert L)  \right]
\nonumber \\
 {\hat {\tilde {\cal C}}}^{[\gamma]}_{s,p=0,q=0} (x \vert x_0)  &&
   =    \frac{ \gamma_s(x \vert x_0)  } 
  {   \frac{1}{ G^{abs(L)}_s (0 \vert 0) {\hat G}^{abs(0)}_s (L \vert L) }  -  c_s^2 }   
  =  \gamma_s(x \vert x_0)   \Delta_s  
\label{calC00}
\end{eqnarray}
The comparison between Eqs \ref{calCdef00}
and \ref{calC00three} \ref{calC00}
allows to rewrite
  \begin{eqnarray}
  \Omega_s (x \vert x_0)
&& =  
 \alpha_s^{[00]}(x \vert x_0) \frac{  \Delta_s }{ {\hat G}^{abs(0)}_s (L \vert L) }   
   +  \beta_s^{[LL]}(x \vert x_0)   \frac{  \Delta_s }{ {\hat G}^{abs(L)}_s (0 \vert 0) }  
 +   \gamma_s(x \vert x_0)   \Delta_s  
 \nonumber \\
 && = \alpha_s^{[00]}(x \vert x_0)  G_s (0 \vert 0)    
   +  \beta_s^{[LL]}(x \vert x_0)    G_s (L \vert L)   
 +  \gamma_s(x \vert x_0)   \Delta_s  
\label{omegaabc}
\end{eqnarray}

 
 \section{Calculation of the conditioned drift $\mu^{[\Sigma_{\infty}^*]}_{\infty}( x,\Sigma=\Sigma_{\infty}^*)$  when $S^{abs(0,L)}_{\tau} (  x)$ vanishes for large time $\tau$}

\label{app_taboo} 

In this appendix, we provide technical details for the asymptotic analysis of the long-time of $S^{abs(0,L)}_{\tau} (  x) $ when it vanishes. As described in the main text, when $S^{abs(0,L)}_{\infty} (  x) $ vanishes in Eq. \ref{survivallimitfromlaplacefree},
one should use the asymptotic behavior of $S^{abs(0,L)}_{\tau} (  x) $ for large time $\tau$ to obtain 
the conditioned drift $\mu^{[\Sigma_{\infty}^*]}_{\infty}( x,\Sigma=\Sigma_{\infty}^*)$ of Eq. \ref{mustarbridgepitinftyregions2}. Recall that Eq. \ref{mustarbridgepitinftyregions2}
\begin{eqnarray}
\mu^{[\Sigma_{\infty}^*]}_{\infty}( x,\Sigma=\Sigma_{\infty}^*) = \mu(x) +  \lim_{\tau \to +\infty} \partial_x    \ln S^{abs(0,L)}_{\tau} (  x)
\label{app_mustarbridgepitinftyregions2} 
 \end{eqnarray}
where the Laplace transform of $S^{abs(0,L)}_{\tau} (  x)$ is given by Eq. \ref{laplacesurvival}
 \begin{eqnarray}
 {\hat S}^{abs(0,L)}_s (x) =  \frac{1}{s} \left[ 1
  - \frac{ {\hat G}^{abs(L)}_s (0 \vert x)}{{\hat G}^{abs(L)}_s (0 \vert 0)} 
  - \frac{{\hat G}^{abs(0)}_s (L \vert x)}{{\hat G}^{abs(0)}_s (L \vert L)}
  \right]
\label{app_laplacesurvival}
\end{eqnarray}
where the expressions of ${\hat G}^{abs(L)}_s (x \vert x_0)$ and ${\hat G}^{abs(0)}_s (x \vert x_0)$ are given by Eq. \ref{Gabsbonlyfree} and  Eq. \ref{Gabsaonlyfree} respectively. There are two cases depending on whether $x \in ]-\infty,0[$ or $x \in ]0,L[$.

\subsection{ Case $x \in ]-\infty,0[$} 
When $x<0$ the expression of Eq. \ref{app_laplacesurvival} reduces to

\begin{eqnarray}
 {\hat S}^{abs(0,L)}_s (x) =  \frac{1}{s} \left[ 1 - e^{x(\sqrt{2s + \mu^2} -\mu)} \right]
\label{app_laplacesurvival_case_x_neg}
\end{eqnarray}
From the inverse Laplace transform given in \cite{ogata} (Eq.21a)
\begin{eqnarray}
    \mathcal{L}^{-1}\left\{\frac{1}{s}e^{\frac{u x}{2 D}-x\sqrt{\frac{u^2}{4 D^2}+\frac{s}{D} }} \right\} = \frac{1}{2} \erfc \left(\frac{x - u \tau}{2 \sqrt{D \tau}} \right) + \frac{1}{2} e^{\frac{u x}{D}} \erfc \left(\frac{x + u \tau}{2 \sqrt{D \tau}} \right)
\end{eqnarray}
where $\erfc(x)$ is the complementary error function, we immediately obtain
\begin{eqnarray}
    {S}^{abs(0,L)}_{\tau} (x) = \frac{1}{2} \left[e^{-2 x \mu} \left(-2 + \erfc \left(\frac{x - \mu \tau}{\sqrt{2 \tau}} \right) \right) + \erfc \left(\frac{x + \mu \tau}{\sqrt{2 \tau}} \right) \right]
\end{eqnarray}
The asymptotic behavior at leading order for large time $\tau$ is given by
\begin{eqnarray}
   \displaystyle {S}^{abs(0,L)}_{\tau} (x) \opsimeq_{\tau \to +\infty} -\frac{1}{\mu^2 \tau^{3/2}} \sqrt{\frac{2}{\pi}} x e^{-\frac{(x+ \mu \tau)^2}{2 \tau}}
\end{eqnarray}
and leads to
\begin{eqnarray}
\mu^{[\Sigma_{\infty}^*]}_{\infty}( x,\Sigma=\Sigma_{\infty}^*) = \mu +  \lim_{\tau \to +\infty} \partial_x    \ln S^{abs(0,L)}_{\tau} (  x) = \frac{1}{x} - \frac{x}{\tau} = \frac{1}{x}
\label{app_mustarbridgepitinftyregions2final} 
 \end{eqnarray}
as stated in Eq. \ref{mustarbridgepitinftyregions2taboo}.

\subsection{ Case $x \in ]0,L[$} 
When $x \in ]0,L[$ the expression of Eq. \ref{app_laplacesurvival} reads
\begin{eqnarray}
 {\hat S}^{abs(0,L)}_s (x) =  \frac{1}{s} \left[ 1 + \frac{e^{-\sqrt{2 s + \mu^2}(L+x)+ \mu(L-x)}(1- e^{2 \sqrt{2 s + \mu^2} x} ) - e^{-x (\sqrt{2 s + \mu^2} +\mu)} (1- e^{2 \sqrt{2 s + \mu^2} (x-L)} )}{1-e^{-2 \sqrt{2 s + \mu^2} L} }  \right]
\label{app_laplacesurvival_case_x_in_0_L}
\end{eqnarray}
or in a more symmetrical form
\begin{eqnarray}
 {\hat S}^{abs(0,L)}_s (x) =  \frac{1}{s} \left[ 1 - e^{\mu (L-x)} \frac{\sinh(\sqrt{2 s + \mu^2} x )}{\sinh(\sqrt{2 s + \mu^2} L )} + e^{- \mu x} \frac{\sinh(\sqrt{2 s + \mu^2} (L-x))}{\sinh(\sqrt{2 s + \mu^2} L )}  \right]
\label{app_laplacesurvival_case_x_in_0_L_sym}
\end{eqnarray}
The inverse Laplace transform of $\sinh(\sqrt{2 s + \mu^2} x)/(s \sinh(\sqrt{2 s + \mu^2} L ))$ can be computed exactly thanks to the residue theorem. First, there is a simple pole at $s=0$ and the residue at this point is
\begin{eqnarray}
    \oplim_{s \to 0} \frac{s e^{s \tau} \sinh(\sqrt{2 s + \mu^2} x)}{s \sinh(\sqrt{2 s + \mu^2} L )} =  \frac{\sinh(\mu x)}{\sinh(\mu L)}
\end{eqnarray}
In addition, the denominator vanishes when $\sinh(\sqrt{2 s + \mu^2} L ) =0$, i.e. when $L \sqrt{2s + \mu^2} = n \pi i$, $n=0, \pm 1, \pm2, ...$ thus for
\begin{eqnarray}
   s = a_n = -\frac{\mu^2}{2} - \frac{n^2 \pi^2}{2 L^2} ~~~~ n=0,1,2, ...
\end{eqnarray}
The residue at the point $s=a_n$ is
\begin{eqnarray}
    \oplim_{s \to a_n} \frac{(s-a_n) e^{s \tau} \sinh(\sqrt{2 s + \mu^2} x)}{s \sinh(\sqrt{2 s + \mu^2} L )} && =  \oplim_{s \to a_n} \frac{(s-a_n)}{\sinh(\sqrt{2 s + \mu^2} L )}  \oplim_{s \to a_n}   \frac{e^{s \tau} \sinh(\sqrt{2 s + \mu^2} x)}{s} \\ \nonumber 
            && = \frac{n \pi i (-1)^n}{L^2} e^{( -\frac{\mu^2}{2} - \frac{n^2 \pi^2}{2 L^2})\tau} \frac{i \sin \left(\frac{n \pi x}{L}\right)}{ -\frac{\mu^2}{2} - \frac{n^2 \pi^2}{2 L^2}}   \\ \nonumber 
            && = \frac{n \pi (-1)^n}{L^2 \left( \frac{\mu^2}{2} + \frac{n^2 \pi^2}{2 L^2} \right)}e^{-( \frac{\mu^2}{2} + \frac{n^2 \pi^2}{2 L^2})\tau} \sin \left(\frac{n \pi x}{L}\right)
\end{eqnarray}
Adding all the residues, we obtain
\begin{eqnarray}
    \mathcal{L}^{-1}\left\{\frac{\sinh(\sqrt{2 s + \mu^2} x)}{s \sinh(\sqrt{2 s + \mu^2} L )} \right\} = \frac{\sinh(\mu x)}{\sinh(\mu L)} + \sum_{n=1}^{\infty} \frac{n \pi (-1)^n}{L^2 \left( \frac{\mu^2}{2} + \frac{n^2 \pi^2}{2 L^2} \right)}e^{-( \frac{\mu^2}{2} + \frac{n^2 \pi^2}{2 L^2})\tau} \sin \left(\frac{n \pi x}{L}\right)
\end{eqnarray}
and then
\begin{eqnarray}
  {S}^{abs(0,L)}_{\tau} (x) && = \mathcal{L}^{-1}\left\{  \frac{1}{s} \left[ 1 - e^{\mu (L-x)} \frac{\sinh(\sqrt{2 s + \mu^2} x )}{\sinh(\sqrt{2 s + \mu^2} L )} + e^{- \mu x} \frac{\sinh(\sqrt{2 s + \mu^2} (L-x))}{\sinh(\sqrt{2 s + \mu^2} L )}  \right]  \right\} \\ \nonumber
   &&=  1 -e^{\mu (L-x)} \left[  \frac{\sinh(\mu x)}{\sinh(\mu L)} + \sum_{n=1}^{\infty} \frac{n \pi (-1)^n}{L^2 \left( \frac{\mu^2}{2} + \frac{n^2 \pi^2}{2 L^2} \right)}e^{-( \frac{\mu^2}{2} + \frac{n^2 \pi^2}{2 L^2})\tau} \sin \left(\frac{n \pi x}{L}\right)   \right]  \\ \nonumber
   && ~~~~~ + e^{-\mu x} \left[  \frac{\sinh(\mu (x - L))}{\sinh(\mu L)} + \sum_{n=1}^{\infty} \frac{n \pi (-1)^n}{L^2 \left( \frac{\mu^2}{2} + \frac{n^2 \pi^2}{2 L^2} \right)}e^{-( \frac{\mu^2}{2} + \frac{n^2 \pi^2}{2 L^2})\tau} \sin \left(\frac{n \pi (x - L)}{L}\right)   \right] \\ \nonumber
   && =  e^{-\mu x} \sum_{n=1}^{\infty} \left(1 - e^{\mu L} (-1)^n \right) \frac{2 n \pi }{L^2 \mu^2 +n^2 \pi^2 }e^{-( \frac{\mu^2}{2} + \frac{n^2 \pi^2}{2 L^2})\tau} \sin \left(\frac{n \pi x}{L}\right)
\end{eqnarray}
Keeping only the term with $n = 1$, one gets the long-time asymptotic behavior
\begin{eqnarray}
   \displaystyle {S}^{abs(0,L)}_{\tau} (x) \opsimeq_{\tau \to +\infty} \frac{2 \pi \left(1 + e^{\mu L} \right)}{L^2 \mu^2 + \pi^2 }e^{-( \frac{\mu^2}{2} + \frac{\pi^2}{2 L^2})\tau} e^{-\mu x}  \sin \left(\frac{\pi x}{L}\right)
\end{eqnarray}
And finally, we obtain
\begin{eqnarray}
\mu^{[\Sigma_{\infty}^*]}_{\infty}( x,\Sigma=\Sigma_{\infty}^*) = \mu +  \lim_{\tau \to +\infty} \partial_x    \ln S^{abs(0,L)}_{\tau} (  x) = \frac{\pi}{L} \cot \left(  \frac{\pi x}{L} \right)
\end{eqnarray}
as announced in Eq. \ref{mustarbridgepitinftyregions2taboo}.


\end{document}